\documentclass[prx,twocolumn,amsmath,amssymb,superscriptaddress,floatfix,nofootinbib,aps]{revtex4-2}
\usepackage{bm,psfrag,graphicx,setspace}
\usepackage[T1]{fontenc}
\usepackage[left]{lineno}
\usepackage[dvipsnames]{xcolor}
\usepackage{braket}
\usepackage[draft]{changes}
\usepackage{bbold}
\usepackage{quantikz}
\usepackage{hyperref}
\usepackage{arydshln}
\usepackage{array}
\usepackage{nicematrix}
\usepackage{tcolorbox}
\usepackage{multirow}
\usepackage{amsthm}
\usepackage{enumitem}
\usepackage{tabularx}
\usepackage{float}
\makeatletter
\let\newfloat\newfloat@ltx
\makeatother
\usepackage{lipsum}

\newcommand\blfootnote[1]{%
  \begingroup
  \renewcommand\thefootnote{}\footnote{#1}%
  \addtocounter{footnote}{-1}%
  \endgroup
}

\newtheorem{theorem}{Theorem}
\newtheorem{subtheorem}{Theorem}
\newtheorem{corollary}{Corollary}[theorem]

\newcommand{\subsubsubsection}[1]{{\textbf{#1.}}}

\makeatletter
\newcommand{\customlabel}[2]{%
   \protected@write \@auxout {}{\string \newlabel {#1}{{#2}{\thepage}{#2}{#1}{}} }%
   \hypertarget{#1}{}
}
\makeatother

\definecolor{tealgreen}{HTML}{00856b}
\definecolor{purple}{HTML}{9403fc}

\hypersetup{colorlinks=true, allcolors=tealgreen}
\selectfont
\newcommand*{\map}[1]{\bar{\mathcal{#1}}}
\newcommand{\tr}[1]{\textrm{Tr}\left({#1}\right)}

\begin{document}
\pagenumbering{arabic}
\title{Constant-depth preparation of matrix product states with adaptive quantum circuits}
\author{Kevin C. Smith}
\thanks{These authors contributed equally to this work.}
\affiliation{Brookhaven National Laboratory, Upton, New York 11973, USA}
\affiliation{Yale Quantum Institute, Yale University, New Haven, Connecticut 06520-8263, USA}
\affiliation{Department of Physics, Yale University, New Haven, Connecticut 06511, USA}

\author{Abid Khan}
\thanks{These authors contributed equally to this work.}
%\email{aakhan3@illinois.edu}
\affiliation{Department of Physics, University of Illinois Urbana-Champaign, Urbana, IL, United States 61801}
\affiliation{IQUIST and Institute for Condensed Matter Theory and NCSA Center for Artificial Intelligence Innovation, University of Illinois at Urbana-Champaign, IL 61801, USA}
\affiliation{NASA Ames Research Center, Moffett Field, CA, 94035, USA}

\author{Bryan  K. Clark}
%\email{bkclark@illinois.edu}
\affiliation{Department of Physics, University of Illinois Urbana-Champaign, Urbana, IL, United States 61801}
\affiliation{IQUIST and Institute for Condensed Matter Theory and NCSA Center for Artificial Intelligence Innovation, University of Illinois at Urbana-Champaign, IL 61801, USA}

\author{S.M. Girvin}
\affiliation{Yale Quantum Institute, Yale University, New Haven, Connecticut 06520-8263, USA}
\affiliation{Department of Physics, Yale University, New Haven, Connecticut 06511, USA}

\author{Tzu-Chieh Wei}
\affiliation{C. N. Yang Institute for Theoretical Physics, State University of New York at Stony Brook, Stony Brook, NY 11794-3840, USA}
\affiliation{Department of Physics and Astronomy, State University of New York at Stony Brook, Stony Brook, NY 11794-3800, USA}

\begin{abstract}

 Adaptive quantum circuits, which combine local unitary gates, midcircuit measurements, and feedforward operations, have recently emerged as a promising avenue for efficient state preparation, particularly on near-term quantum devices limited to shallow-depth circuits. Matrix product states (MPS) comprise a significant class of many-body entangled states, efficiently describing the ground states of one-dimensional gapped local Hamiltonians and finding applications in a number of recent quantum algorithms. Recently, it was shown that the AKLT state -- a paradigmatic example of an MPS -- can be exactly prepared with an adaptive quantum circuit of constant-depth, an impossible feat with local unitary gates alone due to its nonzero correlation length [Smith \emph{et al}., PRX Quantum 4, 020315 (2023)]. In this work, we broaden the scope of this approach and demonstrate that a diverse class of MPS can be exactly prepared using constant-depth adaptive quantum circuits, outperforming theoretically optimal preparation with unitary circuits. We show that this class includes short- and long-ranged entangled MPS, symmetry-protected topological (SPT) and symmetry-broken states, MPS with finite Abelian, non-Abelian, and continuous symmetries, resource states for MBQC, and families of states with tunable correlation length. Moreover, we illustrate the utility of our framework for designing constant-depth sampling protocols, such as for random MPS or for generating MPS in a particular SPT phase. We present sufficient conditions for particular MPS to be preparable in constant time, with global on-site symmetry playing a pivotal role. Altogether, this work demonstrates the immense promise of adaptive quantum circuits for efficiently preparing many-body entangled states and provides explicit algorithms that outperform known protocols to prepare an essential class of states.
\end{abstract}

\maketitle

\blfootnote{\href{kcsmith@ibm.com}{Present address: kcsmith@ibm.com}}
\blfootnote{\href{mailto:abid.a.khan@jpmchase.com}{Present address: abid.a.khan@jpmchase.com}}

{\hypersetup{hidelinks}\tableofcontents}
\section{Introduction}

One of the most important problems intersecting many-body physics and quantum computation is the efficient preparation of interesting and useful entangled states on quantum hardware. For instance, state preparation is a key subroutine for many quantum algorithms~\cite{Dalzell2023, araujo2021divide} and an essential first step for a variety of promising applications of quantum hardware, including quantum simulation~\cite{childs2018toward, daley2022practical}, quantum machine learning~\cite{marin2023quantum}, and quantum sensing~\cite{Degen_2017}. Moreover, the efficient preparation of nontrivial entangled states offers a promising pathway for studying exotic phases of matter not realized in nature~\cite{iqbal2024non, xu2024non}. 

However, while available quantum processors have rapidly increased in qubit count in recent years, the faithful preparation of large entangled states remains an outstanding challenge. A central obstacle is the speed with which correlations spread via local unitary gates: at a given circuit depth, only qubits within a causal light cone can possibly be correlated and, consequently, spreading correlations among \emph{many} qubits generically requires a circuit depth that scales with the system size~\cite{Bravyi2006}. This scaling is particularly prohibitive in the current noisy intermediate-scale quantum (NISQ) era as, due to noise and gate imperfections, current quantum processors are limited to fairly shallow circuit depths~\cite{lau2022nisq}, impeding the preparation of large quantum states with even modest entanglement. Consequently, algorithms that can faithfully prepare nontrivial entangled states with constant-depth circuits are highly desirable, particularly for near-term applications.

Given their fundamental and practical relevance, efficient protocols to prepare matrix product states (MPS) are of particular recent interest~\cite{Schoen2005, PerezGarcia2006,Ran2019,Wei2022c, Malz_2024,Smith_2023}. MPS are low-entanglement many-body states that efficiently parameterize the ground states of gapped one-dimensional local Hamiltonians~\cite{hastings2007area, verstraete2006matrix,Cirac_2021}, and have furthermore played an integral role in the classification of quantum phases~\cite{ chen2011classification, Schuch2010}, offering crucial insights into the nature of symmetry protected topological (SPT) order~\cite{chen2013symmetry, haegeman2012order, pollmann2012symmetry}. They can be generated sequentially by linear-depth unitary circuits~\cite{Schoen2005, PerezGarcia2006, Ran2019}; their unitary circuit complexity, therefore, sits between that of quantum states generated by shallow (constant-depth) and deep (exponential-depth) quantum circuits. While the advantages of MPS representations for classical computation are well-established~\cite{schollwock2011density}, their use cases on quantum hardware are in their advent and rapidly growing. Recent developments include MPS-inspired (and, more generally, tensor-network-inspired) approaches to variational quantum algorithms~\cite{haghshenas2022variational, rudolph2023synergistic, FossFeig2021a, khan2023preoptimizing}, time dynamics~\cite{Chertkov2022, Martin2023}, and loading classical data for quantum machine learning~\cite{Martyn2020, Holmes2020, Dilip2022, iaconis2023tensor, Jobst2023}. In addition, certain MPS are resources for measurement-based quantum computation (MBQC)~\cite{Gross2007a, Else2012, Wang2017, Raussendorf_2017, stephen2017computational} and other information processing tasks such as blind quantum computation~\cite{morimae2015ground} and remote state preparation~\cite{liu2014controlled}. To realize the promise of these applications, a crucial ingredient is the efficient preparation of MPS on quantum hardware.

Noted above, it is well known that MPS of constant bond dimension can be \emph{exactly} prepared with a unitary circuit whose depth scales linearly with the number of sites $N$~\cite{Schoen2005, PerezGarcia2006, Ran2019}. Several works have improved upon this $O(N)$ scaling by restricting to the class of so-called \emph{normal} MPS that exhibit only short-range correlations. In particular, by allowing for an error $\epsilon$, it is possible to faithfully prepare translationally-invariant normal MPS using approximate circuit-based and adiabatic schemes, requiring $O(\textrm{log}(N/\epsilon))~$\cite{Malz_2024} and $O(\textrm{polylog}(N/\epsilon))$~\cite{Wei2022c} time, respectively. A pertinent question is then whether it is possible to improve upon these methods. To that end, it was shown in Ref.~\cite{Malz_2024} that the faithful preparation of normal MPS requires a local unitary circuit of minimum depth $\Omega(\log N)$\footnote{More precisely, this lower bound only applies to states with nonzero correlation length, as normal MPS with zero correlation length can be prepared via a constant-depth local unitary circuit.}. Intuitively, this lower bound stems from the fact that normal MPS typically have nonzero correlation length \cite{Lancien2021}, ruling out the possibility for constant-depth preparation with strictly local unitary gates~\cite{Chen2023}. Furthermore, for \emph{non-normal} MPS that exhibit GHZ-like long-range correlations, linear-depth preparation is provably optimal~\cite{Bravyi2006,yu2023learning}. Thus, the preparation of MPS with local unitary circuits generically requires a depth scaling with system size, limiting their utility for near-term applications on quantum hardware.

In parallel to this effort, there has been considerable recent interest in the bolstered capabilities of so-called adaptive or dynamic circuits~\cite{Piroli_2021, Buhrman2023, FossFeig2023, Piroli2024, Baeumer2024}, which augment unitary circuits with non-unitary resources such as mid-circuit measurements and classical feedforward operations -- capabilities that are now supported on several cloud-based quantum computing platforms~\cite{Corcoles_2021, Moses2023}. In particular, it has been shown that adaptive circuits are capable of preparing long-range entanglement and topological order in constant depth~\cite{Raussendorf_2005, Tantivasadakarn2021, Verresen2021, Lu2022, Bravyi2022, Tantivasadakarn_2023a, Li_2023, Tantivasadakarn_2023}, an impossible feat with local unitary evolution alone~\cite{Bravyi2006,yu2023learning}. Perhaps the simplest and most well-known example of such a speed-up is for the preparation of the GHZ state, achievable with a constant-depth adaptive circuit yet requiring either a linear- or log-depth unitary circuit, depending on the connectivity of the device (see, for example, Ref.~\cite{baumer2023efficient}). Separately, it was recently shown that the Affleck-Kennedy-Lieb-Tasaki (AKLT) state can be prepared exactly using a constant-depth adaptive circuit~\cite{Smith_2023}, a feat unachievable with local unitary gates alone due to its nonzero correlation length. Notably, both the GHZ and AKLT states are simple examples of MPS. This raises the question as to whether there is a common mechanism underlying the reduction in time complexity for preparing these states with adaptive circuits, and to what extent generalization to other MPS is possible.

In this work, we address this question and present a unified framework for exactly preparing a diverse class of translationally-invariant\footnote{Here and throughout this work, ``translational-invariance'' refers only to the bulk, as we can prepare states with arbitrary (periodic or open) boundary conditions.} MPS with constant-depth adaptive quantum circuits. Within this framework, we present two explicit protocols: one for preparing so-called normal MPS with short-range correlations and another for non-normal MPS with long-range correlations. In both cases, we consider MPS of arbitrary (but constant) bond dimension and provide sufficient conditions they must satisfy to be preparable in constant time via our scheme. Furthermore, we discuss two special classes of MPS  that guarantee these conditions, including (i) fixed-point MPS with zero correlation length and, separately, (ii) MPS with global on-site symmetry. Regarding this latter class, one of our key results is that, independent of correlation length, all normal MPS with global on-site symmetry can be exactly prepared via our constant-depth scheme if the symmetry manifests as an irreducible representation on the virtual level and is finite (or has a finite subgroup, in the case of continuous symmetries). In addition, we show that any non-normal MPS can be prepared if it can be decomposed into independently preparable normal MPS. Altogether, these results demonstrate that a broad class of MPS can be exactly prepared using constant-depth adaptive circuits. Furthermore, this class includes short- and long-range entangled MPS that cannot be faithfully prepared in constant depth using local unitary circuits alone, thus illustrating the tremendous promise for adaptive quantum circuits for preparing nontrivial entangled states in the near term.

To illustrate the diversity of physically interesting many-body states preparable with our scheme, we provide a variety of representative examples tabulated in Table~\ref{tab:examples}. These include SPT and symmetry-broken states, MPS with discrete Abelian, non-Abelian, and continuous symmetries, parameterized families of MPS with tunable correlation length, and resource states for MBQC. Among these examples, we discuss the preparation of many paradigmatic MPS, including the Majumdar-Ghosh, AKLT, and generalized qudit GHZ states. Moreover, we demonstrate that our framework enables constant-depth sampling of random MPS and, more generally, facilitates the design of sampling protocols for MPS with specific properties, such as from an SPT phase. Finally, in addition to these examples, we provide a flexible method to construct general families of symmetric MPS of any bond dimension that can be prepared in constant depth using our framework. 

Before continuing to our results, we note that several prior works have discussed the preparation of translationally-invariant MPS via quantum circuits that combine unitary gates, measurements, and feedforward. In particular, Ref.~\cite{Piroli_2021} showed that all quantum phases in 1D collapse to the trivial phase if local unitary circuits are supplemented with Local Operations and Classical Communication (LOCC). Furthermore, Ref.~\cite{Gunn2023} studied how restricting to symmetric unitaries, measurements, and feedforward operations alters this classification, finding that phases protected by finite Abelian groups trivialize via a protocol similar to the one introduced in Ref.~\cite{Smith_2023} that we generalize here. However, in both of these works, only the constant-depth preparation of fixed point states (i.e., those with zero correlation length \cite{Cirac_2017}) was demonstrated; preparation of a non-fixed-point MPS in the same phase then requires subsequent application of a $O(\textrm{polylog}(N))$ depth circuit \cite{Gunn2023}. In contrast, here our goal is to generalize the protocol of Ref.~\cite{Smith_2023}, which showed that a non-fixed-point MPS (i.e., one with nonzero correlation length) could be prepared in constant depth. Thus, our attention is on a comparatively broader class of states and not just the fixed-point representatives of quantum phases.

The remainder of this paper is organized as follows: in Section~\ref{sec:ingredients}, we define a number of useful concepts that will be used throughout this work (Section~\ref{ssec:preliminaries}), and follow this with a high-level outline of our framework (Section~\ref{ssec:outline}). We then provide further details on the core algorithmic ingredients that underlie our protocol, including sequential preparation of MPS (Section~\ref{ssec:sequential}), measurement-based fusion (Section~\ref{ssec:fusion}), and operator pushing (Section~\ref{ssec:pushing}). We then discuss the special role of correlation length and global on-site symmetries toward the latter ingredient. In Section~\ref{sec:preparation}, we demonstrate how these ingredients can be combined to prepare certain MPS in constant depth, first for normal MPS with short-range correlations (Section~\ref{ssec:normal}), followed by generalization to non-normal MPS with long-range correlations (Section~\ref{ssec:nonnormal}). We then present a variety of pedagogical examples (Section~\ref{ssec:examples}) that include states with SPT order, resource states for MBQC, families of MPS with tunable correlation length, and MPS with both discrete on-site symmetries (e.g., $\mathbb{Z}_2$, $A_4$, $\mathbb{Z}_4\times\mathbb{Z}_2$) and continuous symmetries (e.g., $SO(3)$, $SU(n)$, $Sp(2n)$). Finally, we show that our scheme also enables the sampling of random MPS and MPS from a specific phase. We then conclude in Section~\ref{sec:conclusion}.

\section{Ingredients for constant-depth preparation}\label{sec:ingredients}

In this section, we provide a brief introduction to MPS and discuss conventions and language used throughout this work. We then present a brief outline of our framework and follow this with an in-depth discussion of its underlying ingredients.

\subsection{Preliminaries}
\label{ssec:preliminaries}
We begin by considering an $N$-site MPS of physical dimension $d$ and constant bond dimension\footnote{Throughout this work, we will focus on MPS with a bond dimension $D$ that is independent of $N$.} $D$,
\begin{equation}
    \ket{\Psi} = \sum_{\vec{m}} \tr{A^{m_1} A^{m_2} \ldots A^{m_{N}}X}\ket{\vec{m}}.
    \label{eq:MPS}
\end{equation}
Here, $m_i = \{0,1,\ldots d-1\}$ indexes the $d$ possible states of a physical spin at the $i$th site and $\vec{m}$ collects all physical indices $\{m_1,m_2,...\ldots m_N\}$. The state $\ket{\Psi}$ is parameterized by the rank-3 tensor $A$, which can equivalently be interpreted as a set of $D\times D$ matrices $\{A^{m}\}_{m=0}^{d-1}$. This latter viewpoint is particularly useful for invoking the valence bond picture for MPS, where each matrix $A^{m}$ encodes the state of a pair of $D$-dimensional virtual spins underlying a single $d$-dimensional physical spin \cite{PerezGarcia2006}. In that vein, it will later prove beneficial to work directly with the linear map 
\begin{equation}
    \map{A} = \sum_{ijm}A^m_{ij}\ket{m}\bra{ij},
    \label{eq:Amap}
\end{equation}
which takes virtual states in $\mathcal{H}_D\otimes\mathcal{H}_D$ to physical states in $\mathcal{H}_d$. Separately, due to the close relationship between MPS and quantum channels~\cite{ruiz2011tensor}, we will often refer to the matrices $A^{m}$ as Kraus operators throughout this work.

In writing the MPS in Eq.~(\ref{eq:MPS}), we have enforced translational invariance in the sense that each site is described by the same tensor $A$, but with arbitrary boundary conditions specified by the matrix $X$. For example, $X=I$ yields an MPS with periodic boundary conditions, while the case of open boundary conditions corresponds to the selection $X=\ket{R}\bra{L}$, with $\ket{L}$ and $(\ket{R})^*$ the state of the left and right virtual edge spins, respectively. Throughout this work, we will implicitly assume all MPS to be translationally invariant up to boundary conditions unless stated otherwise.

Importantly, the matrices $A^m$ define a unique MPS only up to a gauge freedom, i.e., the state is invariant under $A^m\rightarrow V^{-1} A^m V$ up to modified boundary conditions. To fix this redundancy, we adopt the convention of left-canonical form, where the gauge is chosen such that
\begin{equation}
    \sum_m A^{m\dagger} A^m = \mathbb{1}.
    \label{eq:LCF}
\end{equation}
As we will see in the next section, this choice plays an important role in mapping a given MPS to a sequential quantum circuit. We emphasize, however, that similar to Kraus operators for quantum channels, this does not completely fix the gauge as we can still conjugate $A^m$ by a general unitary operator (i.e., $A^m\rightarrow U^{\dagger} A^m U$) while preserving left-canonical form. 

Furthermore, we can always apply a local change of basis on each physical spin via the replacement $A^m \to \sum_{n}U_{mn}A^{n}$ for some unitary $U$. Without loss of generality\footnote{This can always be made true by choosing $U$ to diagonalize the Hermitian matrix $C$ with elements $C_{mn} = \textrm{Tr}(A^{m\dagger} A^n)$.}, we choose this basis such that the Kraus operators $A^m$ are orthogonal under the Hilbert-Schmidt norm, 
\begin{equation}
    \textrm{Tr}(A^{m \dagger}A^{n})\propto \delta_{mn},
    \label{eq:HilbertSchmidt}
\end{equation}
a choice that ensures that $\map{A}$ maps orthogonal virtual states onto orthogonal physical states. Moreover, throughout this work we assume $\map{A}$ to be surjective (i.e., $\textrm{rank}(\map{A})=d\leq D^2$). If this is not the case, one can always discard a local isometry from $\map{A}$ without altering the MPS \cite{Cirac_2017}. This property will later become important in Section~\ref{ssec:pushing}.

In presenting our constant-depth preparation protocol, it will prove beneficial to distinguish between two distinct classes of states: normal MPS and non-normal MPS. The former are the class of \emph{unique} ground states of gapped one-dimensional Hamiltonians with local interactions \cite{PerezGarcia2006, Cirac_2021}, and exhibit short-range, exponentially decaying correlations. In contrast, the latter describes \emph{degenerate} ground states of gapped local Hamiltonians and are thus closely related to phases with symmetry breaking \cite{PerezGarcia2006, Schuch2010}. Unlike their normal counterparts, non-normal MPS can display long-range correlations, with the GHZ state serving as a paradigmatic example. 

% and represent the  All normal MPS are described by normal tensors $A$ such that the 
Whether a given MPS is normal or non-normal is intimately tied to the structure of $A$. We defer to Refs.~\cite{Cirac_2017, Cirac_2021, Acuaviva2022} for a more detailed discussion on this topic but, in brief, normal MPS are characterized by a tensor $A$ which is itself normal -- i.e., by blocking a finite number of sites\footnote{For a generic normal MPS, one needs to block at most $2D^2(6+\log_2(D))$ sites to achieve injectivity \cite{Cirac_2021}.}, the virtual-to-physical map $\map{A}$ becomes injective (or, equivalently, the set of blocked matrices $\{A^m\}$ span the space of $D\times D$ matrices). A non-normal MPS does not have this property, and no amount of blocking will yield an injective map $\map{A}$. More formally, the normality of a given MPS is evidenced by its canonical decomposition \cite{Cirac_2017, Cirac_2021}, where each Kraus operator is expressed in block-diagonal form $A^m = \bigoplus_{\alpha=0}^{K-1} A^m_{\alpha}$. For normal MPS, $A^m$ contains just a single block ($K=1$), while for non-normal MPS, $A^m$ can be decomposed into multiple blocks ($K>1$) that cannot be further reduced. Importantly, this latter property allows us to write any non-normal MPS as a linear superposition of normal MPS \cite{Cirac_2017}, a feature that we will exploit in Section \ref{ssec:nonnormal} to generalize the constant-depth preparation of normal MPS to the non-normal case.

With the above conventions and definitions in place, we are now ready to present a high-level outline of our strategy to prepare MPS in constant depth. Later, in Section~\ref{sec:preparation}, we will provide a more detailed presentation of explicit protocols to prepare both normal and non-normal MPS, complete with conditions for specific states to be preparable via this scheme.

\subsection{Outline of the preparation strategy}
\label{ssec:outline}

\begin{figure}
\includegraphics[width = \linewidth]{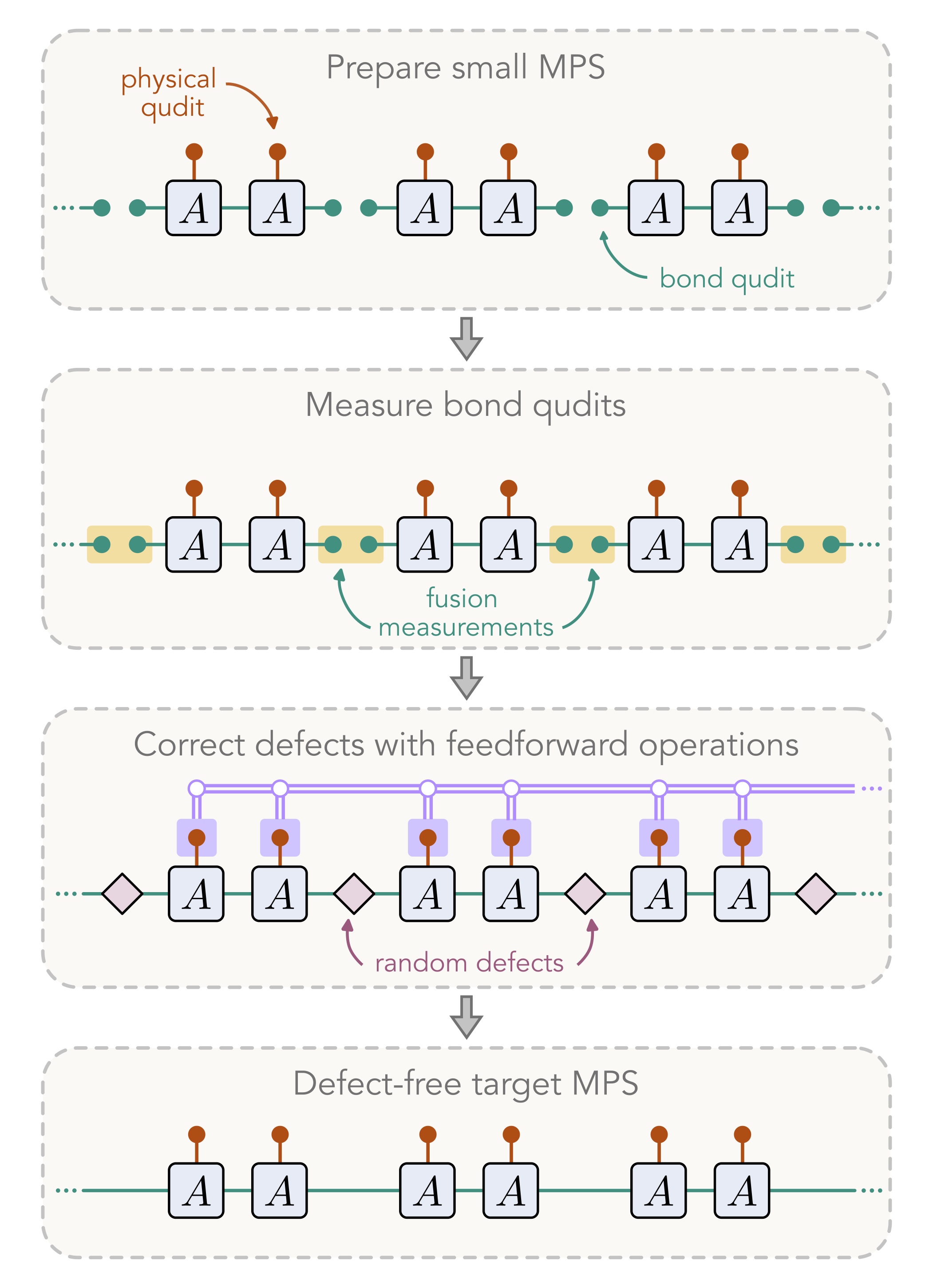}
\caption{Our strategy to prepare a target MPS in constant depth. First, we prepare small MPS using a sequential unitary protocol. Next, we use fusion measurements between edge bond qudits, resulting in random defects at each fusion site. Under certain conditions, these defects can then be deterministically corrected with feedforward operations by leveraging the operator pushing relations of the target state.}
\label{fig:f1}
\end{figure}

Our high-level strategy to prepare MPS in constant depth is illustrated in Fig.~\ref{fig:f1}. It consists of three simple steps:
\begin{enumerate}
\item Prepare multiple small MPS in parallel, using the sequential unitary preparation scheme for each (see Section~\ref{ssec:sequential}).

\item Use mid-circuit \emph{fusion measurements} to merge all independently prepared MPS in parallel, yielding the target state up to random \emph{defects} at each fusion site (see Section~\ref{ssec:fusion}).
\item Correct these defects using feedforward operations and available \emph{operator pushing relations} (see Section~\ref{ssec:pushing}).
\end{enumerate}
In the following sections, we work entirely in terms of qudit resources. However, we note that this is a choice of convenience, and in a practical setting each qudit can be encoded by a logarithmic number of qubits. Crucially, this increases the circuit depth by a constant that depends only on the physical and bond dimensions but not system size, thus preserving our claim of a constant-depth protocol.

Before proceeding, two further comments are warranted. First, we note that our scheme cannot prepare arbitrary MPS, but rather only those with certain properties. This will be discussed further in Section~\ref{sec:preparation}. Furthermore, for MPS where constant-depth preparation is possible, our scheme does not come without trade-offs. In comparison to both the known linear-depth ~\cite{Schoen2005, PerezGarcia2006, Ran2019} and log-depth~\cite{Malz_2024} protocols, we gain a super-exponential advantage in circuit depth at the cost of a constant factor in total qudit count (along with the requirement for mid-circuit measurements and feedforward operations). While this reduction in total spacetime resources is clearly beneficial, the trade-off between temporal and spatial resources may be an important consideration in certain contexts, e.g., for small system sizes or for platforms with limited ancillary resources.

We now continue with an in-depth discussion on the core ingredients of our preparation algorithm, beginning with the sequential preparation of MPS.

\subsection{Sequential preparation}
\label{ssec:sequential}
\begin{figure}
\includegraphics[width = \linewidth]{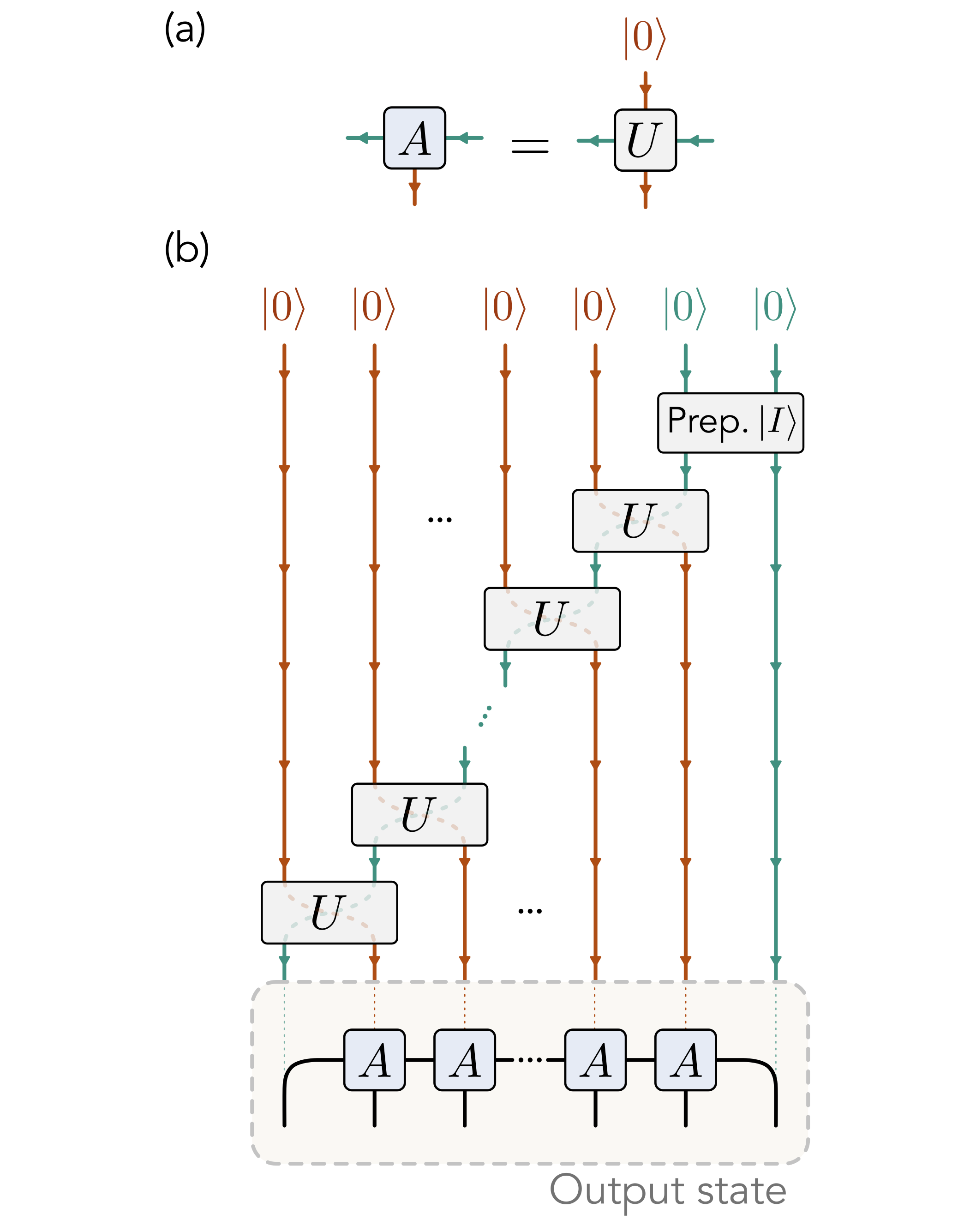}
\caption{Mapping an MPS to a sequential unitary circuit. (a) Leveraging left-canonical form, the tensor $A$ can be embedded in a unitary $U$ acting in a larger Hilbert space. The physical qudit (red) flows vertically, while the bond qudit (green) flows horizontally. (b) Two bond qudits are initially entangled in the Bell state $\ket{I} = \sum_{j}\ket{jj}/\sqrt{D}$. Employing one of these bond qudits, physical sites are prepared via sequential application of $U$. Boundary conditions are determined upon measurement of the bond qudits.}
\label{fig:f2}
\end{figure}
As previously discussed, an arbitrary matrix product state can be prepared using a linear-depth unitary circuit~\cite{Schoen2005, PerezGarcia2006, Ran2019}. This is an important primitive for our constant-depth approach, namely for the initial preparation of small MPS. The key idea is to first identify that, for an MPS in left-canonical form, the map $V_A=\sum_m A^m \otimes \ket{m}$ is an isometry from $\mathcal{H}_D$ to $\mathcal{H}_D\otimes\mathcal{H}_d$. As shown in Fig.~\ref{fig:f2}(a), this enables the definition of the unitary
\begin{equation}
    U = \sum_m A^m\otimes \ket{m}\bra{0} + C_\perp
\end{equation}
via Stinespring dilation.
Intuitively, this unitary prepares a \emph{physical} qudit of dimension $d$ in the state $\ket{m}$ while enacting the Kraus operator $A^m$ on an ancillary subsystem of dimension $D$. We term this ancillary subsystem the \emph{bond} qudit. The operator $C_\perp$ is chosen to ensure the unitarity of the entire operation. By successively acting $U$ on a series of $N$ physical qudits (using the same bond qudit for each), one can prepare an $N$-site MPS via a linear-depth unitary quantum circuit with boundary conditions determined by the initial and final state of the bond qudit. In particular, the right boundary condition is determined by its initial state, while the left boundary condition is \emph{entangled} with the bond qudit at the end of the circuit\footnote{We note that this approach can be modified such that the ancillary bond qudit is unentangled at the end of this sequence, but here we will use the edge-entangled bond qudits to our advantage. See Ref.~\cite{Schoen2005} for more details.}.

In the present work, we adopt a slight modification of this approach that leaves both left and right boundaries each entangled with a bond qudit, a feature that underlies the measurement-based fusion of MPS discussed in Section~\ref{ssec:fusion}. As described in our previous work~\cite{Smith_2023}, the core idea is to preempt the above steps with the initial preparation of a pair of bond qudits in the maximally entangled state $\ket{I} = \sum_j\ket{jj}/\sqrt{D}$. As shown in Fig.~\ref{fig:f2}(b), we can then use one entangled bond qudit to sequentially ``grow'' the MPS via repeated application of $U$, adding a single physical site with each application and ultimately producing the state,
\begin{equation}
    \ket{\Psi'} = \frac{1}{\sqrt{D}}\sum_{ij}\sum_{\vec{m}}\bra{i}A^{m_1} A^{m_2} \ldots A^{m_{N}}\ket{j}\ket{\vec{m}}\otimes\ket{ij},
    \label{eq:mps_ebc}
\end{equation}
corresponding to an $N$-site MPS with the left and right boundaries entangled with the dangling bond qudits indexed by $i$ and $j$, respectively. Throughout the remainder of this work, we will refer to such an MPS as having \emph{entangled boundary conditions}.

Notably, one can use projective measurements to collapse the state onto an MPS with definite boundary conditions: defining $\ket{B} = (1/\sqrt{D})\sum_{ij} B_{ij}^*\ket{ij}$, the act of projecting the dangling qudits of $\ket{\Psi'}$ onto $\ket{B}$ yields the state $\ket{\Psi}\otimes\ket{B}$, where $\ket{\Psi}$ is the MPS in Eq.~(\ref{eq:MPS}) with the replacement $X\rightarrow B^T$. It is important to note that preparing an MPS with a particular boundary matrix is probabilistic with this strategy. However, for normal MPS, this probability scales as $\sim 1/D^2$, requiring on average $\sim D^2$ repetitions to successfully prepare a state with particular boundary conditions\footnote{This assumes that $N$ is large compared to the correlation length $\xi$. For normal MPS, correlations between the $D$-dimensional edge qudits decay as $\sim \exp(-N/\xi)$ (see Ref.~\cite{Smith_2023} for a discussion in the context of the AKLT state). Thus, for $N\gg\xi$, all $D^2$ measurement outcomes are equally likely.}. Finally, while enforcement of periodic boundary conditions (or any other entangling boundary matrix $B$) naively requires either long-range connectivity between the edge qudits or $O(N)$ SWAP gates, we note that this can be remedied by first distributing a qudit Bell pair in $O(1)$ time using measurements and feedforward and, after preparing the MPS, employing gate teleportation to facilitate measurement of the dangling edge qudits in an arbitrary basis.

\subsection{Measurement-based fusion of MPS}
\label{ssec:fusion}

We now discuss the measurement-based fusion of MPS, the second primitive underlying our constant-depth protocol. For illustrative purposes, we specialize to the scenario of fusing two MPS of $n=N/2$ sites each, though we emphasize that our eventual strategy will entail fusing many single- or few-site MPS in parallel.

To illustrate the basic concept, we take two $n$-site copies of an MPS with entangled boundary conditions as in Eq.~(\ref{eq:mps_ebc}), and write the composite (unentangled) MPS pair $\ket{\Phi}=\ket{\Psi'}\otimes\ket{\Psi'}$ as
% \begin{widetext}
\begin{equation}
    \begin{split}
    \ket{\Phi}=&\frac{1}{D}\sum_{ij\ell p}\sum_{\vec{m}}\bra{i}A^{m_1}A^{m_2}\ldots A^{m_n}\ket{j} \\
    &\times\bra{\ell}A^{m_{n+1}}A^{m_{n+2}}\ldots A^{m_N}\ket{p}\ket{\vec{m}}\otimes \ket{ij\ell p}.
    \end{split}
    \end{equation}
% \end{widetext}
Next, we measure the intermediary pair of mutually uncorrelated, dangling bond qudits indexed by $j$ and $\ell$. In particular, we carry out a projective measurement in the basis formed by the states $\ket{B^{k}} = (1/\sqrt{D})\sum_{ij} (B^{k}_{ij})^*\ket{ij}$, where $k = \{0,1,\ldots D^2-1\}$ labels a particular measurement outcome, and $\braket{B^k|B^{k'}} = \delta_{kk'}$ such that the basis is orthonormal. In practice, such a measurement is carried out by first applying the two-qudit unitary $V = \sum_k \ket{k}\bra{B^k}$, followed by two single-qudit measurements in the computational basis $\{\ket{k}\}$. We note that orthogonality between the basis states $\ket{B^k}$ is not strictly necessary. Instead, we only require that the map $V$ is isometric, i.e., $V^\dagger V = \mathbb{1}_{D\times D}$, allowing for a non-orthogonal measurement basis. In this latter case, measurement is facilitated by additional ancillary qubits, corresponding to a positive operator-valued measurement (POVM). We will return to this possibility in Section \ref{sssec:measurement}. For now, we take $V$ to be unitary and the states $\ket{B^k}$ to be mutually orthogonal.

To understand the impact of this measurement and, in particular, its back-action on the rest of the system, it is helpful to rewrite the state $\ket{\Phi}$ in the chosen measurement basis $\{\ket{B^k}\}$,
\begin{widetext}
\begin{equation}
\ket{\Phi}=\frac{1}{D^{3/2}}\sum_k \ket{B^k}\otimes\sum_{ip}\sum_{\vec{m}}\bra{i}A^{m_1}A^{m_2}\ldots A^{m_n}B^kA^{m_{n+1}}A^{m_{n+2}}\ldots A^{m_N}\ket{p}\ket{\vec{m}}\otimes \ket{ip},
\end{equation}
\end{widetext}
where we have rearranged the tensor product ordering to emphasize the structure of the state. Written in this form, it is evident that measuring a particular outcome $k=k_0$ fuses the two unentangled MPS, depositing the corresponding \emph{defect} matrix $B^{k_0}$ at the fusion site (see Fig.~\ref{fig:f3}(a)). This can also be understood graphically (see, for example, Ref.~\cite{bergholm2015diagrammatic}): using the Choi-Jamiolkowski isomorphism, we can define a set of two-qudit projectors 
\begin{equation}
    \begin{split}
        P_k &=(B^{k}\otimes \mathbb{1})^{\dagger}\ket{I}\bra{I}(B^{k}\otimes \mathbb{1}) \\
        & = \ket{B^k}\bra{B^k},
    \end{split}
    \label{eq:projector}
\end{equation}
where $\ket{I}=(1/\sqrt{D})\sum_j\ket{jj}$. Graphically, this projector can be represented as, 
\begin{equation}
\includegraphics[width = 0.85\linewidth]{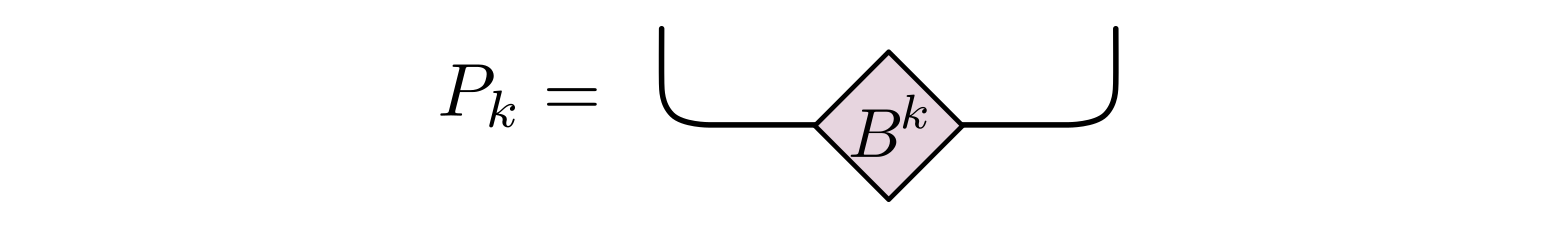},
% \label{eq:VG_to_isometry_nonnormal}
\end{equation}
and may be intuitively thought of as a ``cap'' for the measured dangling bond qudit legs in Fig.~\ref{fig:f3}(a).

In this work, we will be most interested in measurements that project the intermediate bond qudits onto maximally entangled states; in this case, the phenomenon of MPS fusion is easily understood through the principle of entanglement swapping. Furthermore, from Eq.~\eqref{eq:projector}, it is clear that projection onto a maximally entangled state corresponds to a unitary defect matrix $B^k$. This feature will become important in the next section.

\subsection{Operator pushing}
\label{ssec:pushing}

\begin{figure*}
\includegraphics[width = 0.95\linewidth]{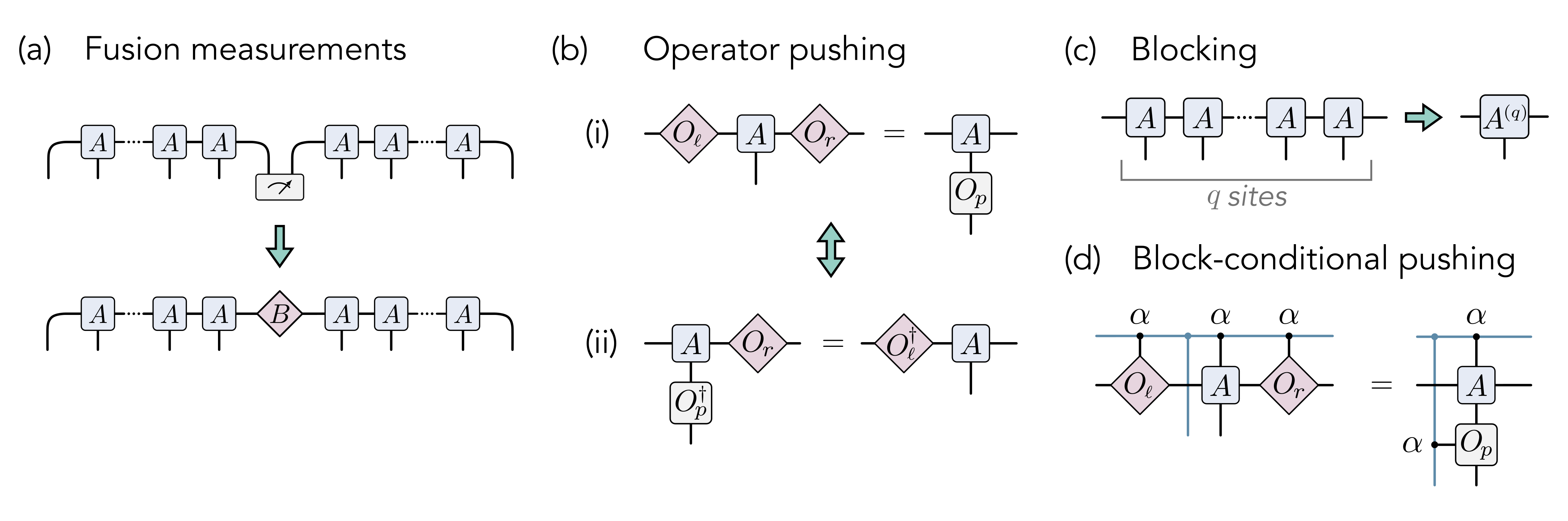}
\caption{Ingredients for constant-depth preparation of MPS. (a) Fusion measurements are employed to merge ``small'' MPS prepared in parallel. Due to the probabilistic nature of measurements, a random defect $B$ is inserted at the fusion site. (b) Operator pushing relations for the tensor $A$. In this work, we define pushing relations as in (i), equating application of the operator $O_p$ on the physical leg to insertion of $O_{\ell}$ and $O_r$ on the left and right virtual legs, respectively. As shown in (ii), we employ these pushing relations to manipulate random defects on the virtual level, ``pushing'' virtual operators from one leg to the other. (c) By blocking $q$ sites of the MPS parameterized by the tensor $A$, we define a new tensor $A^{(q)}$. In general, the pushing relations for $A^{(q)}$ are dependent upon the blocking parameter $q$. (d) To correct defects in non-normal MPS, we decompose such states into a superposition of normal MPS via the block decomposition $A = \bigoplus_{\alpha}A_\alpha$. Pushing relations for $A_{\alpha}$ are independently employed by conditioning $O_p$ on the block index $\alpha$, a notion we term block-conditional operator pushing.}
\label{fig:f3}
\end{figure*}

We now describe the concept of operator pushing, the final primitive for our constant-depth protocol. We emphasize that the idea of operator pushing is not unique to this work, and has previously been applied to construct error correcting codes~\cite{Pastawski_2015, Cao_2022}. Separately, it is closely related to the concept of operational symmetry of entangled states~\cite{Tzitrin2020}. Here, we are interested in using operator pushing to unitarily remove unwanted defects resulting in undesired fusion measurement outcomes. To determine when this is possible, we must first uncover the allowed \emph{pushing relations} derived from the properties of the tensor $A$. 

As shown in Fig.~\ref{fig:f3}(b), operator pushing can be understood through several equivalent angles. In panel (i) we view it as the ability to replicate the action of the operators $O_{\ell}$ and $O_r$ on the left and right virtual legs, respectively, by acting a third operator $O_p$ on the physical leg. As will be discussed in Section~\ref{ssec:symmetry}, this relation bears similarity to the manifestation of global on-site symmetry~\cite{Perez_Garcia_2008, Sanz_2009} in the local tensor $A$. As depicted in panel (ii), this same pushing relation allows us to remove $O_{r}$ from the right virtual leg by acting $O_p^{\dagger}$ on the physical leg and at the expense of applying $O_{\ell}^{\dagger}$ on the left virtual leg. For simplicity, we have additionally assumed $O_r$ and $O_{\ell}$ to be unitary, an assumption that is not strictly necessary but will nonetheless hold for the defects considered in this work. While the relation in panel (ii) foreshadows our eventual strategy of manipulating individual defects on the virtual level, we find that uncovering the allowed pushing relations for a particular tensor $A$ is simpler using the convention of panel (i), and we will therefore favor this viewpoint going forward.\footnote{We also note a third viewpoint, where the virtual and physical operators altogether leave the tensor invariant. This is a symmetry of the local tensor.}

To gain insight into the conditions under which pushing relations exist, it is advantageous to consider the $d\times D^2$ matrix $\map{A}$ that maps from the virtual space $\mathcal{H}_{D}\otimes \mathcal{H}_{D}$ to the physical space $\mathcal{H}_{d}$, defined in Eq.~(\ref{eq:Amap}).

Furthermore, it will prove beneficial to define analogs of the rank-3 tensor $A$ and virtual-to-physical map $\map{A}$ after blocking $q$ sites (see Fig.~\ref{fig:f3}(c)), which we denote by $A^{(q)}$ and $\map{A}^{(q)}$, respectively. In line with the conventions discussed in Section~\ref{ssec:preliminaries}, $\map{A}^{(q)}$ is taken to be a surjective map without loss of generality -- i.e., it is represented by a $d^{(q)}\times D^2$ matrix of rank $d^{(q)} \leq D^2$~\cite{ruiz2011tensor}.

We can then express operator pushing as
\begin{equation}
O_p \mathcal{\bar{A}}^{(q)} = \mathcal{\bar{A}}^{(q)} (O_\ell^T \otimes O_r),
\label{eq:op_pushing}
\end{equation}
where $O_\ell^T$ is the transpose of $O_\ell$, and we have furthermore generalized to the case of pushing through $q$ sites. We now state several formal results related to operator pushing. For simplicity, we drop the superscript $q$, but emphasize that analogous claims hold under the replacement $\map{A}\rightarrow \map{A}^{(q)}$.

\begin{subtheorem}[Existence]\label{thm:existence} For given unitary operators $O_{\ell}$ and $O_r$ acting on the virtual legs of the map $\map{A}$, there exists a corresponding physical operator $O_p$ satisfying the pushing relation Eq.~(\ref{eq:op_pushing}) iff $[O_{\ell}^T~\otimes~O_r,\, \map{A}_R^{-1}\map{A}]~=0$, where $\map{A}_R^{-1}$ is the right-inverse of  $\map{A}$. 
\end{subtheorem}
The proof can be found in Appendix \ref{app:proofs}. As discussed in Section~\ref{ssec:preliminaries}, the map $\map{A}$ is surjective, ensuring the existence of the right inverse $\map{A}_R^{-1}$. However, $\map{A}$ is not left-invertible in general; instead, $\map{A}_R^{-1}\map{A} = \mathcal{P}$ is a projector onto the rowspace of $\map{A}$. As discussed in Appendix~\ref{app:proofs}, the commutator in Theorem~\ref{thm:existence} can be reexpressed as $\mathcal{P}_c (O_{\ell}^T \otimes O_r)\mathcal{P}= 0$, where $\mathcal{P}_c = \mathbb{1}-\mathcal{P}$ is the projector onto the kernel of $\map{A}$. Intuitively, Theorem~\ref{thm:existence} therefore states that the physical operator $O_p$ exists if the virtual operator $O_{\ell}^T \otimes O_r$ does not map elements in the rowspace of $\map{A}$ onto its kernel (i.e., it must map the rowspace onto itself). See Appendix~\ref{app:proofs} for further discussion.

It is instructive to consider the special case where \emph{any} $O_{\ell}$ and $O_r$ can be pushed to the physical level: via Schur's Lemma, this requires that $\map{A}_R^{-1}\map{A} = \mathbb{1}$, i.e., the map $\map{A}$ must be injective such that it is left-invertible. Notably, this can always be achieved for a normal MPS by blocking at most $2D^2(6+\log_2 D)$ sites \cite{Sanz_2010, Cirac_2021} such that the (blocked) physical dimension is equal to the squared bond dimension, $d^{(q)} = D^2$. We emphasize, however, that injectivity of $\map{A}$ is not strictly required for operator pushing, and \emph{particular} operators $O_{\ell}$ and $O_r$ can be pushed through a non-injective map as long as $O_{\ell}^T\otimes O_r$ and the projector $\mathcal{A}_R^{-1}\map{A}$ commute. Intuitively, this requires that the rowspace and kernel of $\map{A}$ are each invariant subspaces of $O_{\ell}^T\otimes O_r$.

While the above theorem provides useful intuition regarding the mere existence (or lack thereof) of $O_p$ given particular virtual operators $O_{\ell}$ and $O_r$, we now narrow our scope to the particular case where $O_p$ is a unitary operator. This will ensure that virtual defect operators can be manipulated deterministically via feedforward operations.

\begin{subtheorem}[Unitarity]\label{thm:unitarity} For given unitary operators $O_{\ell}$ and $O_r$ acting on the virtual legs of the map $\map{A}$, there exists a corresponding unitary physical operator $O_p$ obeying Eq.~(\ref{eq:op_pushing}) iff $[O_{\ell}^T~\otimes~O_r,\, \map{A}^\dagger\map{A}]~=~0$.
\end{subtheorem}
The proof is in Appendix~\ref{app:proofs}. Here, $\map{A}^{\dagger}$ is the conjugate transpose of the $d\times D^2$ matrix $\map{A}$, and $\map{A}^{\dagger}\map{A}$ is therefore a $D^2\times D^2$ matrix. To gain intuition into when the above condition is obeyed, it is helpful to re-express the desired commutation relation in terms of the singular value decomposition $\map{A} = \Sigma V$, where we have without loss of generality chosen the left-hand $d\times d$ unitary to be the identity in accordance with Eq.~(\ref{eq:HilbertSchmidt}). The $j$th row of $V$ encodes the normalized virtual basis state $\ket{v_j} \in \mathcal{H}_D \otimes \mathcal{H}_D$, while the singular value $\Sigma_{i} = \sqrt{\textrm{Tr}( A^{i\dagger}A^i)}$ denotes a relative scaling between $\ket{v_i}$ and the physical basis state $\ket{u_i}\ \in \mathcal{H}_d$ to which it is mapped. 

Using this decomposition, rearrangement of the commutation relation in Theorem~\ref{thm:unitarity} yields
\begin{equation}
[V^\dagger (O_{\ell}^{T}\otimes O_r) V, \,\Sigma^T \Sigma] = 0.
\end{equation}
Above, we found that the existence of $O_p$ requires that $O_\ell$ and $O_r$ leave the rowspace and kernel of $\map{A}$ invariant. In analogy, the above condition further breaks the rowspace into the invariant subspaces $W_k = \text{span}\{\ket{v_i} \mid \Sigma_i = \Sigma_k\}$, where $k$ labels one of the unique singular values. In other words, unitarity of $O_p$ imposes that the subspaces of equal singular value are left invariant by the virtual operation or, equivalently, that the transformed operator $V^\dagger (O_{\ell}^{T}\otimes O_r) V$ is block-diagonal, with block $k$ acting entirely in the subspace $W_k$.

Finally, as it is not immediately obvious, it is worth emphasizing that if $O^T_{\ell}\otimes O_{r}$ commutes with $\map{A}^{\dagger}\map{A}$, it necessarily commutes with the rowspace projector $\map{A}_R^{-1}\map{A}$ when $O_{\ell}$ and $O_r$ are unitary. This implication follows from the above argument concerning invariant subspaces, but we provide an alternative proof in Appendix~\ref{app:proofs}. Thus, the condition supplied by Theorem \ref{thm:unitarity} ensures both the existence and unitarity of the physical operator $O_p$.

We now discuss two particular classes of states for which unitary pushing relations can always be defined: MPS with zero correlation length, and MPS with on-site symmetry. While distinct, both cases lead to MPS characterized by a singular value matrix $\Sigma$ with degenerate values. For the former class, all nonzero singular values are equal to one. For the latter, degenerate singular values characterize invariant subspaces under the action of the symmetry.

\setcounter{subtheorem}{0}

\subsubsection{Special case: Zero correlation length}
\label{ssec:zcl}

We first consider the stringent scenario in which a unitary physical operator $O_p$ exists for \emph{any} choice of unitary operators $O_{\ell}$ and $O_r$. Invoking Schur's Lemma, from Theorem~\ref{thm:unitarity} this requires that $\map{A}$ is unitary, i.e., $\map{A}^{\dagger}\map{A}=\mathbb{1}$ (or, in terms of the singular value matrix, $\Sigma^T\Sigma=\mathbb{1}$). In this case, we can, without loss of generality, choose a physical basis such that $\map{A} = \mathbb{1}$, allowing us to identify the corresponding MPS as a tensor product of $D$-dimensional generalized Bell pairs up to local unitary transformations. Such states correspond to fixed points of the renormalization procedure introduced in Ref.~\cite{Verstraete2005} and, up to blocking, are equivalent to the class of normal MPS with zero correlation length (ZCL\footnote{Here and throughout this work, we adopt the convention of Ref.~\cite{Cirac_2017} that an MPS in canonical form has zero correlation length if and only if $\mathbb{E}^2 = \mathbb{E}$, where $\mathbb{E}$ is the transfer matrix. For a normal MPS, the transfer matrix has a single nonzero eigenvalue of magnitude one.}) \cite{Cirac_2017}.

Though unitarity of $\map{A}$ necessarily implies injectivity and, by extension, that the MPS is normal, an analogous notion can be extended to non-normal MPS. In that case, the renormalization fixed points correspond to GHZ-like superpositions of normal MPS, each with ZCL -- a structure that can be exploited to intuit pushing relations for non-normal fixed points. In brief, each Kraus operator is expressed in a block-diagonal form, $A^m=\bigoplus_{\alpha} A_{\alpha}^m$, where each set $\{A_{\alpha}^m\}_{m=0}^{d-1}$ parameterizes a normal MPS with ZCL; in isolation, each is therefore endowed with pushing relations for arbitrary virtual operators. In terms of the non-normal tensor $A$, this implies that pushing relations can be defined for arbitrary $O_\ell$ and $O_r$ as long as (i) the virtual operators have the same block-diagonal structure as $\{A^m\}$ and (ii) one can condition the physical operator on the block $\alpha$. This is the core idea behind our strategy for correcting defects in non-normal states -- first decompose into normal MPS and then leverage the pushing relations for each by controlling physical operators on the block index $\alpha$ -- see Fig.~\ref{fig:f3}(d).

In the above, we have been somewhat imprecise with the relationship between operator pushing for ZCL states and the requirements for blocking, and a few comments are in order. First, for an MPS with ZCL, it is only after blocking a number of sites $q$ greater than the so-called injectivity length $i(A)$ \cite{Cirac_2021} that the tensor becomes injective and pushing relations can be defined for arbitrary $O_{\ell}$ and $O_r$. In terms of the unblocked sites, this implies that the corresponding physical operator $O_p$ has support on $q$ sites. For example, the cluster state has ZCL, but is injective only after blocking $q=2$ sites. Consequently, arbitrary virtual operators can be pushed through \emph{pairs} of sites via a two-qubit physical unitary, but cannot be pushed through a \emph{single} site. In general, $i(A)$ depends only on the bond dimension $D$ \cite{Sanz_2010, Cirac_2021} and, as a result, the blocked physical operator $O_p$ has support on a constant number of qubits (i.e., independent of $N$). This is an important consideration for our algorithm, as the promise of constant depth requires that all operations ($O_p$ included) have strictly finite support.

Finally, drawing upon our discussion following Theorem \ref{thm:existence}, it is interesting to note that both the injectivity length $i(A)$ and correlation length $\xi$ define important length scales for operator pushing through normal MPS. Roughly speaking, the former sets the physical length scale at which $O_p$ is guaranteed to \emph{exist}. In contrast, $\xi$ sets the length scale at which it is \emph{unitary}, as 
states with nonzero correlation length flow to ones with ZCL after blocking $q\gg \xi$ sites. This suggests a tempting strategy to prepare non-fixed-point MPS: first prepare and fuse MPS of $q\gg \xi$ sites, and subsequently correct all defects using the arbitrary pushing relations of the fixed-point MPS with ZCL. 
However, there is an important subtlety that spoils this strategy from enabling constant-depth preparation -- the fixed point approximates a non-fixed-point MPS to within error $\epsilon$ only after blocking $q = O(\log(N/\epsilon))$ sites~\cite{Piroli_2021, Malz_2024}. To bound the state preparation error, the proposed strategy would therefore require the initial preparation of ``small' MPS of length $O(\log(N/\epsilon))$, spoiling our goal for constant-depth preparation. Consequently, we seek additional structure that enables local operator pushing for states with nonzero correlation length \emph{without} relying on approximation by fixed point states.

\subsubsection{Special case: On-site symmetry}
\label{ssec:symmetry}
We now consider a special class of states that provide such structure: translationally-invariant MPS with global on-site symmetry. As will be shown, such states are endowed with unitary pushing relations without the stringent condition of ZCL. Consequently, we will show that this class provides numerous examples of MPS that can be prepared exactly with a constant-depth adaptive circuit, but otherwise require a log-depth unitary circuit for approximate preparation \cite{Malz_2024}.

\subsubsubsection{Normal MPS} 
We first consider normal MPS symmetric under a group $G$, i.e., there exists some set of unitary operators $U_g$ for $g\in G$ such that, when applied to every physical qudit, the state is invariant up to modified boundary conditions. From the fundamental theorem of MPS \cite{Cirac_2021}, this symmetry manifests in the local tensor $A$ through the relation,
\begin{equation}
    \sum_{n}(U_g)_{mn} A^{n} = e^{i\phi_g}V_g A^m V_g^{\dagger},
    \label{eq:normalsymmetry}
\end{equation}
where the virtual operators $V_g$ form a projective representation of $G$ and the phases $e^{i\phi_g}$ form a 1D irreducible unitary representation of $G$. In terms of the map $\map{A}$, this relation reads $e^{-i\phi_g} U_g\map{A} = \map{A}(V_g^T\otimes V_g^{\dagger})$. Via the choice $O_{\ell}=V_g$, $O_r = V_g^{\dagger}$, and $O_p = e^{-i\phi_g} U_g$,  the condition of Theorem \ref{thm:unitarity} is clearly satisfied, resulting in pushing relations of the form,
\begin{equation}
\includegraphics[width = 0.85\linewidth]{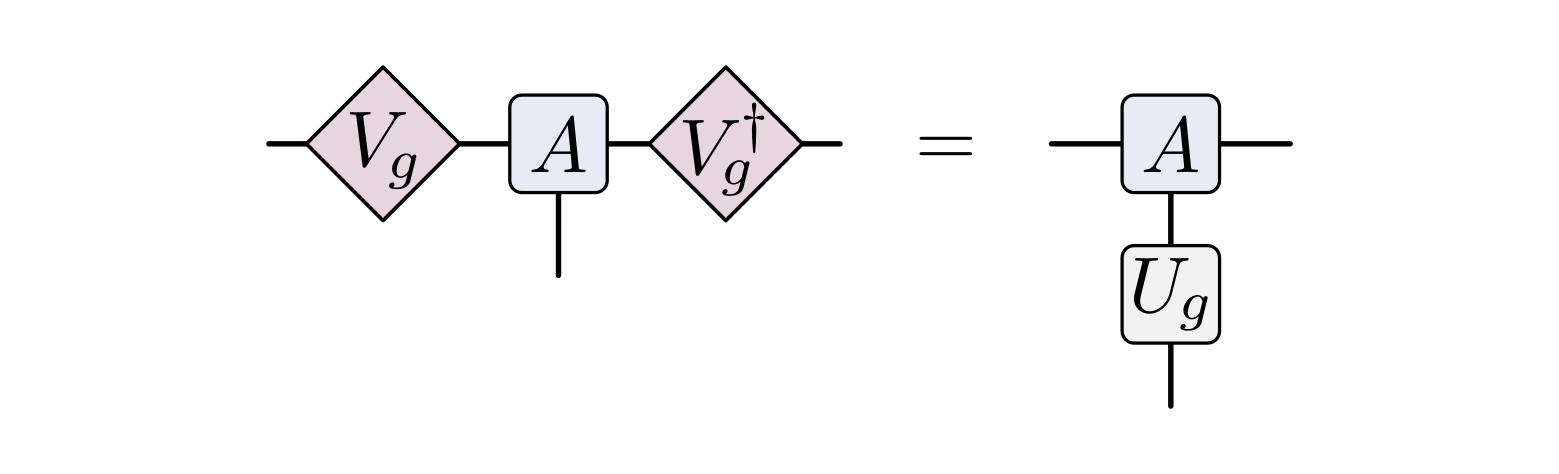}
\label{eq:normal_symmetry}.
\end{equation}
On-site symmetry under the group $G$  therefore guarantees the ability to push projective representations of $G$ through the tensor $A$ via unitary operations on the physical leg.

\subsubsubsection{Non-normal MPS}
For non-normal MPS, the manifestation of symmetry at the virtual level is more complicated. Here, we aim to outline the basic principle, and refer to Refs.~\cite{Schuch2010, Gunn2023} for a more complete discussion. We first recall that any non-normal MPS can be expressed as a superposition of normal MPS, which we index by $\alpha$. This results in a block-decomposition of the form $A^m=\bigoplus_{\alpha=0}^{K-1} A^m_{\alpha}$, where $K$ labels the total 
number of blocks. Assuming periodic boundary conditions, this non-normal MPS physically describes the $K$-fold degenerate gapped ground space of some local parent Hamiltonian \cite{Cirac_2021}. Let us group these blocks into superblocks labeled by $a$. Within each superblock, the symmetry $G$ acts independently, allowing us to further decompose the tensor as \cite{Schuch2010}
\begin{equation}
    A^m=\bigoplus_a^{n_a} \bigoplus_{\alpha\in a} ^{K_a}A^m_\alpha,
\end{equation} 
where $K = \sum_a^{n_a} K_a$. The idea behind this decomposition is that the $K$-fold degeneracy can include both accidental degeneracies (corresponding to distinct superblocks $a$) and degeneracies due to symmetry breaking of $G$ (subdividing superblock $a$ into subblocks $\alpha\in a$). While the symmetry acts independently on each superblock, symmetric operations can permute distinct symmetry-broken states, making the virtual action of the symmetry on the subblocks more complicated.

As illustrated in Fig.~\ref{fig:f3}(d), our strategy to correct defects in non-normal MPS relies on their decomposition into normal MPS, each with pushing relations that can be \emph{independently} applied via block-conditioned physical operations. While the pushing relations inherited from on-site symmetry are naturally decoupled between distinct superblocks, this is not the case among subblocks due to the permutation action. Thus, preempting our protocol to efficiently prepare non-normal MPS, it will prove beneficial to repackage these symmetries into the form of Fig.~\ref{fig:f3}(d).

To achieve this, we simplify to the scenario of a single superblock ($n_a = 1$). Extension to multiple superblocks is then trivial, as the symmetry acts independently on each. As discussed in Refs.~\cite{Schuch2010, Gunn2023}, this situation yields a generalization of Eq.~(\ref{eq:normalsymmetry}) for non-normal MPS:
\begin{equation}
    \sum_{n}(U_g)_{mn} A^{n} = P_g\left[\bigoplus_{\alpha=0}^{K-1} e^{i\phi_{g}^{\alpha}}V_{h(g,\alpha)} A_{\alpha}^m V^{\dagger}_{h(g,\alpha)}\right ](P_{g})^T.
    \label{eq:nonnormalsymmetry}
\end{equation}
Here, the symmetry acts on the virtual level through an interplay of two effects. First, within each block $\alpha$, the matrices $A_{\alpha}^m$ are conjugated by unitary operators $V_{h(g,\alpha)}$, just as in the normal case. These operators form a projective representation of the subgroup $H\leq G$ corresponding to the portion of the full symmetry group $G$ that is unbroken, and therefore act on each block independently. Second, the blocks are permuted through conjugation by $P_g$, operators that form a permutation representation of $G$. The precise relationship between $G$, $H$, and the permutation action can be formalized in the language of induced representations, with $h(g,\alpha)$ uniquely defined by $g$ and $\alpha$ --  we defer to Refs. \cite{Schuch2010, Gunn2023} for details. For our purposes here, it will suffice to highlight a few key features that will play an integral role in the preparation of symmetry-broken states.

First, we emphasize that Eq.~{(\ref{eq:nonnormalsymmetry})} specifies a set of $|G|$ unique pushing relations, diagrammatically expressed as
\begin{equation}
\includegraphics[width = 0.85\linewidth]{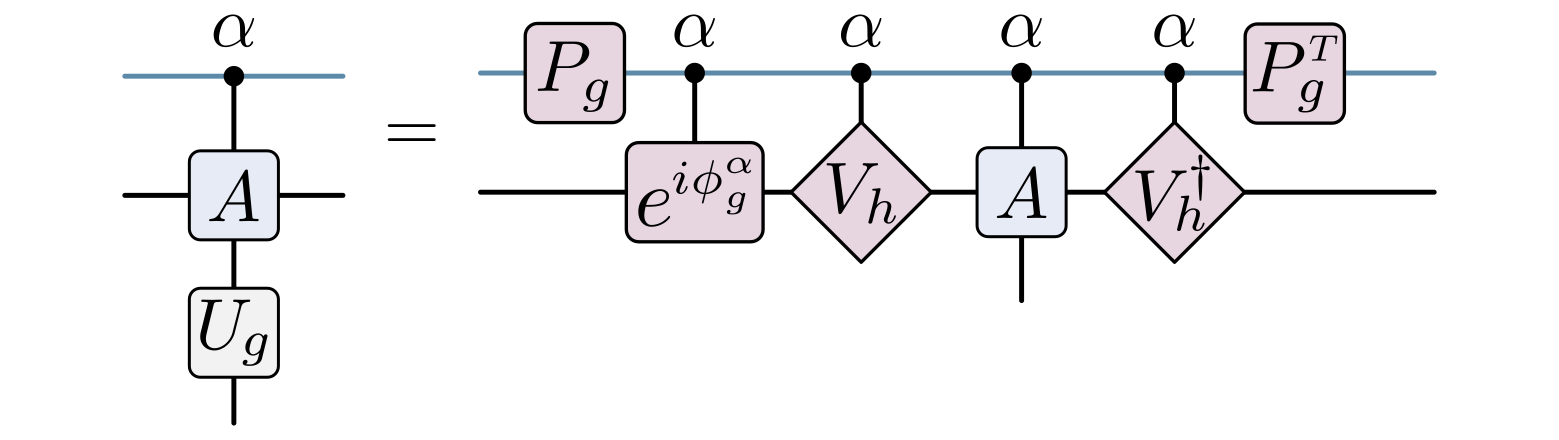},
\label{eq:nonnormalpushing}
\end{equation}
where we have partitioned the bond index into a block index (blue) and an intra-block index (black), and have furthermore adopted ``$\alpha$-controlled'' lines to indicate block-diagonal tensors (i.e., $\bigoplus_{\alpha}A_{\alpha}^m$ and  $\bigoplus_{\alpha}V_{h(g,\alpha)}$). Up to a diagonal phase matrix on the block index, the above can be identified as a pushing relation of the form in Fig.~\ref{fig:f3}(b), with virtual operators $O_{\ell} = V_g$ and $O_r = V_g^{\dagger}$, where $V_g = (P_g\otimes \mathbb{1} )(\bigoplus_{\alpha} V_{h(g,\alpha)})$.

To repackage this relation into the form of Fig.~\ref{fig:f3}(d), we first ``lift'' the block index to the physical space,
\begin{equation}
\includegraphics[width = 0.85\linewidth]{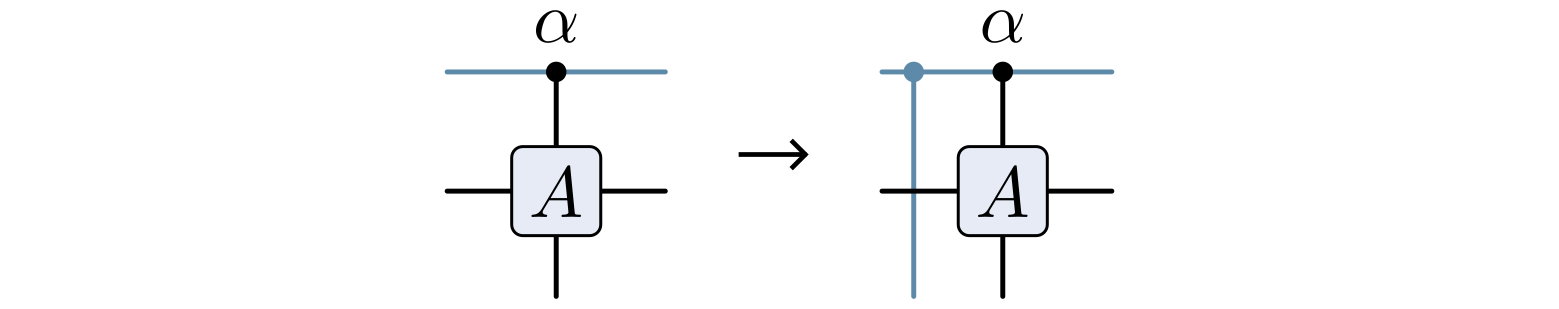}.
\end{equation}
For each $h \in H$ and $\alpha \in K$, we then identify the physical unitary $U_{g(h,\alpha)}$ that obeys Eq.~(\ref{eq:nonnormalpushing}), but with the permutation action on block $\alpha$ trivial (i.e., block $\alpha$ is mapped to itself). As described in Appendix~\ref{app:block-controlled}, we can then conditionally apply $U_{g(h,\alpha)}$ to each block individually, thus establishing for each $h\in H$ block-conditional pushing relations of the form
\begin{equation}
\includegraphics[width = 0.85\linewidth]{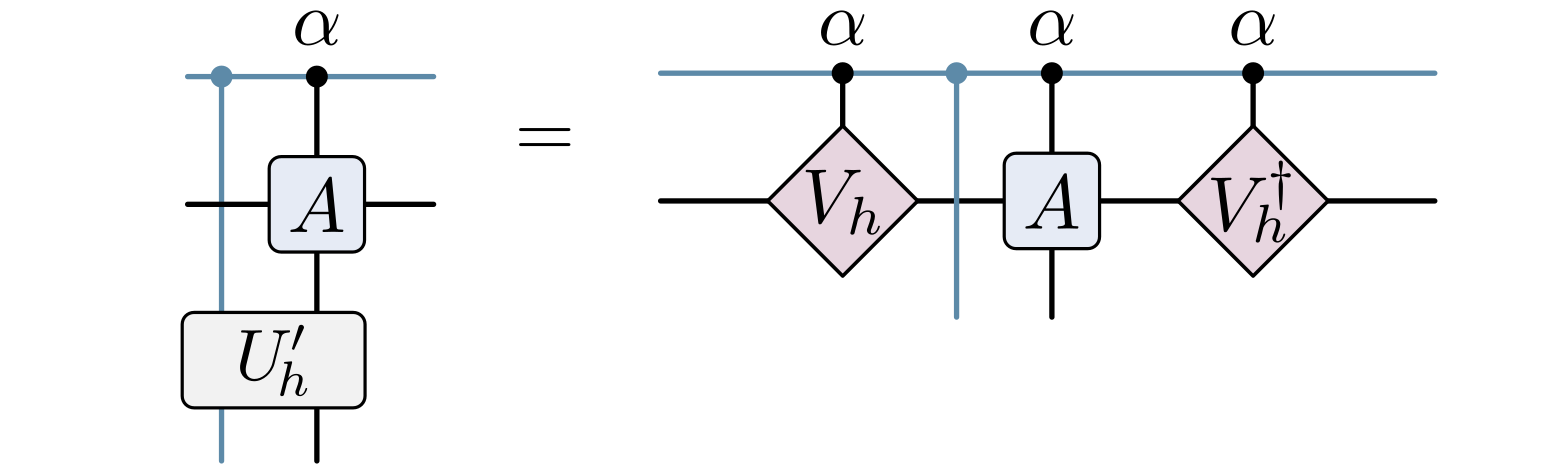}
\label{eq:nonnormalpushing2},
\end{equation}
where the modified physical unitary $U_h'$ is defined as
\begin{equation}
\includegraphics[width = 0.85\linewidth]{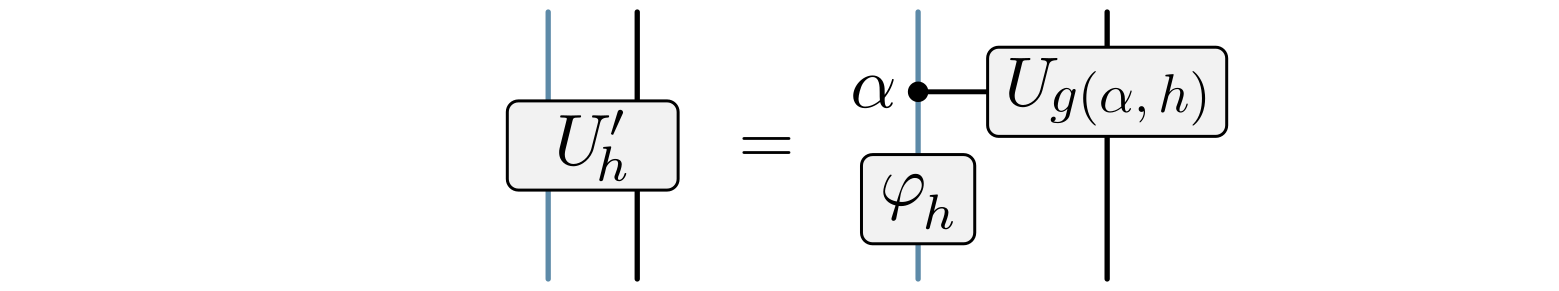}.
\end{equation}
Here, we have adopted a shorthand for the phase matrix $\varphi_h = \bigoplus_\alpha e^{-i\phi_g^{\alpha}}$, with $g=g(h,\alpha)$. For more details on the map $g(h, \alpha)$, we refer to Appendix~\ref{app:block-controlled}. Here, we simply note that for Abelian groups, $g(h, \alpha) = h$, significantly simplifying the pushing relation in Eq.~(\ref{eq:nonnormalpushing2}), as the conditioned $U_{g(\alpha,h)}$ operator can be replaced by an unconditional application of $U_h$.

With Eq.~\eqref{eq:nonnormalpushing2} in hand, we have demonstrated that the $|G|$ pushing relations characteristic of symmetry-broken, non-normal MPS can be reformulated as $|H|$ block-controlled pushing relations of the form in Fig.~\ref{fig:f3}(d), up to an overall phase matrix $\varphi_h$. Here, $H\leq G$ is the unbroken symmetry that is preserved by each block independently. Crucially, this allows us to simplify the complex task of preparing non-normal MPS to the comparatively simpler one of preparing normal MPS, as the former class can be decomposed into superpositions of the latter. In Section~\ref{ssec:nonnormal}, this will allow us to state sufficient conditions for the constant-depth preparation of non-normal MPS. Importantly, these conditions will depend only on the pushing relations of the composing normal MPS, agnostic of the physical mechanism to which they are attributed at the level of the full non-normal MPS.

\section{Constant-depth preparation of MPS}\label{sec:preparation}

We now demonstrate how the above ingredients together enable the constant-depth preparation of translationally-invariant MPS with arbitrary boundary conditions. We begin with the case of normal MPS, and then extrapolate to non-normal MPS. Importantly, we emphasize that this scheme cannot deterministically prepare arbitrary MPS, but is limited to MPS with pushing relations that exhibit certain properties. Below, we discuss the sufficient conditions for a particular MPS to be deterministically preparable with our scheme, and furthermore present several theorems pertaining to important classes of MPS.

Before proceeding, it is worth clarifying that by \emph{deterministic}, here and throughout the remainder of this work we mean that the state
\begin{equation}
    \ket{\Psi} = \sum_{\ell r}\sum_{\vec{m}}\bra{\ell}A^{m_1} A^{m_2} \ldots A^{m_{N}}\ket{r}\ket{\vec{m}}\otimes\ket{\ell r},
    \label{eq:mps_target}
\end{equation}
i.e., with boundary conditions entangled with ancillary qudits, can be prepared exactly and deterministically with a constant-depth adaptive circuit; this state is equivalent to the output of the linear-depth sequential scheme in Section~\ref{ssec:sequential} and, as previously discussed, can be converted into an MPS with particular (open or periodic) boundary conditions via measurement, albeit with a probability that scales as $O(1/D^2)$ in the large $N$ limit\footnote{Note that for the particular case of open boundary conditions, this scaling can be improved to $O(1/D)$ by choosing a definite right boundary condition at the outset (see Appendix~\ref{app:postselection} for details). The left boundary condition is then enforced through measurement as usual.}. One may view this as the deterministic preparation of a ground state of the corresponding parent Hamiltonian, but a random sampling from the degenerate ground state space -- for example, see Ref.~\cite{Smith_2023} for a discussion in terms of the AKLT state. Furthermore, the important feature is that this probability is independent of $N$, and thus on average adds a constant sampling overhead.

\subsection{Normal MPS}
\label{ssec:normal}
\begin{figure*}
\includegraphics[width = 0.95\linewidth]{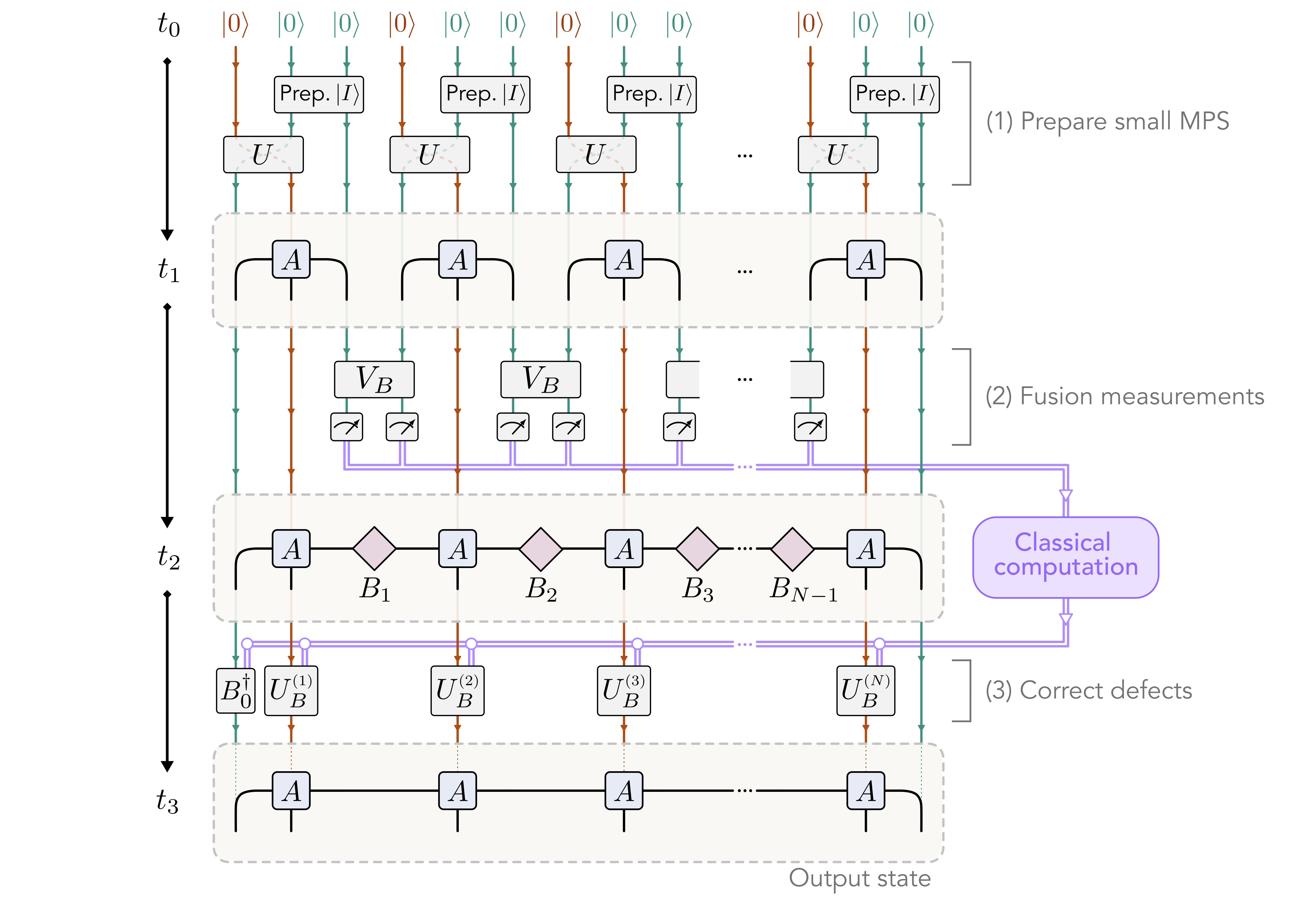}
\caption{Constant-depth preparation of normal MPS using Protocol~\ref{alg:normal}. First, prepare many small MPS in parallel using the sequential unitary protocol outlined in Section~\ref{ssec:sequential}. In the above, we illustrate the preparation of many single-site MPS for simplicity, but note that ``small'' more generally refers to a $q$-site MPS, with $q$ a constant determined by the pushing relations of the target state. Next, carry out fusion measurements in the basis defined by $V_B$, corresponding either to a projective measurement if $V_B$ is unitary (as shown above) or more generally to an ancilla-aided POVM if $V_B$ is an isometry (see Eq.~\eqref{eq:povm}), broadening the set of MPS preparable with this scheme. In either case, measurement yields the desired target state up to a random defect $B_i$ at each fusion site. Finally, leveraging knowledge of measurement outcomes and available pushing relations, correct defects by applying classically-conditioned unitaries $U_B^{(i)}$ at each site $i$ and $B_0^{\dagger}$ at the edge, each dependent upon on all measurement outcomes at sites $j\geq i$ (see Eq.~\eqref{eq:combined_U}). Up to the enforcement of boundary conditions, this constant-depth adaptive quantum circuit exactly and deterministically prepares any target MPS that satisfies the conditions of Theorem~\ref{thm:normal}.}
\label{fig:f4}
\end{figure*}

Combining the previously described ingredients in Section~\ref{sec:ingredients}, the full preparation protocol for preparing normal MPS is shown in Fig.~\ref{fig:f4}. The step-by-step procedure is as follows.

\subsubsection*{Protocol 1: Normal MPS}\customlabel{alg:normal}{1}
\begin{enumerate}[label={(\arabic*)}, wide, labelindent=0pt]
    \item Prepare $n$ $q$-site copies of the target MPS with entangled boundary conditions in parallel, using the sequential unitary preparation for each (Section~\ref{ssec:sequential}). 
    
    \item Employ fusion measurements on all non-edge pairs of bond qudits in an entangling basis defined by $V_B$, in parallel. Through entanglement swapping, this produces the target MPS (with $N=nq$ sites), up to random defects $B_i$ at each fusion site indexed by $i$ (Section~\ref{ssec:fusion}).
    
    \item Combining available unitary pushing relations of the target state and knowledge of the measurement outcomes, remove all defects in parallel using feedforward operations of the form $U_B^{(i)}$ and $B^{0\dagger}$ at each site $i\in \{0,1,\ldots,  N-1\}$ and at the edge, respectively (Section~\ref{ssec:pushing}).

    \item Measure the remaining edge bond qudits to collapse the defect-free MPS onto definite boundary conditions, with each particular outcome having probability $\sim 1/D^2$ (see relevant discussion in Section~\ref{ssec:sequential}).
\end{enumerate}

While this protocol is at face value exceptionally simple, much of the complexity is hidden in the details of each step -- i.e., the choice of $q$, the choice of measurement basis (governed by $V_B$) and, most importantly, when these choices lead to deterministic preparation of the target state. Furthermore, the ``correct'' choices are inextricably informed by the pushing relations of the target state and, as such, should be considered on a case-by-case basis. Nonetheless, general statements concerning certain classes of MPS can be made. To that end, we first extrapolate on the close relationship between the map $V_B$ and the set of defects $\{B^k\}$.

\subsubsection{Constraining the defects $\{B^k\}$}
\label{sssec:measurement}
Closely intertwined with the ability (or lack thereof) to correct the post-measurement state is the choice of measurement basis (or, equivalently, the map $V_B$), which is in one-to-one correspondence with the set of random post-measurement defects $\{B^k\}$, where $k$ labels the possible measurement outcomes (see Section \ref{ssec:fusion}). The choice of measurement basis is therefore equivalent to the choice of defects $\{B^k\}$. For example, in our prior work on preparing the AKLT state~\cite{Smith_2023}, these defects correspond to the 2$\times$2 Pauli operators $\{I, X, Y, Z\}$, and $V_B$ to a map from the Bell basis to the two-qubit computational basis. More generally, we momentarily put aside the question of whether these defects are correctable for a particular MPS, and consider the required properties for a valid basis. To facilitate this discussion, let us take $\eta$ to denote the total number of unique defect types.

First, this measurement basis must define a valid positive operator-valued measurement (POVM). Defining the map, 
\begin{equation}
    V_B = \frac{D}{\sqrt{\eta}} \sum_{k=0}^{\eta} \ket{k}\bra{B^k},
\end{equation}
this imposes the condition $V_B^\dagger V_B = \mathbb{1}$. This is naturally satisfied if $V_B$ is unitary (in which case $\eta = D^2$), but more generally requires that $V_B$ is isometric (corresponding to $\eta \geq D^2$). This latter case can be accounted for in Fig.~\ref{fig:f4} with the replacement
\begin{equation}\label{eq:povm}
\includegraphics[width = 0.85\linewidth]{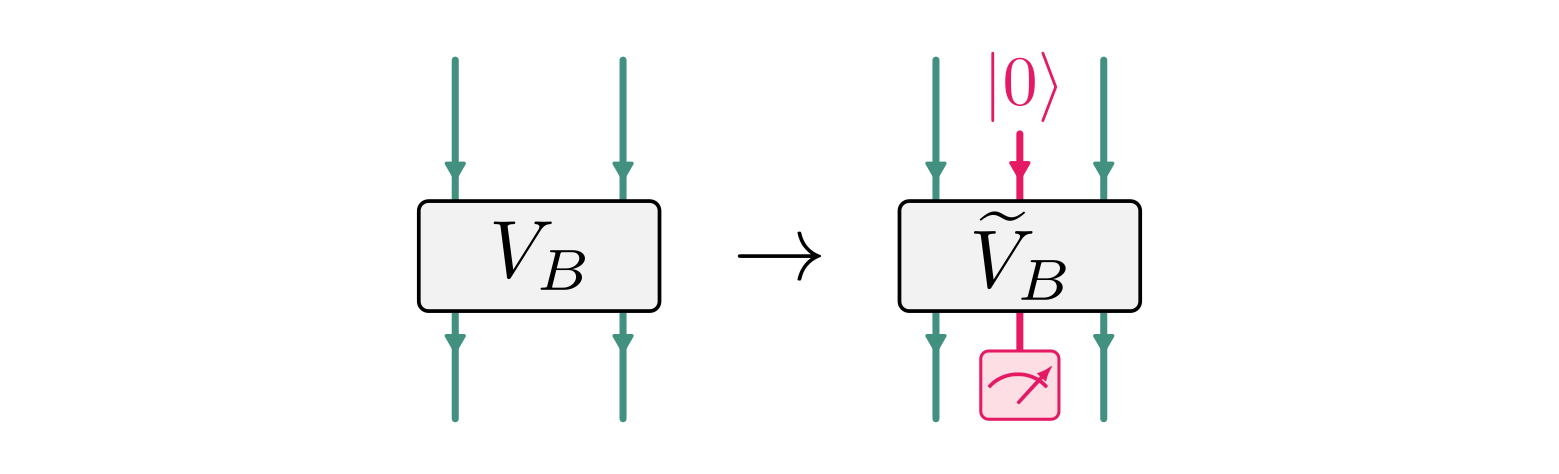},
\end{equation}
where the additional wire represents a $p$-dimensional ancillary qudit\footnote{In particular, this is an ancillary qudit of dimension $p = \textrm{lcm}(\eta,D^2)/D^2$.} that promotes the isometry $V_B$ to a unitary $\widetilde{V}_B = V_B\otimes \bra{0} + C_{\perp}$ in an enlarged Hilbert space, where $C_{\perp}$ is a unitary completion operator. In terms of the defects $\{B^k\}$, this condition implies that they form a (possibly over-complete) basis for the $D\times D$ complex matrices, and requires that the condition
\begin{equation}
\sum_k(B_{ij}^k)^*(B_{\ell m}^k) = \frac{\eta}{D}\delta_{i\ell}\delta_{j m}
\label{eq:isometry_condition}
\end{equation}
is satisfied, where we have recalled the correspondence $\ket{B^k}=(1/\sqrt{D})\sum_{ij}(B_{ij}^k)^*\ket{ij}$.

% $\ket{B^k}=\sqrt{D}\sum_{ij}(B_{ij}^k)^*\ket{ij}$.

Second, we choose the defects $\{B^k\}$ to be unitary, a prerequisite for the existence of unitary pushing relations. Interestingly, we note that imposing that the defect basis $\{B^k\}$ is unitary is equivalent to the condition that all measurement basis states $\ket{B^k}$ are bipartite maximally entangled \cite{Werner_2001}. Thus, we restrict our possible bases to those of maximally entangled states.

Finally, it will prove beneficial to endow the defects with a group structure such that the set $\{B^k\}$ includes the ``defect free'' identity matrix and is closed under multiplication. In particular, we consider bases corresponding to defects that form a projective representation of a finite group $G$, i.e., one that obeys
\begin{equation}
B^g B^h = \omega(g,h)B^{gh} \quad\textrm{for 
 all $g,h\in G$},
\end{equation}
where $\omega(g,h)$ is a phase. We note that it is always possible to choose $B^g$ to be unitary. Further restricting this representation to be irreducible guarantees that Eq.~(\ref{eq:isometry_condition}) is satisfied, as it is exactly equivalent to the grand orthogonality relation with $\eta=|G|$~\cite{Melnikov2022}. Therefore, irreducible projective representations of finite groups provide an ideal defect (and measurement) basis, satisfying all of the above requirements.
We specialize to such bases through the remainder of this work.

As a final general comment on the properties of the measurement basis, we note that for $V_B$ unitary, the defects are orthogonal under the Hilbert-Schmidt norm, i.e., $\tr{B^{k\dagger} B^{k'}}= D\delta_{kk'}$. In this case, our conditions simplify to those of ``nice error bases'', originally introduced in the context of error correction and shown to be equivalent to unitary irreducible projective representations of the group $G$ with dimension $D = |G|^{1/2}$ \cite{knill1996group, knill1996non, klappenecker2003unitary}. In this work, we will explicitly make use of several examples of such bases, including the 2$\times$2 Pauli matrices (which form a projective representation of $\mathbb{Z}_2\times\mathbb{Z}_2$) and their $D\times D$ analogs, the qudit Pauli matrices. The latter are generated by the clock and shift matrices,
\begin{equation}
    \begin{split}
        X&=\sum_{j=0}^{D-1}\ket{j+1 \textrm{ mod $D-1$}}\bra{j} \\ 
        Z&=\sum_{j=0}^{D-1}e^{2\pi i j/D}\ket{j}\bra{j},
    \end{split}
    \label{eq:clockshift}
\end{equation}
and furthermore form a projective representation of $\mathbb{Z}_D\times\mathbb{Z}_D$. However, we emphasize that for the more general case where $V_B$ is merely isometric (and not unitary), the defects will not form an orthogonal basis, and, as a result, our conditions extend beyond nice error bases. While still parameterized by irreducible projective representations of a finite group $G$ (as discussed above), this relaxed condition allows us to decouple the bond dimension $D$ and the group order $|G|$, allowing more generally for POVM-defining bases of $D\times D$ unitary matrices where $D \leq |G|^{1/2}$. In Section~\ref{sssec:A4}, we provide an illustrative example where such a measurement basis is employed, enabling the constant-depth preparation of a continuous family of MPS symmetric under the non-Abelian alternating group $A_4$.

\subsubsection{Conditions for preparing normal MPS}
For a particular target state to be deterministically prepared via Protocol~\ref{alg:normal}, we require that any defect from the set $\{B^k\}$ is correctable. In practice, this amounts to either (i) pushing the defect to the edge and acting with an appropriate unitary that annihilates it (as in Ref.~\cite{Smith_2023} for the case of the AKLT state), or (ii) through ``local removal'', where the defect is corrected via a $k-$local unitary on the physical level (which can also be viewed as pushing a defect $B$ to the identity matrix $I$).

Combining this logic with the discussion of the previous section, we arrive at the following theorem.

\begin{theorem}[Preparation of normal MPS]\label{thm:normal} 
Let $\map{A}^{(q)}$ be the virtual-to-physical map parameterizing a translationally-invariant\footnote{Similar to Section~\ref{sec:ingredients}, here we use translationally-invariant in the sense that each MPS site is parameterized by the same tensor $A$, but with arbitrary boundary conditions.} normal matrix product state $\ket{\Psi_{\textup{MPS}}}$ with bond dimension $D$ and a blocking parameter $q$ that is independent of $N$. Then $\ket{\Psi_{\textup{MPS}}}$ can be exactly and deterministically prepared by Protocol~\ref{alg:normal} if for some finite $q$ there exists a finite group $G$ and a $D$-dimensional irreducible projective representation $B^g$ thereof such that for each $g\in G$, there exists an $h\in G$ such that $[(B^h)^T \otimes B^g, \map{A}^{(q)\dagger}\map{A}^{(q)}]=0$.
\end{theorem}

The proof is a straightforward extension of the results of the previous sections. For clarity of discussion, we specialize to the case of $q=1$ and note that the proof follows analogously for $q>1$, albeit with feedforward operations that are $q$-local (but still carried out in constant depth). First, let the fusion measurement basis correspond to a $D$-dimensional irreducible projective representation of $G$. Next, let $B_i^g$ denote a defect at the $i$th fusion site, with $g\in G$. From Theorem~\ref{thm:unitarity}, the condition $[(B^h)^T \otimes B^g, \map{A}^{\dagger}\map{A}]=0$ guarantees that, for some choice of gauge, there exists a physical unitary $U^{(i)}_{i}$ to push the defect $B_i^g$ through physical site $i$, converting it into a defect $B_{i-1}^h$ at fusion site $i-1$. Because $h\in G$, this process can be iterated until the defect is pushed to the edge. In particular, we apply $U^{(j)}_{i}$ to each physical site $j\leq i$. Crucially, $U^{(j)}_i$ and $U^{(k)}_i$ commute for all $j$ and $k$, and can therefore be applied in parallel. Likewise, the removal of defects originating at all $N-1$ fusion sites can be carried out with an $O(1)$ feedforward step by first classically computing the local unitary,
\begin{equation}
    U^{(i)}_B = U^{(i)}_i U^{(i)}_{i+1} \ldots U^{(i)}_{N-1}.
    \label{eq:combined_U}
\end{equation}
As shown in Figure~\ref{fig:f4}, we then apply $U^{(i)}_B$ to each physical site $i$  and the appropriate group element $B_0^\dagger$ at the boundary in parallel, yielding the exact, defect-free target state.

It is worth emphasizing that Theorem~\ref{thm:normal} provides a set of conditions that are provably sufficient, but not evidently necessary. For example, one might imagine that there are instances of deterministically preparable states where defects are pushed between different representations of the group $G$, or where each defect type requires a different blocking parameter $q$. Separately, the protocol outlined in Figure~\ref{fig:f4} can be extended to non-translationally-invariant states as long as each (possibly blocked) tensor carries the appropriate pushing relations. We leave such broad possibilities as an avenue for future exploration, and here narrow our focus on states characterized by the comparatively manageable set of conditions in Theorem~\ref{thm:normal}.

Leaning on the results of Section \ref{ssec:pushing}, we now state two corollaries of Theorem~\ref{thm:normal} that cover specific cases where the requisite conditions are guaranteed to be satisfied.

\begin{corollary}
Any normal MPS with zero correlation length can be deterministically prepared in constant depth using Protocol~\ref{alg:normal}.
\label{corr:zcl}
\end{corollary}
The proof is a simple extension of Section~\ref{ssec:zcl}, where it was shown that for some blocking parameter $q$, unitary pushing relations can be defined for arbitrary unitary virtual operators $O_{\ell}$ and $O_r$ for MPS with ZCL. In fact, this scenario allows one to choose $O_{\ell}$ to be the identity, enabling the local removal of the defect $O_r = B^g$ for all $g\in G$, regardless of the choice of finite group $G$. As such, we can choose the defects according to convenience, with one option being the $D$-dimensional qudit Pauli matrices which form a representation of $\mathbb{Z}_D\times\mathbb{Z}_D$ and are generated by the clock and shift matrices in Eq.~\eqref{eq:clockshift}.

It is important to note that any normal MPS with ZCL can be prepared in constant depth with a purely \emph{unitary} circuit. Still, the adaptive procedure outlined in Protocol~$\ref{alg:normal}$ may offer advantages for the preparation of such states in, e.g., linear optical platforms where native unitary two-qubit gates are unavailable and entangling operations are instead carried out using joint measurements. In fact, we note a close resemblance between the protocol proposed here and so-called fusion-based quantum computation, where fusion measurements are used to prepare large cluster states from small resource states~\cite{bartolucci2023fusion}. Furthermore, our protocol is an attractive approach for distributed quantum hardware, as separate sections of MPS can be prepared and subsequently fused across multiple local quantum processing units without a direct link between physical sites.

\begin{corollary}
\label{corr:symmetry}
Let $\ket{\Psi_{\textup{MPS}}}$ be a translationally-invariant\footnote{Up to arbitrary boundary conditions (see previous footnote).} MPS characterized by global on-site symmetry under a finite group $G$. If the action of group elements on the physical sites manifests as an irreducible representation on the virtual level, then $\ket{\Psi_{\textup{MPS}}}$ can be deterministically prepared in constant depth using Protocol~\ref{alg:normal}. 
\end{corollary}

The proof is as follows. As discussed in Section~\ref{ssec:symmetry}, normal MPS with on-site symmetry $G$ are endowed with a set of $|G|$ pushing relations with $O_{r} = V_g$ and $O_{\ell} = V_g^{\dagger}$, where the unitary operators $V_g$ form a projective representation of $G$. By virtue of the group axioms, the inverse operations $V_g^{\dagger}$ are also group elements. Consequently, the defects $B^g = V_g$ are correctable for all $g\in G$. However, as discussed in Section~\ref{sssec:measurement}, the set $\{B^g\}$ must also define a POVM. This condition is satisfied if the group $G$ is finite and the representation $V_g$ is irreducible. Thus, states that obey these conditions satisfy the requirements of Theorem~\ref{thm:normal}.

The conditions of Corollary~\ref{corr:symmetry} encompass a variety of physically interesting nontrivial entangled states that can be constructed for any bond dimension (see Appendix~\ref{app:symconst}). For instance, if the virtual operators $V_g$ form a  \emph{projective} representation of $G$, then the target MPS represents a point in a nontrivial symmetry-protected topological (SPT) phase, characterized by elements in the second cohomology group ${\cal H}^2(G,U(1))$~\cite{chen2011complete,pollmann2012detection,singh2015identifying}. States with SPT order exhibit a number of interesting properties such as ground state degeneracy for open boundary conditions, edge modes, and long-range string order~\cite{pollmann2012symmetry, pollmann2012detection, Cirac_2021}. Furthermore, SPT order is intimately linked to utility as a resource for MBQC~\cite{Else2012, stephen2017computational, Raussendorf2023}. 

However, Corollary~\ref{corr:symmetry} is not strictly limited to states with SPT order, but also encompasses certain entangled states in the trivial phase with respect to $G$. This is due to the fact that the symmetry group $G$ can be either Abelian or non-Abelian. In the case of Abelian symmetries, irreducibility of $V_g$ for bond dimension $D>1$ necessarily implies that the representation is projective, as all linear irreducible representations of finite Abelian groups are one-dimensional. MPS satisfying Corollary~\ref{corr:symmetry} with $D>1$ and $G$ Abelian are therefore in a nontrivial SPT phase. In the case of non-Abelian symmetry, however, linear irreducible representations of higher dimensions are possible. Consequently, Corollary~\ref{corr:symmetry} is also satisfied by certain entangled states in the trivial phase that are symmetric under a non-Abelian group. We provide examples of such states in Section~\ref{sssec:A4}, where we demonstrate the preparation of a continuous family of MPS symmetric under the non-Abelian alternating group $A_4$. In addition, we note the these MPS can also be constructed using the method of Appendix~\ref{app:symconst}.

Furthermore, while Corollary~\ref{corr:symmetry} at first glance appears to rule out preparation of MPS with continuous on-site symmetry, we emphasize that this is not the case. As a simple counterexample, it was shown in Ref.~\cite{Smith_2023} that Protocol~\ref{alg:normal} can deterministically prepare the $SO(3)$-symmetric spin-1 AKLT state. There, the key strategy was to employ a measurement basis corresponding to a finite subgroup of $SO(3)$, namely $G=\mathbb{Z}_2\times\mathbb{Z}_2$ (corresponding to the Bell basis). Thus, the implications of Corollary~\ref{corr:symmetry} are not limited to the preparation of MPS within only finite symmetries; we can also prepare MPS with continuous symmetries, given there exists a finite subgroup $G$ with an irreducible projective representation of dimension $D$. This strategy is demonstrated for a variety of MPS with continuous symmetries in Section~\ref{ssec:examples}, including those with $SU(n)$, $SO(2\ell + 1)$, and $Sp(2n)$ symmetry that have previously been identified as resources states for MBQC with qudits \cite{Wang2017}. We refer also to Appendix~\ref{app:ssec:ExamplesSU2} for a tabulation of the finite subgroups of $SU(2)$, which in turn enable the constant-depth preparation of $SU(2)$-symmetric MPS up to $D=6$ (i.e., up to the spin-5/2 representation of $SU(2)$).

Finally, while Corollaries~\ref{corr:zcl} and~\ref{corr:symmetry} provide explicit classes of normal MPS that can be prepared exactly and deterministically using Protocol~\ref{alg:normal}, we emphasize that they by no means fully encompass the states that satisfy Theorem~\ref{thm:normal}. In Section~\ref{sssec:Z2}, we provide an explicit example of a family of states that do not fit within either case, indicating that the complete landscape of MPS that can be exactly prepared with our scheme is not fully captured by these two Corollaries. A complete classification of such states, however, is beyond the scope of this work.

\subsection{Non-normal MPS}
\label{ssec:nonnormal}

\begin{figure*}
\includegraphics[width = 0.95\linewidth]{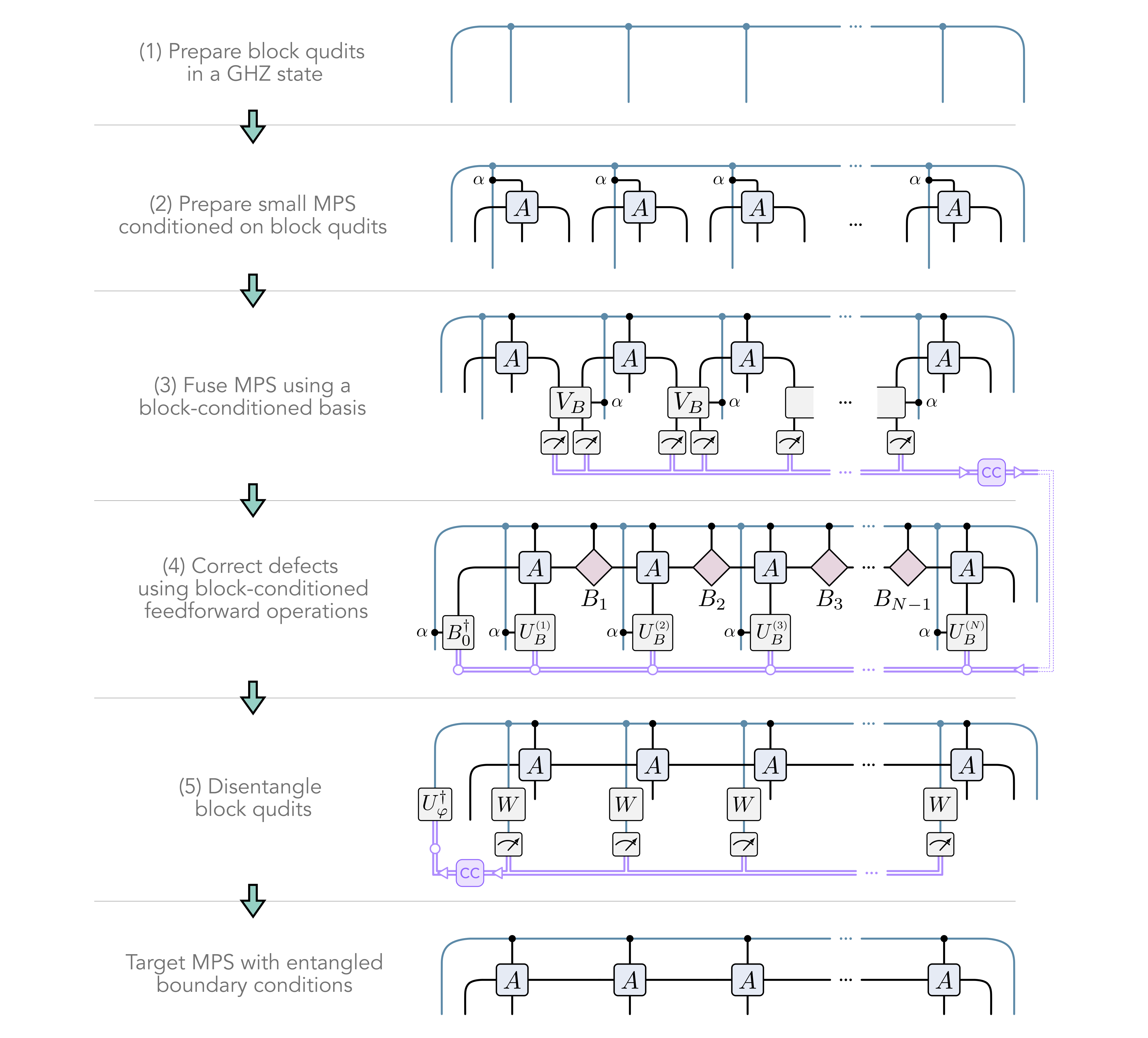}
\caption{Constant-depth preparation of non-normal MPS using Protocol~\ref{alg:nonnormal}. First, prepare a GHZ state of $K$-dimensional block qudits to encode the block index $\alpha$. Next, follow the steps of Protocol~\ref{alg:normal} with each operation locally conditioned on a block-qudit. As in the case of normal MPS, here we illustrate the preparation and fusion of many single-site (block-conditioned) MPS, but more generally allow for small MPS of a constant number of sites $q$. In addition, the block-conditional fusion measurements can be generalized from a projective measurement (as shown) to an ancilla-aided POVM (see Eq.~\eqref{eq:povm_nonnormal}). After correcting defects, all non-edge block qudits are deterministically disentangled using measurements and feedforward (see Appendix~\ref{app:block-controlled}). Up to the enforcement of boundary conditions, this constant-depth adaptive quantum circuit exactly and deterministically prepares any non-normal MPS that satisfies the conditions of Theorem~\ref{thm:nonnormal}.}
\label{fig:f5}
\end{figure*}
Though Protocol~\ref{alg:normal} enables the constant-depth preparation of a variety of normal MPS, it is comparatively limited in the case of non-normal MPS. Intuitively, this is because such states are parameterized by non-injective Kraus operators $A^m$, placing significant constraints on the set of ``pushable'' virtual operators. In turn, this severely complicates (and often outright prohibits) the construction of a complete measurement basis from entirely correctable defects. For example, Corollary~\ref{corr:symmetry} provides no insight into the preparation of non-normal MPS with global on-site symmetry, as such states are characterized by a \emph{reducible} representation of the symmetry group \cite{Sanz_2009}. Though certain non-normal MPS are preparable via Protocol~\ref{alg:normal} (such as the set of generalized qudit GHZ states, described in Section~\ref{sssec:GHZ} and furthermore playing an important role in this section), these limitations motivate a preparation protocol tailored to the block-diagonal structure defining non-normal MPS.

In this section, we present such a protocol, enabling the constant-depth preparation of a broad class of long-range entangled states in constant depth. The overarching strategy is to first ``seed'' the non-normality of the target state by preparing a GHZ state of \emph{block} qudits -- i.e., qudits that encode the block index of the tensor $A$ -- and subsequently apply the general procedure underlying Protocol~\ref{alg:normal}, but now with each step \emph{conditioned} on the block index. Similar to Protocol~\ref{alg:normal} for normal MPS, we emphasize that this scheme cannot prepare arbitrary non-normal MPS but rather only those endowed with particular pushing relations, as will be discussed. We begin by making several preliminary comments regarding the features of non-normal MPS that underpin this strategy. 

First, we recall that any non-normal MPS in canonical form can be expressed as a superposition of normal ones \cite{Cirac_2017}. At the level of the tensor $A$ that parameterizes the non-normal MPS, this is reflected in its block-diagonal structure, $A = \bigoplus_{\alpha=0}^{K-1} \mu_{\alpha} A_{\alpha}$, where $K$ denotes the total number of blocks, each $A_{\alpha}$ independently parameterizes a normal MPS in left-canonical form, and each $\mu_{\alpha}$ is a constant. This decomposition can be used to write
\begin{equation}
    \begin{split}
        \ket{\Psi} &= \sum_{\vec{m}}\tr{A^{m_1}A^{m_2}\ldots A^{m_N}X}\ket{\vec{m}}\\
        &= \sum_{\alpha=0}^{K-1}\sum_{\vec{m}}(\mu_{\alpha})^N \tr{A_{\alpha}^{m_1}A_{\alpha}^{m_2}\ldots A_{\alpha}^{m_N}X_{\alpha}}\ket{\vec{m}},
    \end{split}
\end{equation}
where we have decomposed the boundary matrix $X=\bigoplus_{\alpha=0}^{N-1}X_{\alpha}$ according to the block structure of $A$, leveraging the fact that off-diagonal blocks do not contribute to the trace and can be neglected. As in the normal case, our goal is to first prepare the target non-normal MPS with entangled boundary conditions, Eq.~\eqref{eq:mps_target}, and subsequently convert this state into one with a particular boundary matrix via measurement of edge ancillary qudits.

Second, throughout this section, we will without loss of generality assume each block $\alpha$ to have the same dimension $\bar{D}$ such that the total bond dimension is $D = K \bar{D}$; if this is not the case, it is always possible to ``inflate'' the MPS \cite{singh2015identifying} by mapping $A^m_{\alpha}\rightarrow A_{\alpha}^m\otimes \mathbb{1}_{r_{\alpha}\times r_{\alpha}}$ and $D_{\alpha}\rightarrow r_{\alpha}D_{\alpha}$, where $r_{\alpha} = \textrm{lcm}(D_{0},D_{1},\ldots D_{K-1})/\bar{D}$, where $D_{\alpha}$ is the (uninflated) dimension of block $\alpha$. Furthermore, we will assume that $\mu_{\alpha}= 1$ for all $\alpha$, as it is always possible to absorb $\mu_{\alpha}\neq 1$ into the desired boundary matrix $X$ -- for more details, see Appendix~\ref{app:block-controlled}. Employing these assumptions, Eq.~\eqref{eq:mps_target} can be cast into the form 
\begin{equation}
    \ket{\Psi} = \sum_{\alpha=0}^{K-1}\ket{\Psi_{\alpha}}\otimes\ket{\alpha}^{\otimes 2},
\end{equation} 
where $\ket{\Psi_{\alpha}}$ is exactly a normal MPS with entangled boundary conditions:
\begin{equation}
    \ket{\Psi_{\alpha}} = \sum_{i,j=0}^{\bar{D}-1}\sum_{\vec{m}}\bra{i}A_{\alpha}^{m_1} A_{\alpha}^{m_2} \ldots A_{\alpha}^{m_{N}}\ket{j}\ket{\vec{m}}\otimes\ket{ij},
\end{equation}
where we have decomposed each $D$-dimensional ancillary edge qudit into a \emph{block} qudit of dimension $K$ and a \emph{bond} qudit of dimension $\bar{D}$. The former encodes the block index $\alpha$, while the latter encodes the intra-block boundary conditions for each normal MPS $\ket{\Psi_{\alpha}}$.

This decomposed form of $\ket{\Psi}$ motivates the following preparation strategy: First, prepare a pair of block qudits in the Bell state $\ket{\phi} = (1/\sqrt{K})\sum_{\alpha=0}^{K-1}\ket{\alpha\alpha}$. Then, for each block qudit state $\ket{\alpha}$, conditionally prepare the normal MPS $\ket{\Psi_{\alpha}}$ using Protocol~\ref{alg:normal}. This indeed results in the target state and is the essence of our strategy, but with one important modification -- to enable constant-depth preparation with local unitary gates, we elevate the aforementioned Bell state to a GHZ state of $N+2$ qudits, such that block information is ``distributed'' across the $N$ MPS sites, providing local access to the block index $\alpha$. Crucially, this initial GHZ state can be prepared in constant depth with adaptive circuits via Protocol~\ref{alg:normal} (see Section~\ref{sssec:GHZ}), or through other known measurement-based approaches \cite{Piroli_2021}, ensuring that our overall protocol can be realized with a constant-depth adaptive quantum circuit.

With these ingredients in hand, we now formally outline our protocol (see also Fig.~\ref{fig:f5}). Similar to Protocol~\ref{alg:normal}, we use $n$ to denote the number of small MPS prepared in parallel, each with $q$ sites. $N=nq$ denotes the length of the final MPS. In all, this strategy requires $2n$ $\bar{D}$-dimensional bond qudits\footnote{Constant-depth realization of periodic boundary conditions will require an additional pair of $\bar{D}$-dimensional ancillary qudits -- see the discussion at the end of Section~\ref{ssec:sequential}.}, $N+2$ $K$-dimensional block qudits, and $N$ $d$-dimensional physical qudits.

\subsubsection*{Protocol 2: Non-normal MPS}\customlabel{alg:nonnormal}{2}
% \smg{SMG has read up to here.}

\begin{enumerate}[label={(\arabic*)}, wide, labelindent=0pt]
    \item Prepare $N+2$ qudits of dimension $K$ in the generalized GHZ state, \begin{equation}\ket{\textrm{GHZ}}=\frac{1}{\sqrt{K}}\sum_{\alpha=0}^{K-1}\ket{\alpha}^{\otimes N+2}.
    \end{equation}
    This can be achieved in constant depth using Protocol~\ref{alg:normal} or other measurement-based schemes (see Section~\ref{sssec:GHZ}). In the following steps, this long-range entangled state is used to (locally) enact block-conditional operations, where the operation carried out depends on the block index $\alpha$.
    
    \item Block-conditionally prepare $n$ $q$-site copies of the MPS parameterized by the tensor block $A_{\alpha}$. In particular, employ the strategy of Section~\ref{ssec:sequential} using $\bar{D}$-dimensional bond qudits\footnote{As discussed above, we assume without loss of generality that all blocks have the same internal dimension, i.e., $\bar{D} = D/K$} and condition all operations on the block index $\alpha$ encoded in a block qudit.
    
    \item Employ fusion measurements on all non-edge pairs of bond qudits, projecting each into a maximally entangled state. In general, the measurement basis can be made block-specific by conditioning the pre-measurement isometry $V_B$ on the block index. This produces an $N$-site MPS with $\alpha$-conditioned random defects at each fusion site.

    \item Leveraging the unitary pushing relations specific to each intra-block tensor $A_{\alpha}$, remove all block-conditional defects using a $O(1)$ depth layer of block-conditional feedforward operations.

    \item Disentangle all non-edge block qudits by measuring each in the (qudit) Pauli $X$-basis. Using feedforward, remove the outcome-dependent phase via a phase gate $U^\dagger_{\varphi}$ applied to an edge block qudit (see Appendix~\ref{app:block-controlled} for details).

    \item Measure the remaining edge bond and block qudits to collapse the defect-free MPS onto definite boundary conditions as in Protocol~\ref{alg:normal}, with each particular outcome having probability $\sim K/D^2$.\footnote{Similar to Protocol~\ref{alg:normal}, this assumes that $N\gg \textrm{max}\{\xi_{\alpha}\}$, where $\xi_{\alpha}$ is the correlation length of the normal MPS parameterized by $A_{\alpha}$.}
\end{enumerate}

Similar to Protocol~\ref{alg:normal}, $\alpha$-conditional projective measurement can be generalized to an $\alpha$-conditional POVM through the replacement,
\begin{equation}
\includegraphics[width = 0.85\linewidth]{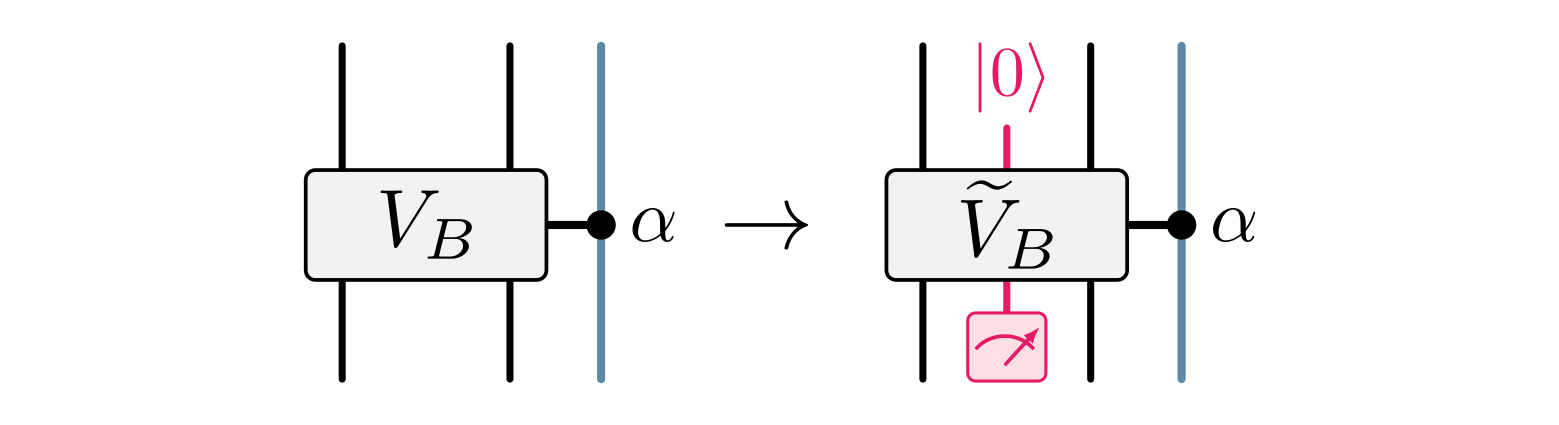},
\label{eq:povm_nonnormal}
\end{equation}
where the additional wire corresponds to a $p$-dimensional ancilla. Furthermore, this protocol can be simplified in many cases of interest. In particular, if each block is endowed with pushing relations for the same group $G$, then we can choose a gauge where each block carries the same irreducible representation such that isometry $V_B$ does not need not be $\alpha$-conditioned, simplifying Step (3) of the protocol.

\subsubsection{Conditions for preparing non-normal MPS}
We now turn to the question of when Protocol~\ref{alg:nonnormal} can deterministically prepare a particular non-normal MPS in constant depth (up to boundary conditions). Viewing non-normal MPS through the lens of its decomposition into normal MPS, we arrive at the following theorem.

\begin{theorem}
    \label{thm:nonnormal}
    Let $\ket{\Psi_{\textup{MPS}}}$ be a non-normal MPS in canonical form, and $\ket{\Psi}=\sum_{\alpha=0}^{K-1}c_{\alpha}\ket{\Psi_{\alpha}}$ be its decomposition, where each state $\ket{\Psi_{\alpha}}$ is a normal MPS. Then $\ket{\Psi_{\textup{MPS}}}$ can be exactly and deterministically prepared by Protocol~\ref{alg:nonnormal} if each composing normal MPS $\ket{\Psi_{\alpha}}$ satisfies Theorem~\ref{thm:normal}.
\end{theorem}
The proof is self-evident from the design of Protocol~\ref{alg:nonnormal} and the conditions of Theorem~\ref{thm:normal}. Naturally, the above theorem guarantees the ability for constant-depth preparation of any non-normal MPS constructed from normal MPS satisfying (but not limited to) Corollary~\ref{corr:zcl}, Corollary~\ref{corr:symmetry}, and mixtures thereof. Furthermore, it enables the constant-depth preparation of the symmetry-broken states previously discussed in Section~\ref{ssec:symmetry} when the subgroup $H\leq G$ is finite and its representation $V_h(g,\alpha)$ is irreducible (see Eq.~\eqref{eq:nonnormalsymmetry}). We provide an example of such a state in Section~\ref{sssec:Z4xZ2}.

In the particular scenario where periodic boundary conditions are enforced and all $\ket{\Psi_{\alpha}}$ satisfy Corollary~\ref{corr:zcl}, the state $\ket{\Psi_{\textrm{MPS}}}$ corresponds to a fixed point of the renormalization procedure introduced in \cite{Verstraete2005}. It has previously been shown that such states can be prepared using constant-depth quantum circuits augmented with mid-circuit measurements and feedforward operations \cite{Piroli_2021}. Furthermore, Ref.~\cite{Gunn2023} recently demonstrated that the fixed point of any non-normal phase protected by finite Abelian symmetry can be prepared using a constant-depth circuit composed of local unitaries, measurements, and feedforward operations that all preserve the symmetry. We note that the procedure employed in Ref.~\cite{Gunn2023} bears similarity to our Protocol~\ref{alg:nonnormal}, though with a measurement strategy that fuses both GHZ and intra-block components in parallel and that is tailored to fixed-point states exhibiting symmetries of the form in Eq.~\eqref{eq:nonnormalsymmetry}. In contrast, the scope of our protocol encompasses both fixed-point and non-fixed-point states with nonzero correlation length. Furthermore, it places no restriction on the relationship between the normal MPS $\ket{\Psi_{\alpha}}$ in different blocks, and thus extends beyond symmetry-broken states (see, for example, the Majumdar-Ghosh states in Section~\ref{sssec:MG}).

\subsection{Examples}
\label{ssec:examples}

\begin{table*}[tbh]
%\bgroup
\def\arraystretch{2.4}
\begin{tabular}{|c|c|c|c|c|c|}
\hline%\hline
& \textbf{Example} & \textbf{Section} & & \textbf{Example} & \textbf{Section} \\
\hline
\multirow{5}{*}{\rotatebox[origin=c]{90}{\parbox[c]{2cm}{\textbf{Normal}}}} & $\mathbb{Z}_2$-symmetric family & \ref{sssec:Z2} &\multirow{3}{*}{\rotatebox[origin=c]{90}{\parbox[c]{2.9cm}{\textbf{Non-Normal}}}} & GHZ states & \ref{sssec:GHZ}\\
\cline{2-3}\cline{5-6}
            & The AKLT state & \ref{sssec:AKLT}  &   & Majumdar-Ghosh states   & \ref{sssec:MG}  \\
\cline{2-3}\cline{5-6}
            & $SU(3)$ symmetry & \ref{sssec:SU3}  &   & $\mathbb{Z}_4\times\mathbb{Z}_2$ symmetry-broken   & \ref{sssec:Z4xZ2} \\
\cline{2-6}
            & $A_4$-symmetric family & \ref{sssec:A4}  &\multirow{2}{*}{\rotatebox[origin=c]{90}{\parbox[c]{2cm}{\textbf{Sampling}}}}   & Random MPS   & \ref{sssec:rmps} \\
\cline{2-3}\cline{5-6}
            & $SU(n)$, $SO(2\ell+1)$, and $Sp(2n)$ & \ref{sssec:highersymm}  &   & Haldane phase   & \ref{sssec:Haldane}\\
\hline
\end{tabular}
\caption{A list of examples presented in Section \ref{ssec:examples}.
}\label{tab:examples}
\end{table*}

In this section, we provide a number of concrete examples to illustrate the diversity of states that are preparable in constant depth with our scheme. As summarized in Table~\ref{tab:examples}, these examples include families of MPS with discrete and continuous symmetries, symmetry-broken states, MPS with non-Abelian symmetry, and resource states for MBQC. We also address the possibility of using our protocols to sample MPS. Namely, we discuss the constant-depth preparation of random MPS, and additionally present a procedure to randomly sample from a nontrivial SPT phase. For each example, we state the pushing relations and measurement basis that enables constant-depth preparation. Except where clarity is needed, we employ a simplified notation for pushing relations, omitting the red diamonds and gray boxes enclosing defects and physical unitaries used throughout this work.

We emphasize that the selected examples are non-exhaustive and were primarily chosen to demonstrate the variety of physically interesting states that can be prepared in constant depth. In fact, it is possible to systematically construct parameterized families of preparable MPS for \emph{any} bond dimension. We present such a construction method in Appendix~\ref{app:symconst}.

% \localtableofcontents

\subsubsection{$\mathbb{Z}_2$-symmetric family}\label{sssec:Z2}
We begin with the $\mathbb{Z}_2$-symmetric family of MPS of physical dimension $d=2$ and bond dimension $D=2$, first introduced in Ref.~\cite{Wolf2006}. It encompasses a set of parameterized states with a continuously ``tunable'' correlation length, interpolating between the cluster state (which has ZCL) and the GHZ state (which has long-range correlations). As such, it is an illustrative case study for our framework, as its primary advantage is in its ability to spread correlations across $N$ sites in constant time, an impossible feat with unitary resources alone. 

Previously, it has been shown that this family of states can be approximately prepared in $O(\textrm{polylog}(N))$ time using an adiabatic scheme \cite{Wei2022c}, and in  $O(\log(N/\epsilon))$ depth with error $\epsilon$  using a circuit-based approach \cite{Malz_2024}. Furthermore, the latter reference demonstrated that this depth can be further reduced to $O(\log\log(N/\epsilon))$ by augmenting a unitary circuit with measurements and feedforward to synthesize all-to-all connectivity. In all of these approaches, the nonzero correlation length of the state presents an unavoidable bottleneck and, consequently, there is a trade-off between precision and preparation time. However, as initially shown in Ref.~\cite{Zhu2022}, this trade-off can be altogether evaded by leveraging mid-circuit measurements and feedforward, enabling an exact, constant-depth preparation. Here, we leverage the MPS representation of this family to demonstrate its preparation using Protocol~\ref{alg:normal}.

This family of states $\ket{\Psi(g)}$ is parameterized by the matrices
\begin{equation}
    A^0=\eta \left(\begin{matrix}
    1 & 0 \\
    \sqrt{-g} & 0
    \end{matrix}\right) \quad A^1=\eta \left(\begin{matrix}
    0 & -\sqrt{-g} \\
    0 & 1
    \end{matrix}\right),
\end{equation}
where $\eta = 1/\sqrt{1+|g|}$. We note that the above form of $A^0$ and $A^1$ differs slightly from that of Ref.~\cite{Wolf2006} as we have cast the matrices into left-canonical form, $\sum_m A^{m\dagger} A^m = \mathbb{1}$. As previously mentioned, $\ket{\Psi(g)}$ captures both the zero-correlation length cluster state ($g=-1$) and the long-range correlated GHZ state ($g=0$). More broadly, $\ket{\Psi(g)}$ has a correlation length that interpolates these two extremes,
\begin{equation}
\xi = \left|\ln\left(\frac{1+g}{1-g}\right)\right|^{-1} .
\end{equation}
Furthermore, it is a normal MPS for all $g$ except at the ``critical point'', $g=0$.

Independent of the parameter $g$, the tensor $A$ obeys the pushing relations
\begin{subequations}
\begin{eqnarray}
\includegraphics[width = 0.85\linewidth]{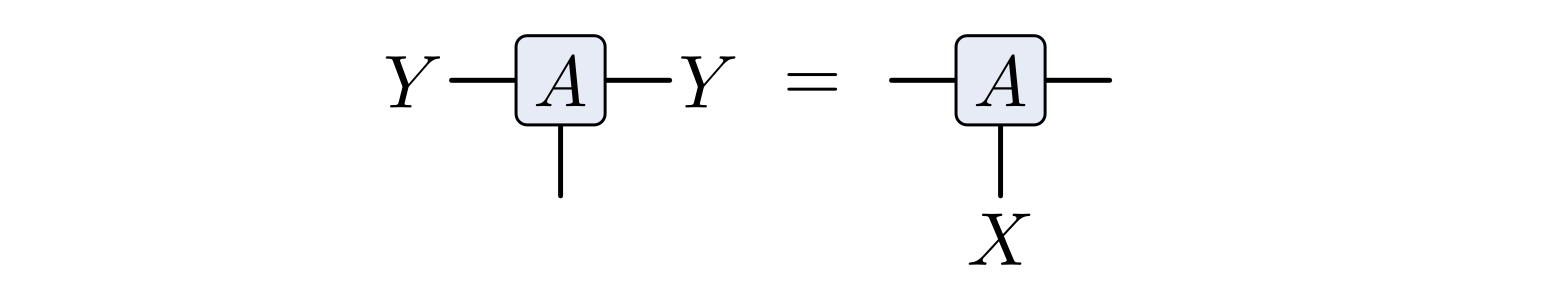} \\
 \includegraphics[width = 0.85\linewidth]{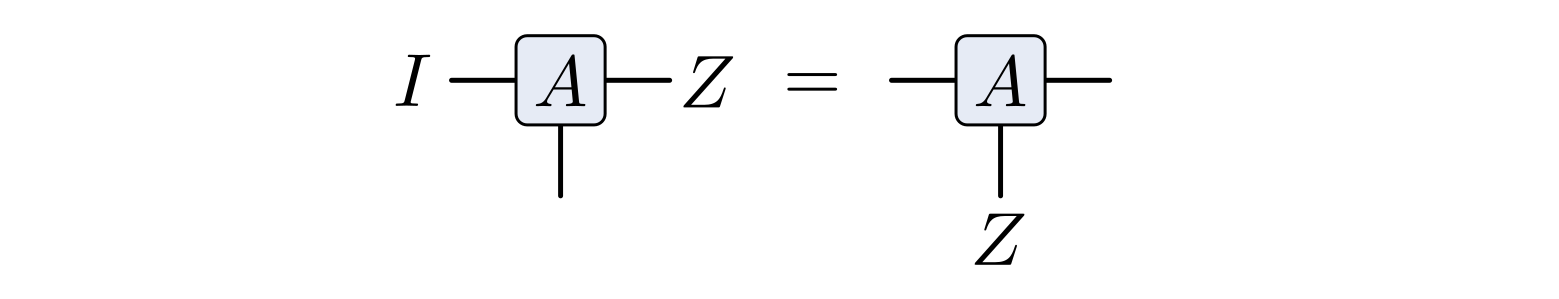}
\end{eqnarray}
\label{eq:Z2pushing}
\end{subequations}
where $X$, $Y$, and $Z$ are Pauli operators. These pushing relations are ``complete'' in the sense that measurement in the Bell basis is guaranteed to yield a pushable defect. Consequently, $\ket{\Psi(g)}$ satisfies all requirements of Theorem~\ref{thm:normal}, and can be exactly prepared in constant depth, independent of $g$ and, by extension, $\xi$.

As a final remark on this family of states, we note that it does not fall under the specifications of either Corollary~\ref{corr:zcl} or Corollary~\ref{corr:symmetry} for general $g$ and, as such, provides a concrete example that shows these corollaries to be non-exhaustive. In particular, Eq.~(\ref{eq:Z2pushing}a) is a manifestation of the $\mathbb{Z}_2$ symmetry, yet with a \emph{reducible} representation $\{V_0=I,V_1=Y\}$ on the virtual level. However, $Z$ defects can be removed ``locally'' via Eq.~(\ref{eq:Z2pushing}b), enabling us to form a complete measurement basis without the irreducibility constraint of Corollary~\ref{corr:zcl}. 

\subsubsection{The AKLT state}
\label{sssec:AKLT}
The spin-1 AKLT state is a historically important instance of a matrix product state~\cite{affleck1988valence, Cirac_2017}, and additionally serves as a paradigmatic example of SPT order \cite{gu2009tensor}. Due to the latter feature, it exhibits a number of exotic properties such as long-range string order and fractionalized edge modes \cite{Nijs1989, Pollmann2010, pollmann2012symmetry}. Furthermore, not unrelated to its SPT order, the AKLT state is a resource for measurement-based quantum computation and quantum teleportation~\cite{Verstraete2004, Gross2007a}. 

The AKLT state can be exactly expressed as an MPS with bond dimension $D=2$ and physical dimension $d=3$. It is parameterized by the matrices
\begin{equation}
    A^+ = \sqrt{\frac{2}{3}}\sigma_+ \quad A^- = -\sqrt{\frac{2}{3}}\sigma_- \quad A^0 = -\sqrt{\frac{1}{3}}Z,
\end{equation}
where $\sigma_{\pm} = (X\pm iY)/2$. As the AKLT state has a nonzero correlation length, it cannot be prepared exactly by a constant-depth unitary quantum circuit. However, as shown in Ref.~\cite{Smith_2023} -- the precursor to this work -- it can be exactly prepared via Protocol~\ref{alg:normal}. 

Enabling this preparation is the $SO(3)\cong SU(2)/\mathbb{Z}_2$ symmetry of the AKLT state, which provides a continuous family of pushing relations,
\begin{equation}
\includegraphics[width = 0.85\linewidth]{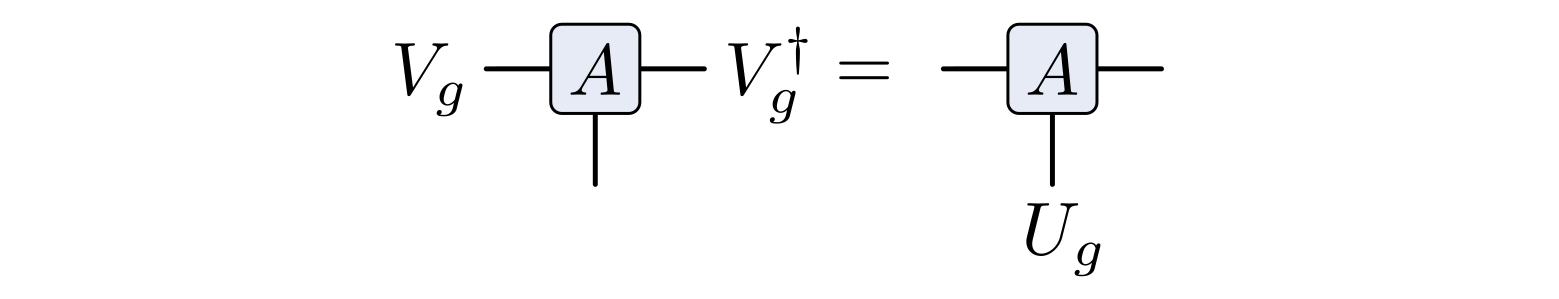}, 
\label{eq:AKLTpushing}
\end{equation}
where $U_g$ ($V_g$) form a linear (projective) representation of $SO(3)$. However, the full $SO(3)$ symmetry is not required for preparation -- in Ref.~\cite{Smith_2023}, measurements were carried out in the Bell basis, corresponding to a selection of defects $B\in\{X,Y,Z,I\}$ that form a projective representation of $\mathbb{Z}_2\times \mathbb{Z}_2\subset SO(3)$. Thus, this example illustrates the previously alluded-to strategy of preparing states with continuous symmetries by choosing a measurement basis in correspondence with a $D$-dimensional irreducible representation of a finite subgroup of the full continuous symmetry. For more details, we defer to Ref.~\cite{Smith_2023}, which additionally includes experimental demonstrations carried out on an IBM Quantum processor.

\subsubsection{$SU(3)$ symmetry}
\label{sssec:SU3}
As an illustrative example of an MPS with $D>2$, we now consider a spin-1 state with 
global on-site $SO(3)$ symmetry. First constructed in Ref.~\cite{Sanz_2009}, this state is parameterized by
\begin{equation}
    \begin{gathered}
        A^+=\frac{1}{\sqrt{2}} \left(\begin{matrix}
        0 & 1 & 0 \\
        0 & 0 & 1 \\
        0 & 0 & 0
        \end{matrix}\right) \quad 
        A^-=\frac{1}{\sqrt{2}}  \left(\begin{matrix}
        0 & 0 & 0 \\
        -1 & 0 & 0 \\
        0 & -1 & 0
        \end{matrix}\right)\\
        A^0=\frac{1}{\sqrt{2}}  \left(\begin{matrix}
        1 & 0 & 0 \\
        0 & 0 & 0 \\
        0 & 0 & -1
        \end{matrix}\right).
    \end{gathered}
    \label{eq:SU3}
\end{equation}
As the virtual degrees of freedom carry a linear (spin-1) representation of $SO(3)$, this state is in the trivial phase with respect to this symmetry. However, in Ref.~\cite{Rao2016} it was shown that the edge modes carry a representation of $SU(3)$, revealing an underlying SPT order with respect to this symmetry, and exhibiting a number of exotic properties as a consequence, such as edges that carry conjugate representations (i.e., quark and anti-quark edge states). 

Separate from the ``single-site'' $SO(3)$ pushing relations, this enlarged $SU(3)$ symmetry gives rise to a continuous set of pushing relations after blocking $q=2$ sites:
\begin{equation}
\includegraphics[width = 0.85\linewidth]{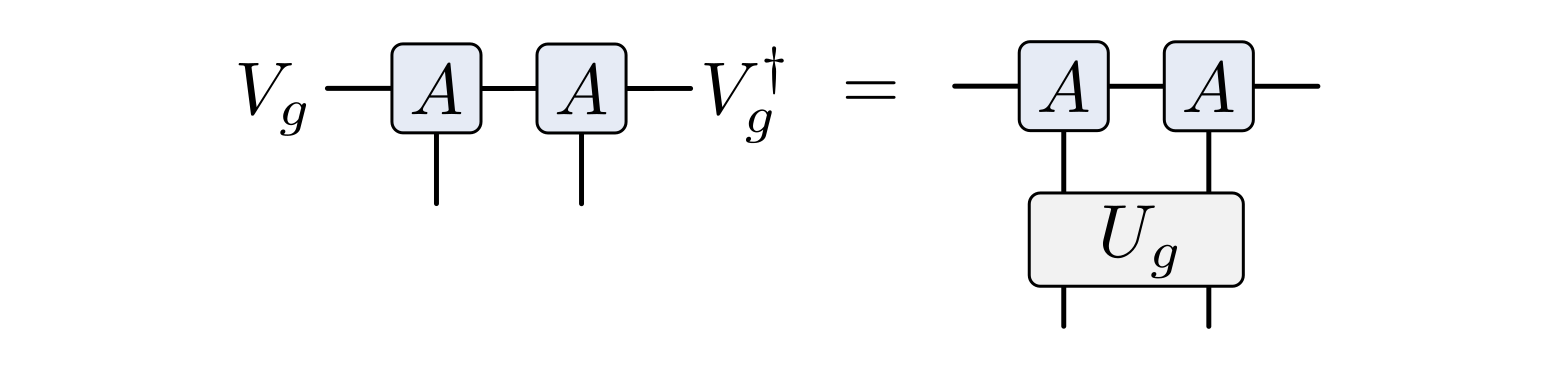}, 
\label{eq:SU3pushing}
\end{equation}
 where $U_g$ ($V_g$) form a linear (projective) representation of  $SU(3)$. We have used a gray box to indicate that the physical unitary $U_g$ is applied to pairs of physical sites. As with the AKLT state, we can prepare this MPS by first identifying a set of defects $B^k\in SU(3)$ that form an irreducible representation of a finite group. To that end, we choose the $3\times3$ qudit Pauli matrices,
\begin{equation}
B^{(i,j)} = X^i Z^j,
\end{equation}
where $Z$ and $X$ are the clock and shift matrices defined in Eq.~\eqref{eq:clockshift}.
This set of defects forms an irreducible projective representation of $\mathbb{Z}_3\times \mathbb{Z}_3$ and provides a natural extension of the Bell basis to qutrits \cite{bennett1993teleporting}. Furthermore, it satisfies Theorem~\ref{thm:normal} for the choice $q=2$, and this MPS is therefore preparable in constant depth.

\subsubsection{The $A_4$-symmetric family}
\label{sssec:A4}
Next, we consider a $d=3$, $D=3$ family of normal MPS with global on-site symmetry under the non-Abelian alternating group $A_4$, constructed using the technique outlined in Appendix~\ref{app:symconst}. This class of states provides an explicit example of the case $D<\sqrt{|G|}$ which, as discussed in Section~\ref{sssec:measurement}, is handled by replacing the projective fusion measurement scheme with a more general (ancilla-aided) POVM -- see Eq.~\eqref{eq:isometry_condition}.

Denoting this family by $\ket{\Psi(\theta, \phi)}$, it can be exactly expressed as an MPS parameterized by the matrices
\begin{equation}
    \begin{gathered}
        A^+=\frac{1}{\sqrt{2}} \left(\begin{matrix}
        0 & u & 0 \\
        v & 0 & u \\
        0 & -v & 0
        \end{matrix}\right) \quad 
        A^-=\frac{1}{\sqrt{2}} \left(\begin{matrix}
        0 & v & 0 \\
        -u & 0 & -v \\
        0 & -u & 0
        \end{matrix}\right) \\ 
        A^0= \frac{1}{\sqrt{2}}\left(\begin{matrix}
        u & 0 & -v \\
        0 & 0 & 0 \\
        v & 0 & -u
        \end{matrix}\right),
    \end{gathered}
    \label{eq:A4fam}
\end{equation}
where we have adopted the shorthand $u = [\cos(\theta/2) + e^{i\phi}\sin(\theta/2)]/\sqrt{2}$ and $v = [\cos(\theta/2) - e^{i\phi}\sin(\theta/2)]/\sqrt{2}$. Notably, the faithful unitary preparation of $\ket{\Psi(\theta, \phi)}$ requires at minimum a log-depth circuit as it has nonzero correlation length for all $\theta$ and $\phi$,
\begin{equation}
    \xi = -1/\ln\left(\frac{1}{2}\sqrt{1+3\cos^{2}\theta}\right).
\end{equation}

In contrast, this family of states can be prepared exactly with Protocol~\ref{alg:normal} by leveraging the on-site symmetry. In particular, $\ket{\Psi(\theta, \phi)}$ is endowed with pushing relations of the usual form,
\begin{equation}
\includegraphics[width = 0.85\linewidth]{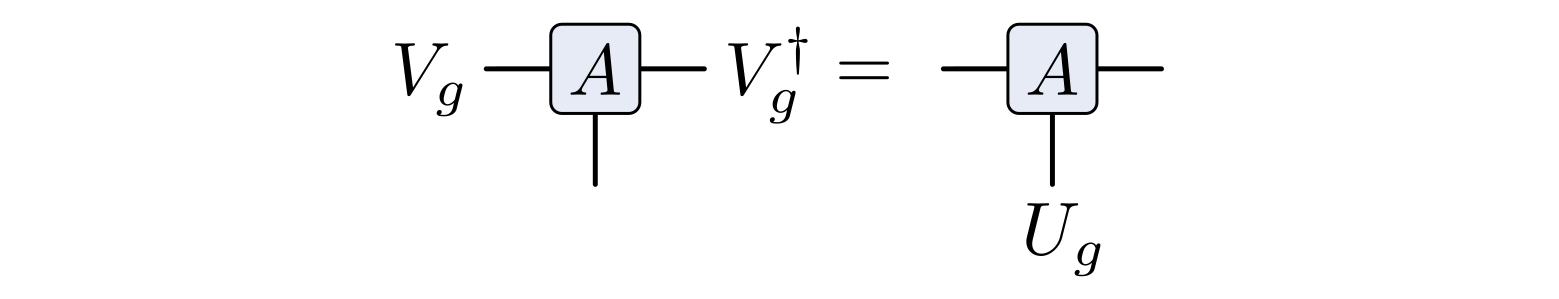}, 
\label{eq:A4pushing}
\end{equation}
where $V_g$ and $U_g$ both form an irreducible linear representation of $A_4$. Therefore, Theorem~\ref{thm:normal} is satisfied by way of Corollary~\ref{corr:symmetry}. More specifically, $A_4$ has the presentation $\left<x,y | x^2=y^3=e, ~yxy=xy^2\right>$. For the virtual representation $V_g$, the generators $x$ and $y$ take the form,
\begin{equation}
    \begin{gathered}
        V_x= \left(\begin{matrix}
        0 & 0 & -1 \\
        0 & -1 & 0 \\
        -1 & 0 & 0
        \end{matrix}\right) \quad
        V_y=\frac{1}{\sqrt{2}}\left(\begin{matrix}
        i/\sqrt{2}        & 1  & -i/\sqrt{2} \\
        -i & 0            & -i \\
        i/\sqrt{2}        & -1   & -i/\sqrt{2}
        \end{matrix}\right).
    \end{gathered}
    \label{eq:A4gen}
\end{equation}
This same representation parameterizes the physical unitaries $U_g$. In contrast to previous examples, $\ket{\Psi(\theta, \phi)}$ is in the trivial phase with respect to the $A_4$ symmetry as the virtual operators $V_g$ form a linear representation. In total, there are $|A_4|=12$ unique defects. Because $|A_4| > D^2 = 9$, fusion measurements for this defect basis require an ancilla-aided POVM (see Eq.~\eqref{eq:povm}). Specifically, this leverages an ancillary qudit of dimension $\textrm{lcm}(12,9)/9 = 4$.

Finally, we note that this family of states includes the $SO(3)$-symmetric state discussed in Section~\ref{sssec:SU3}, corresponding to $\ket{\Psi(\pi/2, 0)}$. This is due to the fact that $A_4$ is a subgroup of $SO(3)$. As a result, $\ket{\Psi(\pi/2, 0)}$ can be prepared either by first preparing many two-site MPS and employing a fusion measurement basis derived from a projective irrep of $\mathbb{Z}_3\times\mathbb{Z}_3$ (as in Section~\ref{sssec:SU3}), or by preparing many single-site MPS and relying on a linear irrep of $A_4$ (as shown here). This demonstrates that for certain target states, there is flexibility in how one leverages pushing relations for constant-depth preparation.

\subsubsection{MBQC resource states with higher symmetries}
\label{sssec:highersymm}
Each of the previous examples concerns the preparation of MPS with a small bond dimension $D\leq 3$. We emphasize, however, that our scheme is also capable of preparing states with higher bond dimensions. This generally requires that the target state is symmetric under a large symmetry group, as the correction of $\eta \geq D^2$ unique random defects necessitates a commensurate number of pushing relations. As an illustration of this principle, we consider a class of states with higher-symmetry SPT orders studied in Ref.~\cite{Wang2017}. In particular, it was shown that AKLT-type states with $SU(n)$, $SO(2\ell+1)$, and $Sp(2n)$ symmetry can be used as resource states for MBQC. In contrast to the $SO(3)$-symmetric spin-1 AKLT state, which encodes a single logical qubit for MBQC, its higher-symmetry variants encode either a single $n$-dimensional qudit (in the case of $SU(n)$ symmetry) or many qubits (in the case of $SO(2\ell+1)$ and $Sp(2n)$ symmetries); see also Refs.~\cite{Tu2008,Ragone2024} and Refs.~\cite{Greiter2007,Gozel2019} for a relevant discussion on $SO(n)$- and $SU(n)$-symmetric generalizations of the spin-1 AKLT state.

We refer to Ref.~\cite{Wang2017} for the explicit form of these MPS. Here, we remark that each is characterized by a suite of pushing relations analogous to those of the AKLT state, Eq.~\eqref{eq:AKLTpushing},
with $V_g$ and $U_g$ both forming an irrep of either $SU(n)$, $SO(2\ell+1)$, or $Sp(2n)$. As with the spin-1 AKLT state, our strategy for preparation is to identify a subset of defects that form a (projective) irrep of a finite subgroup. We list these subgroups in Table~\ref{tab:highersymm}, and furthermore provide an explicit representation of each in Appendix~\ref{app:highersym_details}. See also the similar Table I in Ref.~\cite{Wang2017}, which lists the relevant properties of each MPS family for MBQC. In all cases, the provided basis satisfies the constraints of Corollary~\ref{corr:symmetry}, enabling the exact, constant-depth preparation of this large class of resource states via Protocol~\ref{alg:normal}.

\begin{table}
\centering
\def\arraystretch{1.15}
\setlength{\tabcolsep}{4pt}
\begin{tabular}{|| c | c | c | c ||} 
 \hline
  & $SU(n)$ & $SO(2\ell + 1)$ & $Sp(2n)$ \\
 \hline
 $d$ & $n^2-1$ & $2\ell + 1$ or $\ell(2\ell + 1)$ & $n(2n+1)$ \\
  $D$ & $n$ & $2^{\ell}$ & $2n$ \\
  $B_k$ & $\mathbb{Z}_n\times\mathbb{Z}_n$ & ($\mathbb{Z}_2\times\mathbb{Z}_2)^{\ell}$ & $\mathbb{Z}_2\times\mathbb{Z}_n \times D_n$  \\
 \hline
\end{tabular}
\caption{Generalizations of the spin-1 AKLT state with $SU(n)$, $SO(2\ell+1)$, and $Sp(2n)$ symmetries constructed in Ref.~\cite{Wang2017}. All can be prepared in constant depth using Protocol~\ref{alg:normal}. We tabulate the physical dimension $d$, bond dimension $D$, and appropriate measurement basis $B_k$ for each case. The latter property is reported in terms of the relevant finite subgroup for which the defects $\{B_k\}$ form a projective irrep, with $\mathbb{Z}_n$ the cyclic group of order $n$ and $D_n$ the dihedral group of order $2n$. See Appendix~\ref{app:highersym_details} for more details.}
\label{tab:highersymm}
\end{table}

\subsubsection{Generalized qudit GHZ states}\label{sssec:GHZ}
Turning to the preparation of non-normal MPS, we begin with the $d$-dimensional qudit GHZ state of the form $\ket{\Psi} = (1/\sqrt{d})\sum_{j=0}^{d-1} e^{i\phi_j}\ket{j}^{\otimes N}$.
As discussed in Section~\ref{ssec:nonnormal}, this state plays a pivotal role in Protocol~\ref{alg:nonnormal}, providing an initial ``seed'' for the preparation of non-normal MPS. Furthermore, the preparation of GHZ-like states is an important task in its own right, lending itself to applications such as quantum secret sharing~\cite{hillery1999quantum} and quantum metrology~\cite{toth2014quantum}. While the measurement-based preparation of GHZ-like qudit states has been previously explored~\cite{Piroli_2021, Malz_2024}, here we illustrate the integration of this example into our unified framework for preparing MPS with measurements and feedforward.

For simplicity, we focus on the case $\phi_j=0$ for all $j$, but note that our discussion extends to the more general form. The state $\ket{\Psi}$ can be cast as an MPS of bond dimension $D=d$, described by the tensor $A$ with elements
\begin{equation}
    A_{ij}^m = \delta_{ij}\delta_{im}.
\end{equation}
Importantly, $\ket{\Psi}$ is symmetric under any globally applied on-site permutation operator, including powers of the $D$-dimensional shift operator $X$ defined in Eq.~\eqref{eq:clockshift}.
Furthermore, any diagonal operator (including powers of the clock operator $Z$) can be ``locally'' removed from the virtual level. As a result, the tensor $A$ obeys the pushing relations
\begin{subequations}
\begin{eqnarray}
\includegraphics[width = 0.85\linewidth]{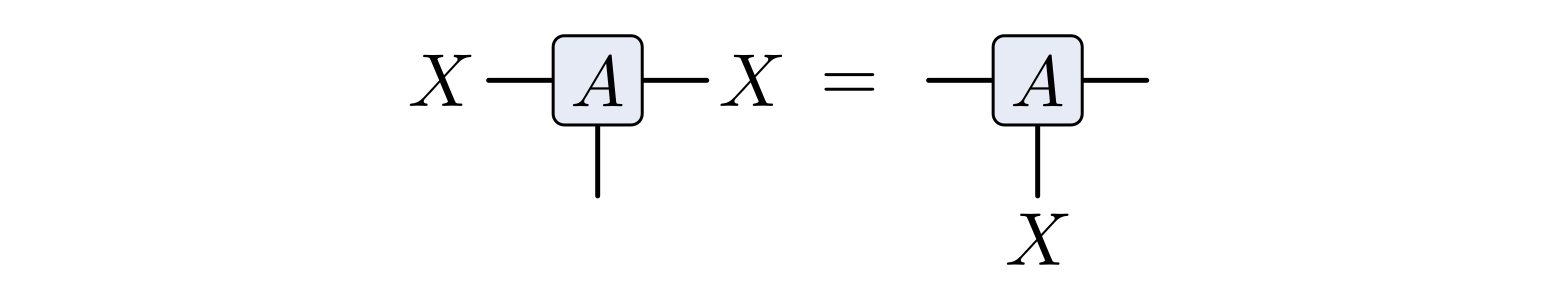} \\
 \includegraphics[width = 0.85\linewidth]{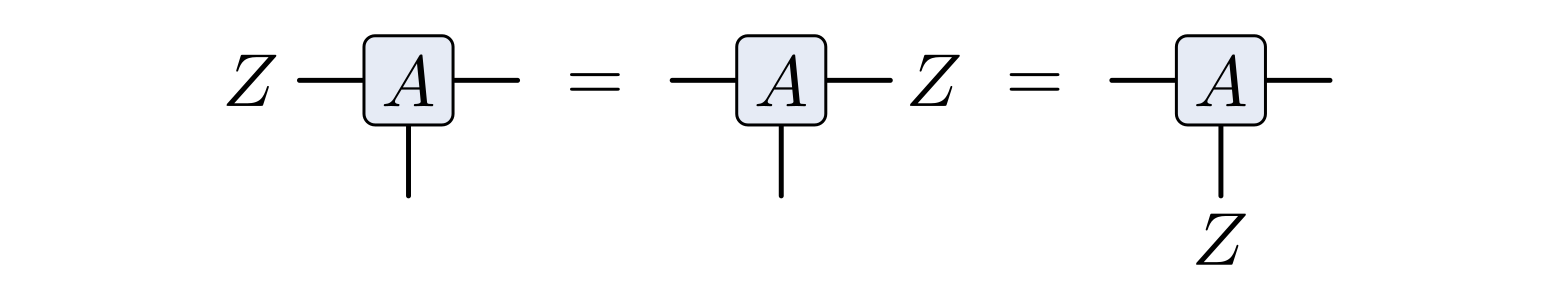}
\end{eqnarray}
\label{eq:GHZpushing}
\end{subequations}
which, in turn, generate $D^2$ pushing relations for the defects $B^{(i,j)}=X^{i}Z^j$. As these defects form a projective irreducible representation of $\mathbb{Z}_D\times \mathbb{Z}_D$, Theorem~\ref{thm:normal} is satisfied and $\ket{\Psi}$ can be prepared via Protocol~\ref{alg:normal}.

We note that for the task of preparing GHZ states, Protocol~\ref{alg:normal} is not optimal, as other measurement-based approaches (for example, see Ref.~\cite{Piroli_2021}) require fewer ancillae. This is due to the fact that generalized Bell measurements involve the determination of both $XX$ and $ZZ$, such that two ``$d$its'' of information are learned, requiring two ancillary bond qudits at each fusion site. However, due to the special form of $A$, the $XX$ information (which produces defects of the form $Z^k$) is not needed for fusion, and one can alternatively measure $ZZ$ alone (see, for example, Refs.~\cite{baumer2023efficient,chen2023realizing} for relevant experiments that prepare the $d=2$ GHZ state). One can view this strategy as a variant of Protocol~\ref{alg:normal} where the defects are not matrices but rank-3 tensors with a ``residual'' physical leg of dimension $d$ (encoding the $XX$ information). It can be shown that these defect tensors are equivalent to $A$ up to a random (correctable) permutation matrix on a virtual leg, enabling the preparation of GHZ-like states with only one $d$-dimensional ancilla per fusion site.

\subsubsection{Majumdar-Ghosh states}
\label{sssec:MG}
Next, we provide a simple example of non-normal MPS that can be prepared in constant depth with  Protocol~\ref{alg:nonnormal}. In particular, we consider the paradigmatic Majumdar-Ghosh (MG) state \cite{Majumdar_1969} and its higher-spin generalizations~\cite{karimipour2008matrix}. The former describes the exact (degenerate) ground state of a spin-1/2 Heisenberg chain with nearest and next-nearest neighbor interactions. A precursor to the AKLT state, the MG state and its parent Hamiltonian have played a historically important role toward the understanding of gapped spin-chains with continuous $SU(2)$ symmetry. Furthermore, this state can be exactly expressed as an MPS with $d=2$, $D=3$, and matrices
\begin{equation}
    \begin{gathered}
        A^0= \left(\begin{matrix}
        0 & 1 & 0 \\ 
        0 & 0 & 0 \\
        \frac{1}{\sqrt{2}} & 0 & 0
        \end{matrix}\right) \quad 
        A^1= \left(\begin{matrix}
        0 & 0 & 1 \\ 
        -\frac{1}{\sqrt{2}} & 0 & 0 \\ 
        0 & 0 & 0
        \end{matrix}\right).\\
    \end{gathered}
    \label{eq:MG}
\end{equation}

The MG state is both non-normal and two-periodic~\cite{PerezGarcia2006}. To handle the latter property, we block pairs of sites and additionally transform the physical index into a convenient basis, yielding
\begin{equation}
    \widetilde{A}^m = \left(\begin{matrix}
        \delta_{m0} & 0 \\ 
        0 & -P_m/2  \\ 
        \end{matrix}\right),
\end{equation}
where $P_m$ denotes a $2\times 2$ Pauli matrix, with $m\in\{0,1,2,3\}$ corresponding to $P_m\in\{I,X,Y,Z\}$, respectively. Importantly, each $\widetilde{A}^m$ is expressed in terms of $K=2$ blocks, allowing us to make the decomposition $\widetilde{A} = \widetilde{A}_0 \oplus \widetilde{A}_1$, with $\widetilde{A}^m_0 = \delta_{m0}$ and $\widetilde{A}^m_1=-P_m/2$. As each block has a different bond dimension, we inflate $\widetilde{A}_0$ (i.e., replace it by $\widetilde{A}_0\rightarrow I\otimes \delta_{m0}$) such that $D_{\alpha}=\bar{D}= 2$ for $\alpha=\{0,1\}$. Crucially, each intra-block tensor $\widetilde{A}_{\alpha}$ parameterizes a normal MPS.

With these preliminaries established, we now describe the preparation of this state using Protocol~\ref{alg:nonnormal}. First, we prepare a GHZ state composed of $N+2$ block qudits of dimension $K=2$. Following this, each block of the target MPS is prepared by block-conditionally applying the steps of Protocol~\ref{alg:normal} using bond (physical) qudits of dimension $\bar{D}=2$ ($d=2)$ and a blocking parameter of $q=2$. Notably, each intra-block MPS has zero correlation length, and arbitrary defects can therefore be corrected without reliance on the $SU(2)$-symmetry of the state. For example, we can apply fusion measurements in the standard Bell basis, leading to $2\times2$ Pauli defects that can be (locally) removed via block-conditioned feedforward operations applied to the physical qubits. Consequently, the Majumdar-Ghosh state satisfies Theorem~\ref{thm:nonnormal} by virtue of Corollary~\ref{corr:zcl}, and can be prepared in constant depth.

As discussed in Section~\ref{ssec:symmetry}, MPS with on-site symmetry are characterized by representations of the symmetry that manifest on the physical and virtual level. In the above example, we have focused on the standard Majumdar-Ghosh state, characterized by spin-$j$ representations of $SU(2)$ with $j_{\textrm{phys}} = \frac{1}{2}$ and $j_{\textrm{virtual}} = 0\oplus \frac{1}{2}$. However, the physical and virtual representations of $SU(2)$ can be generalized to $j_{\textrm{phys}}=s$, $j_{\textrm{virtual}} = 0\oplus s$ for arbitrary $s$, resulting in a family of fully dimerized states known as generalized Majumdar-Ghosh states \cite{karimipour2008matrix}; all can be prepared via Protocol~\ref{alg:nonnormal}. See also Ref.~\cite{Sanz_2009} for relevant discussion pertaining to the construction of general MPS with $SU(2)$ symmetry.

\subsubsection{$\mathbb{Z}_4\times \mathbb{Z}_2 \to \mathbb{Z}_2\times \mathbb{Z}_2$ symmetry-breaking}
\label{sssec:Z4xZ2}
Next, we turn to an illustrative example that highlights the utility of Protocol~\ref{alg:nonnormal} for preparing nontrivial symmetry-broken states. In particular, we leverage recent results from Ref.~\cite{Gunn2023}, where a classification of the phases of non-normal MPS under $G=\mathbb{Z}_4\times\mathbb{Z}_2$ symmetry was presented. Furthermore, it was shown that the fixed points of these phases can be prepared in constant time using symmetric local unitaries, measurements, and feedforward operations. Here, we turn to the preparation of an explicit non-fixed-point MPS belonging to one of these phases, constructed using the technique described in Appendix~\ref{app:symconst}. In particular, we consider the symmetry-broken phase that preserves the subgroup $H=2\mathbb{Z}_2\times\mathbb{Z}_2 \leq G$, and narrow our focus to a $d=3$, $D=4$ non-normal MPS in this phase. This state, which we denote by $\ket{\Psi_{\textrm{SB}}}$, is parameterized by the matrices, 
\begin{equation}
    \begin{gathered}
        A^+=\sqrt{\frac{2}{3}}\left(\begin{matrix}
        \sigma_+ & 0 \\
        0 & i\sigma_-
        \end{matrix}\right) \quad
        A^-=-\sqrt{\frac{2}{3}}  \left(\begin{matrix}
        \sigma_- & 0 \\
         0 & i\sigma_+
        \end{matrix}\right)\\
        A^0=-\sqrt{\frac{1}{3}}\left(\begin{matrix}
        Z & 0 \\
         0 & Z
        \end{matrix}\right).
    \end{gathered}
    \label{eq:Z4xZ2}
\end{equation}

Interestingly, the upper block of this MPS corresponds exactly to the matrices of the AKLT state (see Section~\ref{sssec:AKLT}). The same is true for the lower block up to a unitary transformation on the physical spin-1 degree of freedom. Similar to the AKLT state, $\ket{\Psi_{\textrm{SB}}}$ is characterized by a nontrivial hidden antiferromagnetic ordering, where $+$ and $-$ alternate with any number of intermediate sites in the $0$ state~\cite{Nijs1989, Smith_2023}. Distinct from the AKLT state, however, $\ket{\Psi_{\textrm{SB}}}$ has an additional intriguing property that is due to the relative phase between each block. Assuming periodic boundary conditions and denoting the normal MPS of block $\alpha$ by $\ket{\Psi_{\alpha}}$, the \emph{parity} of total `$+-$' pairs is conserved across all allowed configurations in the non-normal MPS $\ket{\Psi_{\textrm{SB}}^{(\pm)}}=[\ket{\Psi_{0}} \pm \ket{\Psi_{1}}]/\sqrt{2}$. Specifically, the state $\ket{\Psi_{\textrm{SB}}^{(+)}}$ includes only configurations with even pairs, while $\ket{\Psi_{\textrm{SB}}^{(-)}}$ contains only those with odd pairs. Thus, one must acquire nonlocal information in order to distinguish between these two states.

The $\mathbb{Z}_4\times \mathbb{Z}_2$ symmetry of this non-normal MPS manifests on the virtual level via the projective representation~\cite{Gunn2023},
\begin{equation}
    V_{(a,b)} =   \left(\begin{matrix}
        X^b Z^{\lfloor a/2 \rfloor} & 0 \\
        0 & X^b Z^{\lfloor (a+1)/2\rfloor} \\
        \end{matrix}\right)\left(X^{a}\otimes \mathbb{1}\right),
    \label{eq:V_Z4xZ2}
\end{equation}
where $a\in \left\{0,1,2,3\right\}$ and  $b\in\left\{0,1\right\}$. Here, the intra-block operators, written in terms of $2\times 2$ Pauli matrices $X$ and $Z$, form an irreducible projective representation of $H=2\mathbb{Z}_2\times \mathbb{Z}_2$. While we do not write them out here, for each $g=(a,b)\in G$, there exists a corresponding symmetry operation $U_g$ on the physical level that together form a linear representation of $G$. As discussed in Section~\ref{ssec:nonnormal}, for $g \notin H$ (i.e., for odd $a$), $U_g$ induces a permutation between the symmetry-broken states within each block. On the virtual level, this is carried out by the operator $\widetilde{P}_{g} = (X^a \otimes \mathbb{1})$ on the right-hand side of Eq.~\eqref{eq:V_Z4xZ2}. 

Graphically, we can represent this symmetry action as a set of $|G|$ pushing relations of the form,
\begin{equation}
\includegraphics[width = 0.85\linewidth]{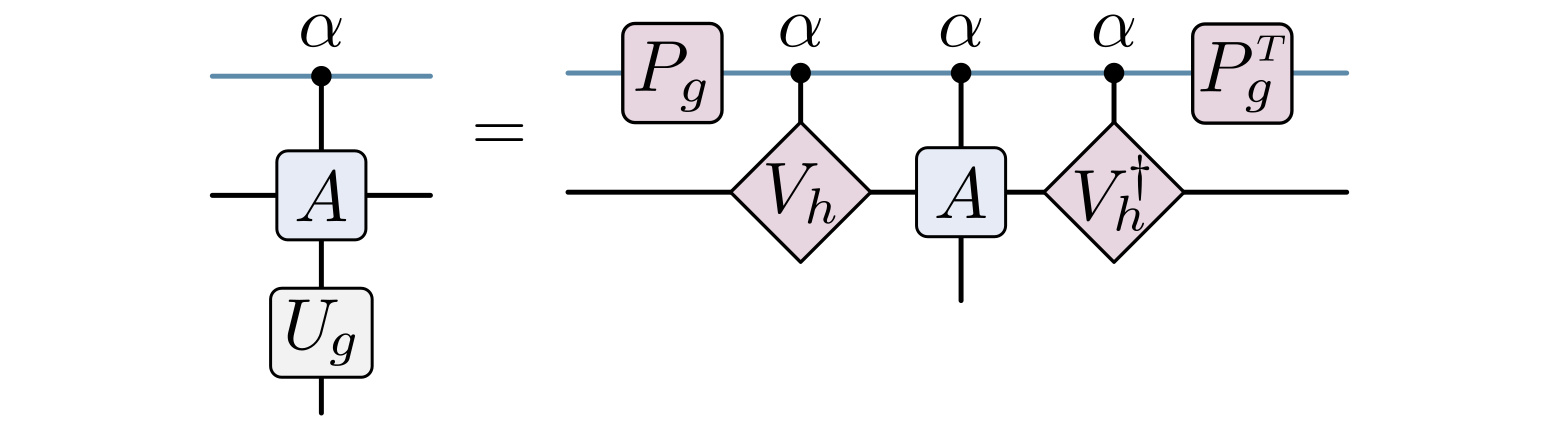}, 
\label{eq:Z4xZ2pushing}
\end{equation}
where $V_{h(g,\alpha)} = X^b Z^{\lfloor (a+\alpha)/2\rfloor}$ and $P_g = X^a$ for $g=(a,b)$. As the virtual operators $V_h$ form an irreducible projective representation of the finite group $H$, this indicates that the normal MPS within each block $\alpha$ can be prepared via Protocol~\ref{alg:normal} -- an unsurprising statement, as each is equivalent to an AKLT state up to a local unitary transformation. This then guarantees that, by way of Theorem~\ref{thm:nonnormal}, the non-normal MPS  $\ket{\Psi_{\textrm{SB}}}$ can be prepared using Protocol~\ref{alg:nonnormal} (with $\bar{D}=2$ and $K=2$). 

The only remaining missing component for Protocol~\ref{alg:nonnormal} is a set of block-controlled operator pushing relations. In general, such relations can be derived by following the procedure outlined in Section~\ref{ssec:symmetry} and Appendix~\ref{app:block-controlled}. However, this is unnecessary in the present case -- due to the simplicity of that target state  $\ket{\Psi_{\textrm{SB}}}$, all defects can be corrected without relying on block-conditional operations. To see this, we first note that both of the normal MPS underlying $\ket{\Psi_{\textrm{SB}}}$ carry the same representation of $H$. As a result, it is not necessary to block-control the measurement basis in Step (3) of Protocol~\ref{alg:nonnormal}, and both blocks will share the same defect after fusion measurements. We can then correct these defects by simply picking out the $|H|$ pushing relations from Eq.~\eqref{eq:Z4xZ2pushing} for which there is no permutation action. This leads to a set of $|H|$ pushing relations that are block-independent:
\begin{equation}
\includegraphics[width = 0.85\linewidth]{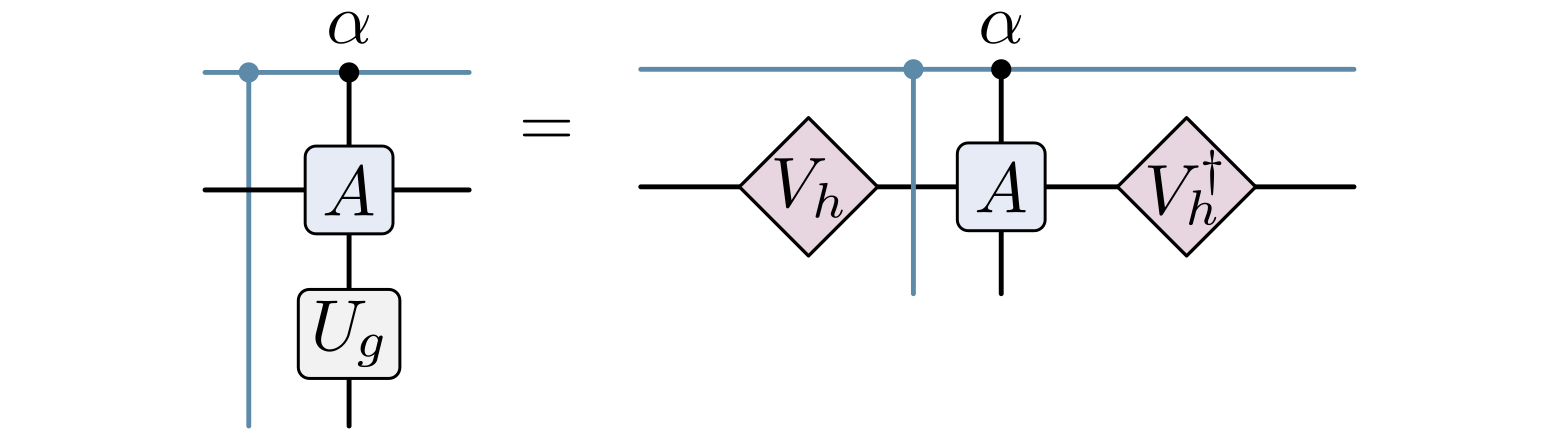}, 
\label{eq:contr2_Z4xZ2pushing}
\end{equation}
Thus, the only component of Protocol~\ref{alg:nonnormal} that requires a block-controlled operation is the initial preparation of small MPS in Step (2). In general, such drastic simplification is not possible, particularly for MPS with non-Abelian symmetries, or those constructed from normal MPS that are unrelated by symmetry -- cases that can be handled by the more general form of Protocol~\ref{alg:nonnormal}.

\subsubsection{Random MPS}
\label{sssec:rmps}
We now explore a compelling application of our protocol: the preparation of random matrix product states (RMPS). RMPS have proven a useful concept in various contexts, including statistical quantum physics \cite{garnerone2010statistical, haferkamp2021emergent} and tensor-network based machine learning \cite{liu2022presence, garcia2023barren}. Furthermore, it was recently shown that, on average, RMPS are highly magical -- i.e., their non-stabilizerness generically grows exponentially with system size \cite{Chen2022a}, and there has been increased interest in their properties and utility as a result ~\cite{lami2024quantum, frau2024non}. Consequently, finding an efficient scheme to prepare RMPS is not only pertinent to the aforementioned applications, but would additionally enable the rapid generation of useful quantum resources. To that end, we now show that our fusion-based strategy provides such a scheme, enabling the preparation of RMPS in constant time.

We first recall the unitary embedding illustrated in Fig.~\ref{fig:f2}(a) and reproduced here for convenience,
\begin{equation}
\includegraphics[width = 0.85\linewidth]{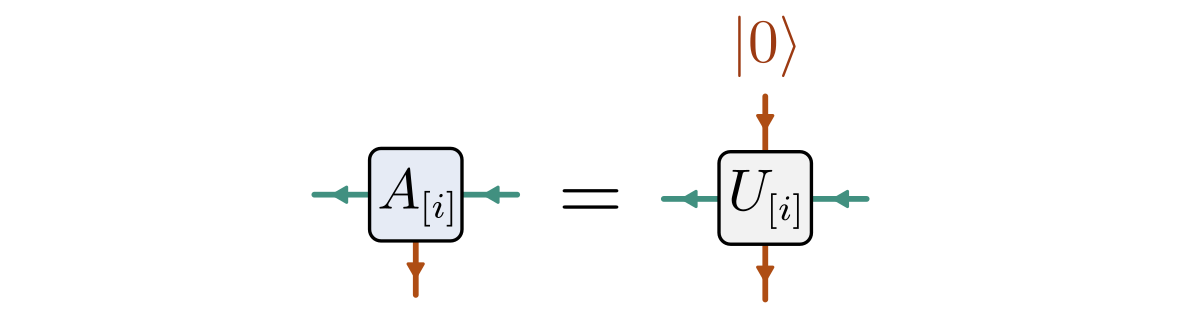},
\label{eq:unitaryembedding}
\end{equation}
where the subscript $[i]$ denotes the $i$th site. Due to the above relation, an MPS can be parameterized either in terms of the rank-3 tensors $A_{[i]}$ or the embedding unitaries $U_{[i]}$. RMPS are typically defined in terms of the latter: for each site $i$, we randomly sample $U_{[i]}\in U(dD)$, where $U(dD)$ is the Haar measure for $dD\times dD$ unitary matrices, with $d$ and $D$ the physical and bond dimensions of the MPS, respectively.

Our strategy to prepare RMPS is simple -- we follow the first few steps of Protocol~\ref{alg:normal}, beginning with the preparation of small RMPS and following with fusion measurements in a basis of maximally entangled states (such as the generalized Bell basis, yielding $D\times D$ qudit Pauli defects). In contrast to the scenario where we wish to prepare a \emph{particular} MPS, however, it is not necessary to correct the defects: if $U$ is a Haar random unitary and $B$ any unitary, then $UB$ is itself a Haar random unitary by the translational invariance of the Haar measure \cite{meckes2019random}. Consequently, one can just ``absorb'' random defects into adjacent sites, and measure the dangling edge bond qudits to project into definite boundary conditions. The resulting state is an RMPS.

Finally, we note that the above strategy can be generalized for the constant-depth preparation of higher-dimensional random tensor network states, which have found applications toward the study of holography \cite{hayden2016holographic} and entanglement phase transitions \cite{vasseur2019entanglement, Yang2022}.

\subsubsection{Random sampling from an SPT phase}
\label{sssec:Haldane}
Our strategy for preparing RMPS can be incorporated into other MPS sampling protocols, such as from a nontrivial SPT phase. As an example, consider the spin-1 Haldane phase. Away from the AKLT point, it has been shown that the MPS tensors factorize into \emph{protected} and \emph{junk} subsystems \cite{Else2012}:
\begin{equation}
A^m = A^m_{\textrm{AKLT}}\otimes A^m_{\textrm{junk}}.
\end{equation}
Here, $A^m_{\textrm{AKLT}}$ denotes the $2\times2$ matrices of the AKLT state and encodes the subsystem protected by $\mathbb{Z}_2\times\mathbb{Z}_2$ symmetry. Separately, the matrices $A^m_{\textrm{junk}}$ encode the junk space, within which the symmetry acts trivially (i.e., the virtual operators decompose into trivial representations of $\mathbb{Z}_2\times\mathbb{Z}_2$). Then defects are correctable within the protected subspace, but not within the junk space. As a result, we cannot deterministically prepare an arbitrary MPS within the Haldane phase using Protocol~\ref{alg:normal}. Instead, we can adapt our scheme to randomly sample from the Haldane phase by (i) choosing the matrices $A_{\textrm{junk}}^m$ at random, (ii) choosing a fusion measurement basis with defects that factorize as $B_{g,k} = V_g\otimes B_{\textrm{junk}}$, where the operator $V_g$ form an irreducible projective representation of $\mathbb{Z}_2\times\mathbb{Z}_2$, and (iii) correcting (absorbing) defects within the protected (junk) subspace. Though this scheme cannot guarantee translational invariance, the Haldane phase does not require this symmetry to be respected. The extension of this sampling strategy to other SPT and non-normal phases is an interesting direction for future exploration.

\section{Conclusion}\label{sec:conclusion}
In this work, we have introduced a framework for the exact preparation of certain MPS using constant-depth adaptive quantum circuits. Building upon the protocol introduced in Ref.~\cite{Smith_2023} to prepare the AKLT state, this framework relies on an extremely simple concept -- using measurements to fuse together small MPS prepared in parallel and subsequently employing feedforward operations to correct random defects in the post-measurement state. Leveraging this strategy, we have presented two explicit preparation protocols: one for normal MPS with short-range correlations and another for non-normal MPS with GHZ-like long-range correlations. Despite their reliance on non-unitary resources, we have shown that for target states with certain well-defined properties, both protocols are deterministic in the bulk (i.e., up to a probabilistic selection of boundary conditions), thus enabling the exact preparation of a broad set of physically interesting, nontrivial entangled states in a time independent of system size. 

Notably, this constitutes a significant (i.e., super-exponential) improvement in preparation time over exact linear-depth~\cite{Schoen2005} and approximate log-depth~\cite{Malz_2024} circuit-based protocols to prepare MPS, while incurring only a constant factor in total qudit count. Of course, it also requires mid-circuit measurements and feedforward operations, capabilities that are supported on several available cloud-based quantum computing platforms~\cite{Corcoles_2021, Moses2023}. Perhaps most interesting is the fact that the class of states preparable with our scheme encompasses instances of MPS with either nonzero correlation length and/or long-range correlations -- both scenarios for which faithful constant-depth preparation is provably impossible with local unitary gates alone~\cite{Bravyi2006,yu2023learning, Malz_2024}. Thus, our approach not only provides a significant speed-up over existing protocols for preparing a variety of physically interesting MPS but, more generally, underscores a key advantage of adaptive quantum circuits over their unitary counterpart: the ability to prepare non-local quantum correlations in constant time.

It is important to emphasize that our scheme does not enable the preparation of \emph{arbitrary} MPS in constant time, but is instead limited to \emph{certain} target states that are endowed with particular ``pushing relations'' -- a key concept that we leverage to correct random post-measurement defects. In that vein, we have presented a set of sufficient and easily verifiable conditions for particular MPS to be prepared in constant depth via our framework. Furthermore, we have delved into special cases that guarantee these conditions, including (i) fixed-point MPS with zero correlation length and (ii) MPS with global on-site symmetry. While it has previously been shown that the former class can be prepared in constant depth using measurements and feedforward~\cite{Piroli_2021, Gunn2023}, preparation of the latter class constitutes one of our key results. 

In particular, we have shown that independent of correlation length, any normal MPS with global on-site symmetry can be exactly prepared in constant depth via our scheme, as long as that symmetry manifests as an \emph{irreducible} representation on the virtual level and is finite (or, alternatively, has a finite subgroup). Furthermore, we have demonstrated that it is possible to construct preparable non-normal MPS by taking arbitrary superpositions of preparable normal MPS. Altogether, this indicates that, for any bond dimension, there exist broad families of both normal and non-normal MPS with on-site symmetries that can be prepared in constant depth via our scheme. To that end, we have additionally provided a recipe for constructively generating such states -- see Appendix~\ref{app:symconst}.

To highlight the diversity of nontrivial states preparable with our scheme, we have provided a variety of illustrative examples that include symmetry-protected topological and symmetry-broken states, MPS with finite Abelian, non-Abelian, and continuous symmetries, resource states for MBQC, and families of states with tunable correlation length. Moreover, we have demonstrated the applicability of our framework toward the design of sampling protocols, showcasing a capability to sample both random MPS and those from a particular SPT phase in constant depth.

This work opens up a number of promising avenues for future research. For one, it would be interesting to incorporate our constant-depth protocols into the emerging applications for MPS on quantum hardware. These include, for example, tensor-network-inspired variational quantum algorithms~\cite{haghshenas2022variational, rudolph2023synergistic, FossFeig2021a, khan2023preoptimizing}, quantum simulation of time dynamics~\cite{Chertkov2022, Martin2023}, and loading of classical data for quantum machine learning~\cite{Martyn2020, Holmes2020, Dilip2022, iaconis2023tensor, Jobst2023}. In addition, our scheme provides an extremely efficient route to prepare families of resource states for MBQC and, in some sense, generalizes the core idea underlying fusion-based quantum computation~\cite{bartolucci2023fusion} beyond the cluster state. It would be interesting to pursue this connection further, particularly in the context of linear optical platforms where one relies on joint measurements to carry out entangling operations, in close alignment with our scheme. Similarly, we note that our protocol constitutes a particularly attractive approach for the efficient preparation of large entangled states spanning distributed quantum hardware, as sections of MPS can be independently prepared and subsequently fused across multiple local quantum processing units without a direct link between physical sites.

Furthermore, this work leaves open several intriguing questions. In particular, a complete classification of MPS that can be faithfully prepared with constant-depth adaptive quantum circuits is beyond the scope of this work. Here, we have presented a general framework that enables the preparation of a wide variety of MPS, and have furthermore provided conditions for target states that are sufficient, but not strictly necessary. For example, Corollary~\ref{corr:symmetry} requires that a global on-site symmetry presents as an irreducible representation on the virtual level. However, in Section~\ref{sssec:Z2} we provided an explicit example of a family of $\mathbb{Z}_2$-symmetric MPS that can be prepared via our scheme, despite carrying a reducible representation on the virtual level. Thus, our conditions do not encompass all states that can be prepared, and we leave open the task of complete classification to future work. Moreover, it may be possible to generalize the protocols presented here in a fashion specifically tailored to reducible representations. Separately, in this work we have considered MPS with constant bond dimension $D$ that is independent of system size $N$. For the more general scenario where $D\sim \textrm{poly}(N)$, our approach naively allows for the constant-depth preparation of such states, but with the unrealistic requirement for (bond) qudits of dimension $\textrm{poly}(N)$, hindering practicality. If we instead encode each bond qudit using $\textrm{log}(D)$ qubits, our scheme requires a circuit depth of $O(d^2D^2)$~\cite{khan2023preoptimizing}. Furthermore, our approach requires a $O(D^2)$ sampling overhead for the selection of particular boundary conditions, and is therefore constant-time only for constant $D$. Thus, alternative strategies to efficiently prepare more general MPS with measurement and feedforward constitutes an interesting direction for future study.

An important practical direction is to benchmark our constant-depth approach on quantum hardware, and furthermore compare it to purely unitary linear-depth (or log-depth) preparation of MPS. Throughout this work, we have implicitly assumed classical feedforward to be a ``free'' and instantaneous resource. In practice, however, integration into quantum hardware can be challenging -- for example, in the context of preparing the GHZ state, Ref.~\cite{baumer2023efficient} found that linear-depth unitary circuits outperformed constant-depth dynamic circuits on an IBM Quantum processor; this was primarily attributed to an inefficient implementation of classical feedforward, with a temporal cost scaling with the number of possible mid-circuit measurement outcomes (i.e., exponentially in system size). However, this limitation is not a fundamental one, and it is expected that this scaling will soon be improved to either linear or constant in system size~\cite{baumer2023efficient}. Furthermore, for small system sizes, the question of whether there is an advantage in replacing a unitary circuit with an adaptive circuit that includes many mid-circuit measurements depends inextricably on details specific to that platform, such as qubit coherence times and mid-circuit measurement fidelities. Given these nuanced considerations, it would be interesting to investigate how such hardware-specific details influence the ``break-even'' point of our scheme -- i.e., the system size for which constant-depth adaptive preparation outperforms its unitary counterpart.

Finally, an especially intriguing next step would be to investigate whether our framework can be extended to efficiently prepare other tensor network states beyond MPS. Seemingly, this extension is straightforward in the case of tree tensor networks, as one can employ the same strategies here as long as the target state is characterized by a sufficient set of pushing relations. In contrast, extension to higher-dimensional projected-entangled pair states (PEPS) and the multiscale entanglement renormalization ansatz (MERA) is less trivial. While the constant-depth preparation of certain PEPS such as the 2D toric code ground state~\cite{Verresen2021, FossFeig2023} can be expressed in a language similar to the one developed here~\cite{Lu2022}, extension to PEPS with on-site symmetry (e.g., such as the spin-3/2 AKLT state on a honeycomb lattice) presents additional challenges. In particular, closed loops can effectively ``trap'' random defects, inhibiting the ability to push them to the edge for removal. We expect similar challenges to arise in the case of MERA. Thus, it remains an open problem as to whether this framework is useful for preparing higher dimensional, non-fixed-point tensor network states. We leave these questions for future work.

\vspace{10pt}
\noindent\emph{Note added} -- The posting of this preprint to the arXiv was coordinated with simultaneous postings by Sahay \emph{et al.}~\cite{sahay_mps, sahay_mps2} and Stephen \emph{et al.}~\cite{stephen_mps}. Both discuss the measurement-based preparation of matrix product states, and were developed independently from the work here.

\section*{Acknowledgements}
We would like to thank Ben McDonough, Paul Anderson, Tyler Ellison, Norm Tubman, Efekan Kokcu, Omar Alsheikh, and Alexander Kemper for helpful conversations pertaining to this work. T.-C.W. thanks Misha Litvinov and Yabo Li for useful discussions. This project was funded by the U.S. Department of Energy, Office of Science, National Quantum Information Science Research Centers, Co-design Center for Quantum Advantage under contract number DE-SC0012704. C2QA led in this research. A.K. and B.K.C acknowledge support from the NSF Quantum Leap Challenge Institute for Hybrid Quantum Architectures and Networks (NSF Award 2016136).
T.-C.W. acknowledges the support of the National Science Foundation under Award No. PHY 2310614 for the part on the AKLT state and its generalization. 

External interest disclosure: SMG is a consultant for, and equity holder in, Quantum Circuits, Inc.

\appendix

\section{Proofs of Theorem \ref{thm:existence} and 
\ref{thm:unitarity}}\label{app:proofs}
We begin by proving Theorem~\ref{thm:existence}. Starting from Eq.~(\ref{eq:op_pushing}) and dropping the superscript $q$ for clarity, we can first note that $\map{A}$ is by construction surjective and is therefore guaranteed to have a right-inverse $\map{A}_R^{-1}$. Right-multiplying Eq.~(\ref{eq:op_pushing}) by $\map{A}_R^{-1}$, we find
\begin{equation}
O_p = \map{A}O\map{A}_R^{-1},
\label{app:eq:Otilde}
\end{equation}
where we have adopted the shorthand $O = (O_\ell^T\otimes O_r)$. While this yields a definition for $O_p$, it is not guaranteed to provide a solution to Eq.~(\ref{eq:op_pushing}) as $\map{A}$ is not left-invertible unless it is also injective. To check the conditions under which it is a valid solution, we substitute this definition into Eq.~(\ref{eq:op_pushing}) to find
\begin{equation}
\map{A}O \map{A}_R^{-1}\map{A} = \map{A}O.
\end{equation}
Identifying $\mathcal{P} = \map{A}_R^{-1}\map{A}$ as the projector onto the rowspace of $\map{A}$ (i.e., the set of virtual states in $\mathcal{H}_{D}\otimes \mathcal{H}_{D}$ mapped onto the physical basis states spanning $\mathcal{H}_d$), we can rewrite the above equation as 
\begin{equation}
    \map{A}O\mathcal{P}_c=0,
    \label{eq:app:projector_relationship}
\end{equation}
where $\mathcal{P}_c = \mathbb{1} - \mathcal{P}$ is the projector onto the kernel of of $\map{A}$. Already, this provides the necessary condition for existence of $O_p$: left-multiplying by $\map{A}_R^{-1}$, we find $\mathcal{P}O\mathcal{P}_c = 0$. This result is intuitively sensible -- $O_{\ell}$ and $O_r$ can induce transitions between elements within the rowspace and kernel, and can additionally map elements from the rowspace $\emph{to}$ the kernel (as these will be annihilated by $\map{A}$ and can similarly be enacted by projecting out those same elements at the level of $O_p$). However, noting that elements of the physical space $\mathcal{H}_d$ and rowspace of $\map{A}$ are in one-to-one correspondence, $O_p$ cannot possibly map elements from the kernel of $\map{A}$ to its rowspace, leading to the determined imposed condition on the virtual operations.

In the case where $O$ is unitary, $\mathcal{P}O\mathcal{P}_c = 0$ implies that $\mathcal{P}_c O \mathcal{P} = 0$. To see this, note that $O$ can be expressed as a lower-triangular block matrix, with $\mathcal{P}O\mathcal{P}$, $\mathcal{P}_cO\mathcal{P}$, $\mathcal{P}_cO\mathcal{P}_c$ corresponding to the upper-left, lower-left, and lower-right blocks. The inverse of a lower-triangular block matrix is itself lower-triangular block matrix. However, $O^\dagger$ is clearly an upper-triangular matrix, and $O$ must therefore be block-diagonal (i.e., $\mathcal{P}_c O \mathcal{P} = 0$). Then $O = \mathcal{P}O\mathcal{P} + \mathcal{P}_cO\mathcal{P}_c$, equivalent to the commutation condition $[O, \mathcal{P}] = 0$ which, cast back in terms of $\map{A}$, becomes
\begin{equation}
    [O, \map{A}_R^{-1}\map{A}] = 0,
    \label{app:eq:existence_commutator}
\end{equation}
concluding our proof of Theorem~\ref{thm:existence}.

Turning to Theorem~\ref{thm:unitarity}, we first assume the existence of an operator $O_p$ defined by Eq.~(\ref{app:eq:Otilde}) such that Eq.~(\ref{eq:op_pushing}) holds via Theorem~\ref{thm:existence}. Further demanding that $O_p$ is unitary then requires
\begin{equation}
    (\map{A}_R^{-1})^\dagger O^\dagger \map{A}^\dagger \map{A} O \map{A}_R^{-1} = \mathbb{1}.
    \label{app:eq:unitarity_requirement}
\end{equation}
Noting that $\map{A}$ and $\map{A}^{\dagger}$ are right- and left-invertible, respectively, we can right-multiply by $\map{A}$ and left-multiply by $\map{A}^{\dagger}$ to find
\begin{equation}
    \begin{split}
        \map{A}^\dagger \map{A} &= (\map{A}_R^{-1}\map{A})^\dagger O^{\dagger} \map{A}^{\dagger}\map{A} O \map{A}_R^{-1}\map{A} \\
        &= O^{\dagger} \map{A}^{\dagger}\map{A} O,
    \end{split}
\end{equation}
where we have used the fact that $[O,\map{A}_R^{-1}\map{A}]=0$ in going from the first to the second line. Using the fact that $O$ is unitary, simple rearrangement gives
\begin{equation}
    [O,\map{A}^\dagger \map{A}] = 0,
    \label{app:eq:unitary_commutator}
\end{equation} concluding our proof of Theorem~\ref{thm:unitarity}.

Finally, we note that the commutator in Eq.~(\ref{app:eq:unitary_commutator}) implies that of Eq.~(\ref{app:eq:existence_commutator}) when $O$ is unitary, which is not immediately obvious at first glance. To see this, it is simplest to prove that Eq.~(\ref{app:eq:unitary_commutator}) implies Eq.~(\ref{eq:app:projector_relationship}), which in turn ensures commutation of $O$ and $\map{A}_R^{-1}\map{A}$, as we have shown. Using the fact that $(\map{A}\map{A}_R^{-1})^\dagger$ is the identity, we insert this into the left-hand side of Eq.~(\ref{eq:app:projector_relationship}) to find
\begin{equation}
    \begin{split}
        \map{A} O \mathcal{P}_c &= (\map{A}_R^{-1})^{\dagger}\map{A}^\dagger \map{A} O \mathcal{P}_c \\
        &= (\map{A}_R^{-1})^{\dagger}O \map{A}^\dagger \map{A} \mathcal{P}_c =0 \\
    \end{split}
\end{equation}
where we have used the commutation relation of Eq.~(\ref{app:eq:unitary_commutator}) in going from the first line to the second, and have furthermore used the fact that $\mathcal{P}_c$ is the projector onto the kernel of $\map{A}$. Therefore, for a unitary virtual operator $O$, the condition Eq.~(\ref{app:eq:unitary_commutator}) alone ensures that $O_p$ exists and is unitary.

\section{Block-controlled pushing relations for symmetry-broken states}
\label{app:block-controlled}

In this appendix, we aim to provide additional details on the derivation of the pushing relations in Eq. \ref{eq:nonnormalpushing2}. As mentioned in the main text, the details surrounding the action of the symmetry on the virtual level can be understood through the language of induced representations -- for a complete discussion, we refer to Refs.~\cite{Schuch2010, Gunn2023}. Here, we will provide the necessary details to understand the connection between the $|G|$ pushing relations of the form in Eq.~(\ref{eq:nonnormalpushing}) and the $|H|$ pushing relations in  Eq.~(\ref{eq:nonnormalpushing2}), with $|H|\leq |G|$. To do this, we will lean heavily on the arguments of the aforementioned references.

First, we make a few remarks regarding the subgroup $H$ and its relation to $G$. In brief, the basic idea is this: as shown in Eq. \ref{eq:nonnormalsymmetry}, the physical symmetry operators $U_g$ manifest on the virtual level as a combination of two effects: conjugation by operators $V_{h(g,\alpha)}$ within each block (up to a phase $e^{i\phi_g^\alpha}$, as in the normal case), and permutations among the blocks. Let us denote the permutation action by $\pi_g(\alpha) = \gamma$, i.e., the physical symmetry $U_g$ maps block $\alpha$ to block $\gamma$. Clearly, $\gamma$ is a function of both $g$ and $\alpha$, and we therefore define $\gamma\equiv \gamma(g,\alpha)$. 

We choose a particular block $\alpha_0$ and identify the elements $g\in G$ for which $\gamma(g,\alpha_0) = \alpha_0$, i.e., those whose permutation action map $\alpha_0$ to itself. As discussed in Ref.~\cite{Schuch2010, Gunn2023}, this naturally defines the subgroup \begin{equation}
    H~= \{h:h\in G\, | \,\pi_h(\alpha_0) = \alpha_0\}\leq G.
\end{equation}
The virtual intra-block operators $V_{h(g,\alpha)}$ form a projective representation of $H$. 

Next, we note that $G$ can be broken into disjoint left-cosets $k_{\alpha} H$, where we have chosen a set of representatives $k_{\alpha}\in G$ such that $\pi_{k_{\alpha}}(\alpha_0) = \alpha$. In other words, for each block $\alpha$, $k_\alpha$ specifies the element such that the corresponding physical operators $U_{k_{\alpha}}$ send our selected block $\alpha_0$ to $\alpha$. These representatives can then be used to uniquely determine $\gamma(g,\alpha)$ and $h(g,\alpha)$ via the relation \cite{Schuch2010}
\begin{equation}
    g k_{\alpha} = k_{\gamma(g,\alpha)} h(g,\alpha).
    \label{eq:inducedmaps}
\end{equation}
Crucially, this clarifies the permutation action -- each element $g\in 
G$ maps elements of the coset $k_\alpha H$ to the coset $k_{\gamma(g,\alpha)} H$, and the permutation operators mirror this mapping, sending block $\alpha$ to $\gamma(g,\alpha)$.

Returning to the topic of pushing relations, the determination of $\gamma(g,\alpha)$ and $h(g,\alpha)$ allows us to completely specify $|G|$ distinct instances of Eq.~(\ref{eq:nonnormalpushing}). However, we now show that if one can conditionally apply the physical unitary $U_g$ to each block, it is possible to derive a second set of pushing relations that act invariantly on the block structure with no permutation action. 

To see this, we first emphasize that the subgroup $H$, by construction, defines a set of elements such that acting $U_h$ on the physical leg maps block $\alpha_0$ to itself. Notably, this does not necessarily guarantee that $U_h$ leaves all other blocks invariant, as $U_h$ can still permute a subset of blocks. However, if we \emph{conditionally} apply $U_h$ only to the block $\alpha_0$, this leads to the pushing relation
\begin{equation}
\includegraphics[width = 0.85\linewidth]{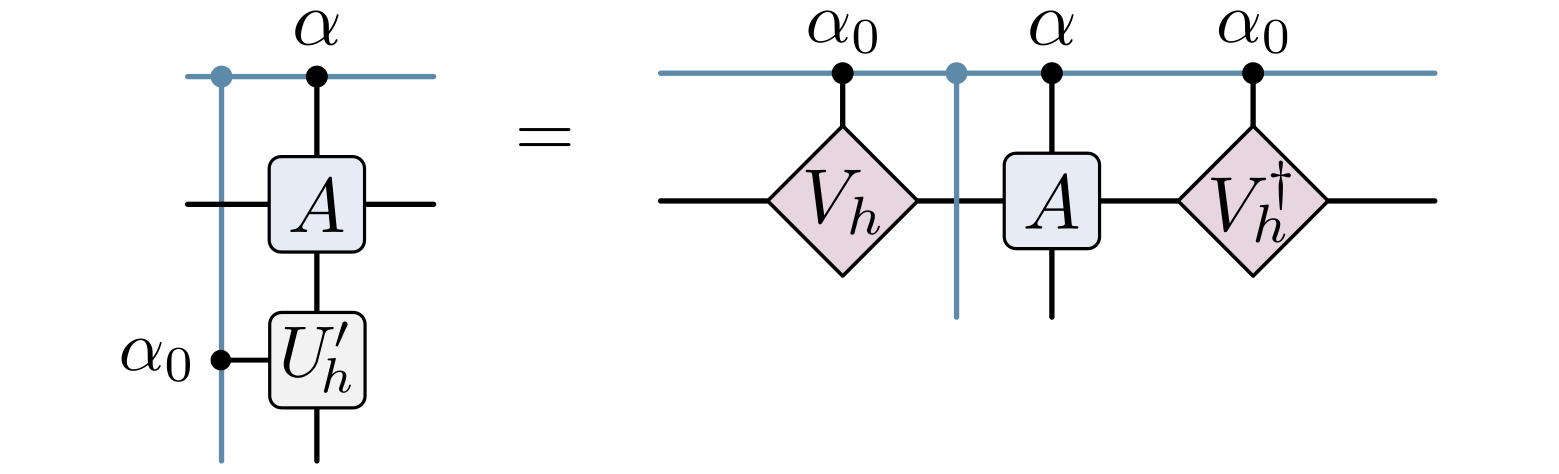},
\label{eq:cond_pushing_alpha0}
\end{equation}
where we have used a control on $\alpha$ to indicate a tensor with nontrivial components for all $\alpha$ (i.e., $A = \bigoplus_{\alpha}A_{\alpha}$), and have used a control on $\alpha_0$ alone to denote a tensor that has a nontrivial component only for block $\alpha_0$. Furthermore, we have redefined the physical unitary such that there is no residual phase, $U_h' = U_h e^{-i\phi_h^{\alpha_0}}$. This gives us the freedom to push an irrep of $H$ through block $\alpha_0$ without causing permutations.

We now aim to extrapolate this ability to all of the blocks. In analogy to the above exercise with block $\alpha_0$, this is achieved by finding the subgroup $H_{\alpha}\leq G$ that maps a general block $\alpha$ to itself. We emphasize that these groups are all isometric and, because the choice of $\alpha_0$ was arbitrary, Eq.~(\ref{eq:cond_pushing_alpha0}) is already sufficient to show that the pushing relations we seek exist. For clarity of explanation, however, we will make explicit their construction.

For a given block $\alpha$, we can define the subgroup
\begin{equation}
H_\alpha = \{k_{\alpha} h k_{\alpha}^{-1} : h\in H\} = k_{\alpha} H k_{\alpha}^{-1}.
\end{equation}
By inserting the elements of this group into Eq.~(\ref{eq:inducedmaps}), it is straightforward to see that this group provides the desired property: $\gamma(k_{\alpha} h k_{\alpha}^{-1},\alpha) = \alpha$, and thus can be used to write a pushing relation analogous to Eq.~\eqref{eq:cond_pushing_alpha0}, but for an arbitrary block $\alpha$. By applying many such pushing relations in parallel, each conditioned on a distinct block $\alpha$, we arrive at the desired result:

\begin{equation}
\includegraphics[width = 0.85\linewidth]{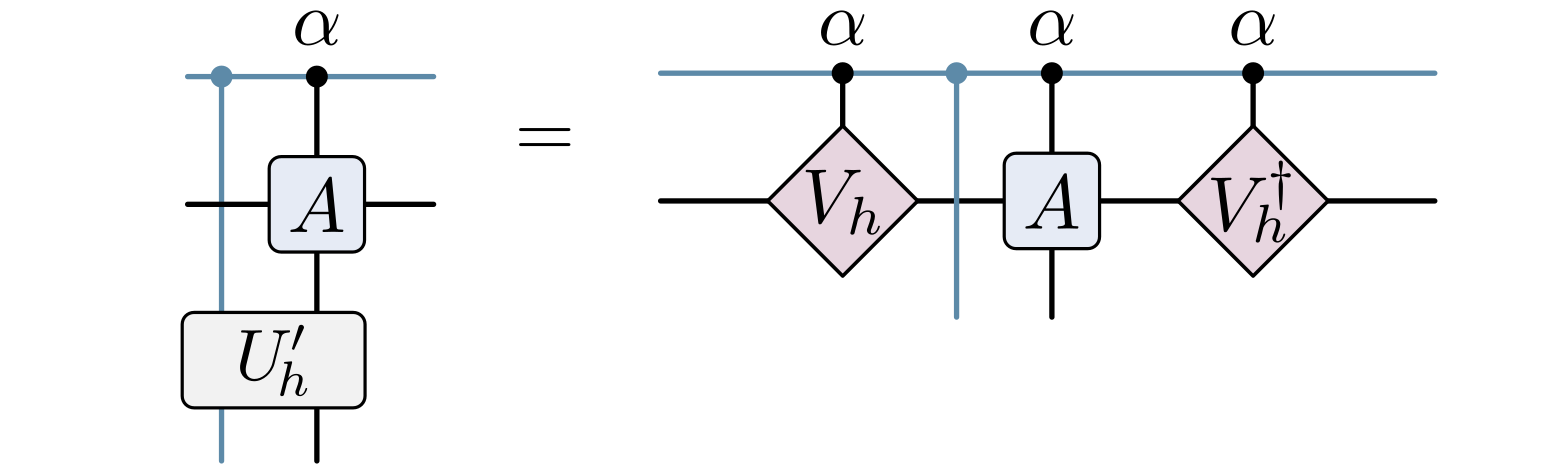},
\label{eq:cond_pushing_alpha}
\end{equation}

where we have again redefined $U_h'$ as
\begin{equation}
\includegraphics[width = 0.85\linewidth]{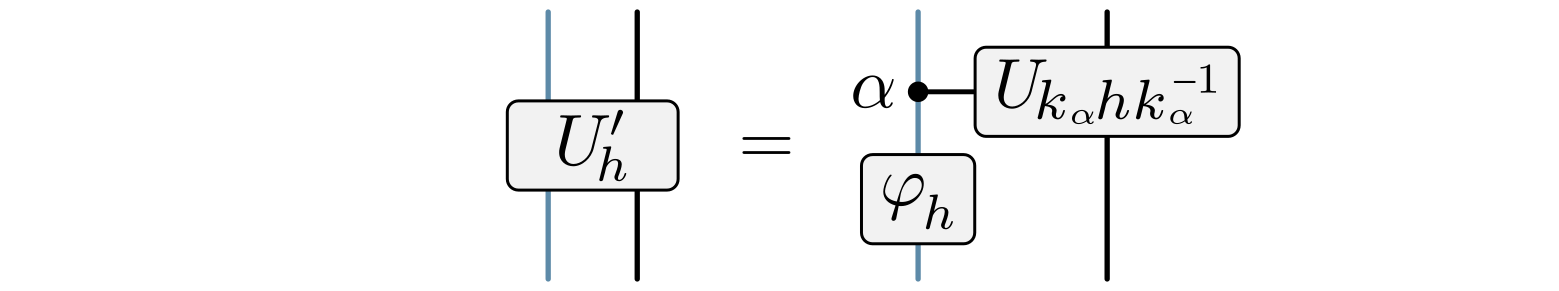},
\end{equation}
where we have additionally defined the phase matrix $\varphi_h = \bigoplus_\alpha e^{-i\phi_g^{\alpha}}$ with $g = k_{\alpha} h k_{\alpha}^{-1}$. Thus, we have arrived at the set of desired pushing relations, enabling the manipulation of projective representations of $H$ \emph{within} each block without incurring nontrivial operations on the block index. We note that for the special case where $G$ is an Abelian group, Eq.~(\ref{eq:cond_pushing_alpha}) simplifies significantly as $k_{\alpha} h k_{\alpha}^{-1} = h$ up to a possible phase. Consequently, $U_{k_{\alpha} h k_{\alpha}^{-1}} = U_h$, and it is unnecessary to condition the physical operations on the lifted block index. Similarly, further simplifications are gained if each block carries the same virtual representation of $H$, allowing us to replace the $\alpha$-conditioned virtual operations with unconditional ones. However, for generality we have here opted to present the most general result, suitable for both Abelian and non-Abelian groups, and furthermore allowing for different representations within each block.

\section{Further details on Protocol \ref{alg:nonnormal}}
In this Appendix, we expand upon several key details underlying Protocol~\ref{alg:nonnormal}. First, we clarify the sufficiency of considering the limit $\mu_\alpha\rightarrow 1$ for the more general preparation of non-normal MPS with arbitrary block amplitudes $\mu_{\alpha}$. Following this, we expand upon an important step in our preparation protocol where the block qudits are disentangled from the site qubits via measurement.

\subsection{Sufficiency of the limit $\mu_{\alpha}\rightarrow 1$}
First, let us recall that the goal of Protocol~\ref{alg:nonnormal} is to first prepare the state $\ket{\Psi}$, defined as
\begin{equation}
    \begin{split}
        \ket{\Psi} &= \sum_{\ell r}\sum_{\vec{m}}\bra{\ell} A^{m_1}A^{m_2}\ldots A^{m_N}\ket{r}\ket{\vec{m}} \otimes\ket{\ell r},\\
    \end{split}
    \label{app:eq:mps_EBC}
\end{equation}
up to an overall normalization constant. Following this, the subsystems indexed by $\ell$ and $r$ are measured, collapsing the state onto an MPS with particular (but probabilistically determined) boundary conditions. For concreteness, let us imagine that our goal is to measure the edge qudits in the state $\ket{X} = \sum_{ij} (X^*)_{ji}\ket{ij}$, such that the final prepared MPS is
\begin{equation}
\ket{\Psi'} = \sum_{\vec{m}}\tr{A^{m_1}A^{m_2}\ldots A^{m_N}X}\ket{\vec{m}}
\label{app:eq:mps_X}
\end{equation}

For a non-normal MPS, the matrices $A^m$ can be decomposed as 
\begin{equation}
    A^m = \bigoplus_{\alpha=0}^{K-1} \mu_{\alpha} A^m_{\alpha},
\end{equation}
where the intra-block tensors are in left-canonical form, $\sum_m A_\alpha^{m \dagger} A_\alpha^m = \mathbb{1}$. Rewriting Eq.~\eqref{app:eq:mps_X} in this light (and leaving the bounds on $\alpha$ implied to simplify notation), we find
\begin{equation}
    \begin{split}
        \ket{\Psi'}= \sum_{\vec{m}}\tr{\bigoplus_{\alpha} \mu_{\alpha}^N A_{\alpha}^{m_1}A_{\alpha}^{m_2}\ldots A_{\alpha}^{m_N} X}.
    \end{split}
\end{equation}
Noting that only the block-diagonal entries of $X$ contribute to the trace, we can without loss of generality assume $X=\bigoplus_{\alpha} X_{\alpha}$ such that it has the same block-diagonal structure as the matrices $A^m$. Furthermore, let us define the modified boundary matrix
\begin{equation}
    \widetilde{X} = \frac{1}{\eta}\bigoplus_{\alpha} \mu_\alpha^N X_{\alpha},
\end{equation}
with $\eta$ a normalization factor that ensures $\tr{\widetilde{X}^\dagger \widetilde{X}} = \mathbb{1}$. Then our target state can be re-expressed as
\begin{equation}
        \begin{split}  \ket{\Psi'}&=\sum_{\vec{m}}\tr{\bigoplus_{\alpha}A_{\alpha}^{m_1}A_{\alpha}^{m_2}\ldots A_{\alpha}^{m_N} \widetilde{X}} \\
        &=\sum_{\vec{m}}\tr{ \widetilde{A}^{m_1}\widetilde{A}^{m_2}\ldots \widetilde{A}^{m_N} \widetilde{X}}, \\
    \end{split}
\end{equation}
where we have defined the modified matrices $\widetilde{A}^m = \bigoplus_{\alpha} A^m_{\alpha}$. But these correspond exactly to the original matrices $A^m$ in the limit $\mu_{\alpha}\rightarrow 1$. Consequently, we can either prepare the amplitudes $\mu_{\alpha}\neq 1$ at the outset (i.e., by incorporating them into the initial GHZ state preparation) or, alternatively, we can without loss of generality prepare a variant of Eq.~\eqref{app:eq:mps_EBC} parameterized by the modified tensor $\widetilde{A}$ and incorporate the amplitudes $\mu_{\alpha}^N$ in the final measurement step by projecting onto the modified boundary condition $\widetilde{X}$ (with probability $p\simeq 1/KD^2$ in the large $N$ limit, where $D$ is the maximum dimension of the individual blocks).

\subsection{Disentangling the block qudits}

Next, we expand upon step (5) of Protocol \ref{alg:nonnormal}, where ``lifted'' ancillary block qudits are disentangled from the target state via measurements. After step (4), the system is in the state
\begin{equation}
\ket{\Psi_4} = \sum_{ij}\sum_{\alpha}\ket{\Psi^{ij}_\alpha}\otimes\ket{\alpha}^{\otimes N} \ket{L_{i\alpha}}\ket{R_{j\alpha}},
\end{equation}
where we have adopted the shorthand $\ket{L_{i\alpha}} = \ket{i}\otimes \ket{\alpha}$ for the composite left dangling qudit (composed of both a $D$-dimensional bond and a $K$-dimensional block qudit) and, likewise, $\ket{R_{j\alpha}} = \ket{j}\otimes \ket{\alpha}$ for the composite right dangling qudit, and have furthermore defined
\begin{equation}
    \ket{\Psi^{ij}_\alpha} = \sum_{\vec{m}}\bra{i}A_{\alpha}^{m_1}A_{\alpha}^{m_2}\ldots A_{\alpha}^{m_N}\ket{j}\ket{\vec{m}}.
\end{equation}

Our present goal is to disentangle the $N$ ``bulk'' block qudits (not to be confused with the ``edge'' block qudits that partially compose $\ket{L_{i\alpha}}$ and $\ket{R_{j\alpha}}$), leaving the remainder of the system in the form of Eq.~(\ref{app:eq:mps_EBC}). To that end, we first apply a Walsh-Hadamard gate $W$ to each $K$-dimensional (bulk) block qudit,
\begin{equation}
    W = \frac{1}{\sqrt{K}}\left[
    \begin{matrix} 
    1 & 1 & 1 & \cdots & 1 \\ 
    1 & \omega^{K-1} & \omega^{2(K-1)} & \cdots & \omega^{(K-1)^2} \\
    1 & \omega^{K-2} & \omega^{2(K-2)} & \cdots & \omega^{(K-1)(K-2)} \\
    \vdots & \vdots & \vdots & \ddots & \vdots \\
    1 & \omega & \omega^2 & \cdots & \omega^{K-1}
    \end{matrix}\right]
\end{equation}
where $\omega=e^{2\pi i/K}$, and subsequently measure each qudit in the computational basis. Regardless of the measurement outcome, this collapses $\ket{\Psi_4}$ into the form
\begin{equation}
\ket{\Psi_5} = \sum_{ij}\sum_{\alpha}e^{i\phi_{\alpha,\vec{k}}}\ket{\Psi_{\alpha}}\otimes \ket{L_{i\alpha}}\ket{R_{j\alpha}},
\end{equation}
where we have discarded the measured bond qudits, and have furthermore made the definition
\begin{equation}
\phi_{\alpha,\vec{k}} = -\frac{2\pi \alpha}{K}\sum_{j=0}^{N-1} k_j,
\end{equation}
where each $k_j\in\{0,1,\ldots K-1\}$ labels the measurement outcome for the $j$th block qudit. These phases are known from the measurement result, and as such, can be removed using a (feedforward) diagonal phase gate on the remaining edge block qubits. Alternatively, one can adaptively incorporate this phase into the basis for the final measurement of the composite dangling edge qudits.

Applying the appropriate unitary $U_{\varphi}^{(\vec{k})\dagger}$ to remove these phases, we arrive at the desired outcome:
\begin{equation}
    \begin{split}
        \ket{\Psi_5'} &= U_{\varphi}^{(\vec{k})\dagger}\ket{\Psi_5} \\
        &=\sum_{ij}\sum_{\alpha}\ket{\Psi^{ij}_\alpha} \otimes\ket{L_{i\alpha}}\ket{R_{j\alpha}}\\
        &=\sum_{\ell r}\sum_{\vec{m}}\bra{\ell}A^{m_1}A^{m_2}\ldots A^{m_N}\ket{r}\ket{\vec{m}}\otimes\ket{\ell r} \\
    \end{split}
\end{equation}
where we have defined composite indices $i,\alpha\rightarrow \ell$ and $j,\alpha\rightarrow r$, leveraging the fact that the matrices $A^m$ are block-diagonal. The only remaining step is to measure the composite dangling edge qudits to collapse the state onto one with particular boundary conditions.

\section{Reducing the post-selection overhead for open boundary conditions}
\label{app:postselection}
While both Protocol~\ref{alg:normal} and Protocol~\ref{alg:nonnormal} deterministically prepare the bulk of the target MPS, achieving particular boundary conditions requires a post-selection overhead of $\mathcal{O}(D^2)$. We stress that this overhead depends only on the bond dimension $D$, which we assume to be indendent of system size $N$ in this work. However, for small system systems and particular experimental platforms, this overhead may present an important trade-off with the linear- and log-depth unitary preparation schemes. In this Appendix, we discuss how one can reduce this overhead for the case of open boundary conditions.

To explain, let us for simplicity let us consider the case of normal MPS. For periodic boundary conditions, the $\mathcal{O}(D^2)$ post-selection overhead is seemingly unavoidable. This can be understood via Fig.~\ref{fig:f4} where, in the final stage, all defects have been corrected. To realize periodic boundary conditions, we perform a projective, generalized Bell basis measurement between the remaining bond qubits at the edge (either nonlocally, or using a distributed ancillary Bell pair). If this measurement results in a non-identity defect, then it is uncorrectable -- any operations on the physical qubits will merely push the defect cyclically around the state. Thus, we post-select on a defect-free final measurement outcome, which has a probability of success $p \sim 1/D^2$ under the assumption $N\gg \xi$, where $\xi$ is the correlation length of the target state. 

On the other hand, for open boundary conditions, we can reduce the post-selection overhead to $\mathcal{O}(D)$. In this case, the target state in Eq.~\eqref{eq:MPS} becomes
\begin{equation}
    \ket{\Psi} = \sum_{\vec{m}} \bra{L}A^{m_1} A^{m_2} \ldots A^{m_{N}}\ket{R}\ket{\vec{m}},
    \label{app:targetstate}
\end{equation}
where $\ket{L}$ and $\ket{R}$ are $D$-dimensional unit vectors. The procedure then can be simplified by, in Fig.~\ref{fig:f4}, replacing the right-most small MPS with one having a definite right virtual edge-state. Specifically, this amounts to replacing the right-most two-qudit ``Prep. $\ket{I}$'' gate with a single qudit gate that prepares the state $\ket{R}$. The rest of the procedure is identical, with the caveat that one must push all defects to the left (as there is no right dangling bond qudit). After the correction of defects, this yields the state,
\begin{equation}
    \ket{\Psi'} = \sum_{i}\sum_{\vec{m}}\bra{i}A^{m_1}A^{m_2}\ldots A^{m_N}\ket{R}\ket{\vec{m}}\otimes\ket{i}.
\end{equation}
Therefore, to realize the target state in Eq.~\eqref{app:targetstate}, one must post-select upon measuring the left dangling qudit (indexed by $i$) in the state $\ket{L}$. In the limit where $N\gg \xi$, this result has probability $p\sim 1/D$. It is unclear whether this can be further improved. We leave this as a direction for future work.

\section{More details on Table~\ref{tab:highersymm}}
\label{app:highersym_details}

Here, we expand on the defect bases denoted in Table~\ref{tab:highersymm}, providing an explicit representation for each of the indicated $SU(n)$, $SO(2\ell+1)$ and $Sp(2n)$ subgroups. The first two are straightforward: as discussed in the text surrounding Eq.~\eqref{eq:clockshift}, the clock and shift matrices generate a projective irrep of $\mathbb{Z}_n\times \mathbb{Z}_n\subset SU(n)$, typically referred to as the qudit Pauli basis. Separately, the set of $4^{\ell}$, weight-$\ell$ ($2\times 2$) Pauli operators forms a projective irrep of $(\mathbb{Z}_2\times \mathbb{Z}_2)^{\ell}\subset SO(2\ell+1)$. We note that measurement in this basis is particularly straightforward -- if each bond qudit is encoded using $\log_2(D)$ qubits, then fusion measurements can be carried out in a pairwise manner using the qubit Bell basis.

Due to the symplectic constraint on the correctable defects, the relevant irrep for the $Sp(2n)$-symmetric states is more subtle. First, we define
\begin{equation}
r=\left(\begin{matrix} Z & 0 \\ 0 & Z^*
\end{matrix}\right) \quad s=i\left(\begin{matrix} 0 & \mathbb{1} \\ \mathbb{1} & 0 \\
\end{matrix}\right),
\end{equation}
where $Z$ is the $n\times n$ clock matrix. Together, $r$ and $s$ generate a projective irrep of $D_n$, the dihedral group of order $2n$. Separately, we define the following two matrices,
\begin{equation}
a=i\left(\begin{matrix} \mathbb{1} & 0 \\ 0 & -\mathbb{1}
\end{matrix}\right) \quad b=\left(\begin{matrix} X & 0 \\ 0 & X \\
\end{matrix}\right),
\end{equation}
where $X$ and $\mathbb{1}$ are the $n\times n$ shift and identity matrices, respectively. Together, $a$ and $b$ generate a projective irrep of $\mathbb{Z}_2 \times \mathbb{Z}_n$. Taking the direct product between these two representations, the resulting matrices
\begin{equation}
    \begin{split}
        B_{i,j,k,\ell} = a^{i}b^{j}r^{k}s^{\ell}\in Sp(2n) \quad &i,\ell\in\{0,1\} \\ &j,k\in\{0,n-1\}
    \end{split}
\end{equation}
form an irrep of $\mathbb{Z}_2\times\mathbb{Z}_n\times D_n\subset Sp(2n)$, and thus provide a correctable defect basis for the preparation of the SPT-ordered MPS with $Sp(2n)$ symmetry in Ref.~\cite{Wang2017}. 

\section {Constructing MPS with on-site symmetry}\label{app:symconst}

In this appendix, we describe a method for constructing families of MPS with global on-site symmetry. We note that this problem has been studied previously~\cite{Sanz_2009}. Here, our goal is the following: given a group $G$ and a $D$-dimensional (projective) representation $V$, we would like to construct a parameterized family of MPS that satisfies Eq.~(\ref{eq:normalsymmetry}) (or, equivalently, Eq.~\ref{eq:nonnormalsymmetry} in the case of symmetry-broken non-normal MPS). Naturally, such a family of MPS would be preparable using Protocol~\ref{alg:normal}, independent of the bond dimension corresponding to the chosen representation. Likewise, the physical dimension of the MPS is constrained to a set of possible values determined by the choice of $V$. 

We begin by constructing a linear representation of $G$ from the projective representation $V$ by defining $\bar{V}$, where
\begin{equation}
    \bar{V}_g \equiv V_g\otimes V_g^*.
\end{equation}
Since $\bar{V}$ is a linear representation, it can be decomposed as a direct sum of irreducible representations (irreps) of $G$:
\begin{equation}\label{eq:usum}
    V_g\otimes V_g^*=\bigoplus_{J}\left(\bigoplus_{n=1}^{n_J}u_g^{J}\right),
\end{equation}
where $J$ labels the irrep of $G$ and $n_J$ is the number of times irrep $J$ appears in $\bar{V}$. In other words, $\bar{V}_g$ is a $D^2$-dimensional matrix that can be block-diagonalized via a unitary $W$:
\begin{equation}\label{eq:wuw}
\bar{V}_g = W \bar{U}_g W^\dagger,
 \end{equation}
where
\begin{equation}\label{eq:U}
\bar{U}_g=
\begin{pNiceMatrix}[margin]
\ddots\\
 & \Block[draw]{3-3}{} \\
 & & u_g^J&\\
 & & & &\Block[l]{1-1}{\scriptstyle d_J\times d_J}\\
 & & & & \ddots \\
\end{pNiceMatrix}.
\end{equation}
Here, $d_J$ is the dimension of irrep $J$. To construct the parameterized tensor $A$ that defines the (translationally-invariant) MPS, we require two elements: the unitary $W$ and an isometric operator $P$ that picks out one or more of the irreps in Eq.~(\ref{eq:U}) that determines the physical dimension. The unitary $W$ can be determined by solving the system of linear equations,
\begin{equation}
    \bar{V}_g W - W\bar{U}_g = 0, \quad \forall g\in G.
\end{equation}
Since there are $|G|$ linear equations and the $D^2$-dimensional unitary matrix $W$ has $D^4$ independent real parameters, the tensor $A$ (and, by extension, the family of symmetric MPS) will be parameterized by 
\[
\text{max}\left(0,D^4-|G|\right) \leq N \leq D^4
\] 
real variables.

After determining $W$, we next choose particular irrep blocks in $\bar{U}$. This fixes the physical dimension $d$ of the MPS, corresponding to the sum of the dimensions of the chosen irrep blocks. To extract the chosen blocks, we define an isometric operator $P$ that is a $D^2\times d$ matrix with ones along the diagonal of the chosen blocks and zeros otherwise. For example, to select just one irrep $J$, the appropriate isometry $P$ is a $D^2\times d_J$ matrix with ones along the diagonal of the block containing $u^{J}$ in $\bar{U}_g$. The operator $P$ will then obey
\begin{equation}
    \bar{U}_g P = P U_g, 
\end{equation}
where $U_g$ is the $d$-dimensional block-diagonal representation that was selected from $\bar{U}_g$. The tensor $A$ is then constructed by reshaping the $D^2\times d$ matrix,
\begin{equation}
    \bar{A} = WP,
    \label{eq:Abar=WP}
\end{equation}
where the $m$th column of $\bar{A}$ corresponds to the $D\times D$ matrix $A^m.$
One can verify that $\bar{A}$ satisfies the desired property:
\begin{align}
    \bar{V}(g)\bar{A} &= \left(W\bar{U}_gW^\dagger\right)\left(WP\right)\\
    &= W\bar{U}_gP \\
    &= WPU_g\\
    &= \bar{A}U_g,
\end{align}
which is equivalent to the symmetry condition of Eq.~(\ref{eq:normalsymmetry}). We now illustrate this practical construction method for several simple examples of on-site symmetry.

\subsection{Example: $\mathbb{Z}_{2}\times\mathbb{Z}_2$}
\label{app:ssec:ExamplesZ2xZ2}

Let us consider the 2-dimensional irreducible projective representation $V_g$ of $\mathbb{Z}_{2}\times\mathbb{Z}_2$, the Pauli matrices. For this representation, $V_e = I$ and $V_k=\sigma_{k}$. Because this group is Abelian, Eq.~(\ref{eq:wuw}) yields a diagonal matrix $\bar{U}_g$, with each entry corresponding to one of the four 1D representations of the group. Furthermore, we find
\begin{equation}
W = \frac{1}{\sqrt{2}}    
\begin{pNiceMatrix}
a & 0 &  c &  0\\
0 & b &  0 &  d\\
0 & b &  0 & -d\\
a & 0 & -c &  0\\
\end{pNiceMatrix},
\end{equation}
where $a,b,c,d\in U(1)$. We can then build an MPS of physical dimension $d=2, 3$, or 4 ($d=1$ is not interesting). Choosing the isometry,
\begin{equation}
    P =
\begin{pNiceMatrix}
0 & 0 \\
0 & 0 \\
1 & 0 \\
0 & 1 
\end{pNiceMatrix} ,
\end{equation}
the tensor $A$ resulting from Eq.~\eqref{eq:Abar=WP} corresponds to a $d=2$ MPS of the form
\begin{equation}
A =\frac{1}{\sqrt{2}}
\begin{pNiceMatrix}
c\ket{0} & d\ket{1}\\
-d\ket{1} &-c\ket{0}\\
\end{pNiceMatrix}.
\end{equation}
Here, we are adopting a shorthand notation where the above matrix corresponds to a pair of matrices of the form
\begin{equation}
    A^0 = \frac{1}{\sqrt{2}}\left(\begin{matrix}
        c & 0 \\ 0 & -c
    \end{matrix}\right)\quad 
    A^1 = \frac{1}{\sqrt{2}}\left(\begin{matrix}
        0 & d \\ -d & 0
    \end{matrix}\right).
\end{equation}
We will use this shorthand throughout the remainder of this Appendix. The above isometry is not the only option, however. We could instead choose, for example,
\begin{equation}
    P =
\begin{pNiceMatrix}
0 & 0 & 0 \\
1 & 0 & 0 \\
0 & 1 & 0 \\
0 & 0 & 1
\end{pNiceMatrix},
\end{equation}
which yields a $d=3$ MPS of the form 
\begin{equation}
A =\frac{1}{\sqrt{3}}   
\begin{pNiceMatrix}
c\ket{1} & b\ket{0}+d\ket{2}\\
b\ket{0}-d\ket{2} &-c\ket{1}\\
\end{pNiceMatrix}.
\end{equation}
This parameterized family of MPS includes the ground state of the AKLT model, corresponding to the selection $(b,c,d) = (i, -1, 1)$.
Finally, for the case $d=4$, we can see $P=\mathbb{1}_{4\times4}$. This yields the family
\begin{equation}
A =\frac{1}{2}   
\begin{pNiceMatrix}
a\ket{0}+c\ket{2} & b\ket{1}+d\ket{3}\\
b\ket{1}-d\ket{3} & a\ket{0}-c\ket{2}\\
\end{pNiceMatrix},
\end{equation}
which encompasses the cluster state with blocked pairs of sites.

\subsection{Example: $SU(2)$}
\label{app:ssec:ExamplesSU2}
Next, we turn to the case of $SU(2)$ symmetry. In this context, the MPS construction procedure is best understood using the language of spin angular momentum. Let $V_g$ be a spin-$S$ irreducible representation of $SU(2)$. Then $U_g$ will be an irrep of a spin $S'$ system, where $0 \leq S' \leq 2S$ from the addition of angular momentum for two spin $S$ systems. The unitary $W$ in Eq.~(\ref{eq:wuw}) becomes a matrix of Clebsch-Gordan coefficients. 
For example, we can let $V_g$ be the spin-$1/2$ representation:
\begin{equation}
    V_g = e^{ia\hat{n}\cdot\vec{\sigma}/2},
\end{equation}
where $a$ is a real parameter, $\hat{n}$ is a unit 3-vector, and $\vec{\sigma}$ is the Pauli 3-vector. Then Eq.~(\ref{eq:usum}) is equivalent to
\begin{equation}
    \mathbf{\frac{1}{2}}\otimes\mathbf{\frac{1}{2}} = \mathbf{0}\oplus \mathbf{1},
\end{equation}
with \begin{equation}
    W = 
\begin{pNiceMatrix}
\frac{1}{\sqrt{2}} & 0 & -\frac{1}{\sqrt{2}} & 0\\
0 &1 & 0 & 0\\
0 & 0 &0 & -1\\
\frac{1}{\sqrt{2}} & 0 & \frac{1}{\sqrt{2}} & 0\\
\end{pNiceMatrix}.
\end{equation}
The only non-trivial option is the spin-1 representation, which can be selected using the isometry,
\[
P =
\begin{pNiceMatrix}
0 & 0 & 0 \\
1 & 0 & 0 \\
0 & 1 & 0 \\
0 & 0 & 1
\end{pNiceMatrix}. 
\]
Together with $W$, this yields the MPS
\begin{equation}
A =\frac{1}{\sqrt{3}}   
\begin{pNiceMatrix}
-\frac{1}{\sqrt{2}}\ket{\bar{0}} & \ket{+}\\
-\ket{-}&\frac{1}{\sqrt{2}}\ket{\bar{0}}\\
\end{pNiceMatrix},
\end{equation}
which is exactly the spin-1 AKLT state expressed in the $S^z$ eigenbasis $\{\ket{+}, \ket{\bar{0}}, \ket{-}\}$. 

More generally, our method allows us to construct $SU(2)$-symmetric MPS of higher bond dimension $D$. As discussed in the main text, preparing such states generally requires that there exists a finite subgroup $H\subset SU(2)$ that has a (projective) irrep of dimension $D$. One can then narrow to a set of defects that form a representation of this subgroup, such that Theorem~\ref{thm:normal} is satisfied and the state can be prepared via Protocol~\ref{alg:normal}. Several examples of this strategy were discussed in the main text, including in Sections~\ref{sssec:AKLT} and ~\ref{sssec:highersymm}

In that vein, Table~\ref{tab:su2} shows finite subgroups of $SU(2)$, each corresponding to a different spin representation. MPS constructed from these spin representations can be prepared using constant-depth adaptive circuits by leveraging the above ``subgroup'' strategy.

\begin{table}[ht]
    \centering
    \begin{tabular}{|c|c|c|c|c|}
    \hline
         spin & irrep dimension & allowed physical spin& $H$  & $p\,$ \\
         \hline
         1/2  & 2 & 1 &$\mathbb{Z}_2\times\mathbb{Z}_2$ & 1 \\
         1    & 3 & 2, 1&$A_4$                            & 4 \\
         3/2  & 4 & 3, 2, 1&$2O$                             & 3 \\
         2    & 5 & 4, 3, 2, 1&$I$                              & 12 \\
         5/2  & 6 & 5, 4, 3, 2, 1&$2I$                             & 10 \\
         \hline
    \end{tabular}
    \caption{Table of spin representations $V_g$, its matrix dimension, and the finite subgroup with an irrep of the same dimension. $2O$ is the binary octahedral group, $I$ is the icosahedral group, and $2I$ is the binary icosahedral group. The right-most column $p$ gives the dimension of the ancilla qudit needed to construct the projective measurements. }
    \label{tab:su2}
\end{table}

\subsection{Example: $\mathbb{Z}_{2}$}
Finally, we turn to perhaps the simplest example of our construction -- the Abelian group $\mathbb{Z}_{2}$. Let us consider a $D=2$ dimensional unitary representation of this group. Without loss of generality, this can be chosen as $V_z = \sigma_{y}$. Then 
\begin{align}
    \bar{V}_z=V_z\otimes V_z &= 
\begin{pNiceMatrix}
 0 & 0 & 0 & 1\\
 0 & 0 &-1 & 0\\
 0 &-1 & 0 & 0\\
 1 & 0 & 0 & 0\\
\end{pNiceMatrix}.\\
& = W \begin{pNiceMatrix}
 1 & 0 & 0 & 0\\
 0 & 1 & 0 & 0\\
 0 & 0 & -1 & 0\\
0 & 0 & 0 & -1\\
\end{pNiceMatrix} 
W^\dagger
\end{align}
Here, 
\begin{equation}
W = 
\begin{pNiceMatrix}
 w_{1} & w_{2} &  w_{3} &  w_{4}\\
 w_{5} & w_{6} &  w_{7} &  w_{8}\\
-w_{5} &-w_{6} &  w_{7} &  w_{8}\\
 w_{1} & w_{2} & -w_{3} &  -w_{4}\\
\end{pNiceMatrix},
\end{equation}
and the $w_i$ parameters are further constrained from $WW^\dagger = W^\dagger W = I$. $\bar{V}$ is decomposed as two 1D irreps, each with multiplicity 2.
Now let us choose the first and third blocks of $\bar{U}_z$, corresponding to the following selection for the isometry: 
\begin{equation}
    P =
\begin{pNiceMatrix}
1 & 0 \\
0 & 0 \\
0 & 1 \\
0 & 0 
\end{pNiceMatrix}.
\end{equation}
This fixes our $d=2$ dimensional representation on the physical level, with $U_z = \sigma_{z}$. Together with $W$,  Eq.~\eqref{eq:Abar=WP} yields,
\begin{align}
A &= 
\begin{pNiceMatrix}
w_1\ket{0} + w_3\ket{1} & w_5\ket{0} + w_7\ket{1} \\
-w_5\ket{0} + w_7\ket{1} & w_1\ket{0} - w_3\ket{1} \\
\end{pNiceMatrix}\\
&= \frac{1}{\sqrt{2}}
\begin{pNiceMatrix}
a\ket{+} + b\ket{-}  & c\ket{+} + d\ket{-} \\
-d\ket{+} - c\ket{-}  & b\ket{+} +a\ket{-} \\
\end{pNiceMatrix}. 
\end{align}
where $a = w_1 + w_3$, $b=w_1 - w_3$, $c=w_5+w_7$, and $d=w_5-w_7$. This corresponds to a family of $D=2$ MPS with on-site $\mathbb{Z}_{2}$ symmetry. In particular, setting
\[
w_1 = w_3 = \frac{1}{\sqrt{2}}\frac{1}{\sqrt{1+|g|}}\\\qquad
w_5 = w_7 = -\frac{1}{\sqrt{2}}\sqrt{\frac{-g}{1+|g|}},
\] and making the basis transformation $\ket{+}\to\ket{0}$, $\ket{-}\to\ket{1}$, we find that this parameterized MPS is encompasses the $\mathbb{Z}_2$-symmetric family described in Section~\ref{sssec:Z2}. 

% \newpage

%\bibliographystyle{apsrev4-2}
\bibliography{references}

\end{document}